\documentclass[superscriptaddress,secnumarabic,twocolumn,
 amssymb,amsmath,nobibnotes,aps,prd,showkeys,showpacs,nofootinbib]{revtex4}
\usepackage{amsmath}
\usepackage{amssymb}
\usepackage{graphicx}
\usepackage{subfig}
\usepackage{color}
\definecolor{darkgreen}{rgb}{0.0, 0.65, 0.31}
\usepackage[normalem]{ulem}

\begin{document}

\title{Dynamics of classes of barotropic fluids in spatially curved FRW spacetimes}
\author{Morteza Kerachian}
\email{morteza.kerachian@gmail.com}
\affiliation{Institute of Theoretical Physics, Faculty of Mathematics and Physics,
Charles University, CZ-180 00 Prague, Czech Republic}
\author{Giovanni Acquaviva}
\email{gioacqua@gmail.com}
\affiliation{Institute of Theoretical Physics, Faculty of Mathematics and Physics,
Charles University, CZ-180 00 Prague, Czech Republic}
\author{Georgios Lukes-Gerakopoulos}
\email{gglukes@gmail.com}
\affiliation{Astronomical Institute of the Academy of Sciences of the Czech Republic,
Bo\v{c}n\'{i} II 1401/1a, CZ-141 00 Prague, Czech Republic}

\begin{abstract}
 In this article we perform dynamical analysis of a broad class of barotropic fluids in the spatially curved Friedmann-Robertson-Walker (FRW) spacetime background without considering the cosmological constant. The first part of our study concerns the dynamics of a fluid with an unspecified barotropic equation of state (EoS) having as the only assumption the non-negativity of the fluid's energy density. After defining a new set of dimensionless variables and a new evolution parameter, we introduce the function $\Gamma$ that encodes the EoS. In this general setup several features of the system are identified: critical points, invariant subsets and the characteristics of the function $\Gamma$, along with their cosmological interpretations.  The second part of our work provides two examples with specific $\Gamma$ functions. In the first example we provide a $\Gamma$ function and then we exhibit how it can be trimmed down to a specific class of EoS through physical arguments, while in the second example we discuss the quadratic EoS studied in Phys.Rev. D {\bf 74}, 023523 (2006) by comparing our approach with their analysis. 
\end{abstract}

\pacs{~~}
\maketitle

\section{Introduction}

For a Friedmann-Robertson-Walker (FRW) cosmology with only a fluid component, the Friedmann equation and the Raychaudhuri equation are respectively given by
\begin{align}
 H^2 + \frac{k}{a^2} =\frac{\epsilon}{3},\label{eq:fried}\\
 2\,\dot{H} +3\, H^2 + \frac{k}{a^2}= -P\, ,\label{eq:raych}
\end{align}
while the continuity equation for the energy density is
\begin{equation}\label{eq:state}
\dot{\epsilon} +3\, H (P+\epsilon)=0\, ,
\end{equation}
where $a$ is the scale factor, $k$ is the spatial curvature, $H=\frac{\dot{a}}{a}$ is the Hubble expansion rate, $\dot{~}$ denotes derivative with respect to the coordinate time, $\epsilon$ is the energy density and $P$ is the pressure of the barotropic fluid.
Once the EoS is given, {\it i.e.} $P=P(\epsilon)$, the system is closed and it is in principle possible to provide the scale factor as function of the energy density:
\begin{align} \label{eq:scaleenerg}
  \displaystyle  a=a_0 \exp{\left(-\frac{1}{3}\int_{\epsilon_0}^{\epsilon}\frac{d\epsilon}{\epsilon+P(\epsilon)}\right)}\, ,
\end{align}
where subscripts $0$ denote integration constants resulting from integrating evolution equations, like the Raychaudhuri and the continuity equations. In this way the problem is reduced to solving a single second order ordinary differential equation (ODE) and it seems trivial. However, the integral in Eq.~\eqref{eq:scaleenerg} is not necessarily easy to calculate and the resulting ODE might be difficult to treat. Of course, in order to address the problem in this manner, one has to know first the EoS.   

In our study we analyse the dynamics of barotropic fluids in spatially curved FRW without specifying the EoS. The only assumption is that $\epsilon \ge 0$, while we do not impose any restrictions on the pressure $P$. Allowing pressure to attain negative values allows us to describe cosmological models driven by one fluid with a phenomenological EoS which can end with an accelerated expansion. During the evolution of the models such a fluid can have a standard behavior, {\it i.e.} the speed of sound  $0 \leq c_s^2 \leq 1$, but also more exotic cases. Analysis of cosmological models including fluids with rather general EoS can be found in the literature.  In \cite{linder2009aetherizing,bielefeld2014dark} the authors study the general properties of barotropic EoS: although a linear relation between energy density and pressure is implied, the proportionality parameter has a generic dependence on the scale factor. In \cite{ivanov2018dynamical} a dynamical analysis of a generic real gas is performed in the parameter space spanned by the Hubble function $H$, the number density $n$ and the temperature $T$ of the gas.  A quite general equation of state has been considered in \cite{nojiri2005properties} with the aim of studying future cosmological singularities. On the other hand, attempts to determine the form of EoS from observations can be found, for example, in \cite{nakamura1999determining} (see also \cite{copeland2006dynamics} for a review).

A general functional form of dark energy has been considered in various settings: for instance, in \cite{Aldrovandi2005} the cosmological consequences of a time-dependent $\Lambda$ were discussed.  A varying cosmological constant can be interpreted as a particle creation process, which has been discussed for example in \cite{Supriya19}, while an attempt to reconstruct dark matter/dark energy interaction through particle creation has been presented in \cite{yang2019reconstructing}.

The analysis that we carry out in the present paper takes inspiration from dynamical analysis of scalar fields in cosmology: in the simplest models describing a scalar field with an unspecified potential, general features regarding the dynamics of the system can be inferred by inspecting the properties of the so-called ''tracker parameter", which depends on second derivatives of the potential (see \cite{steinhardt1999cosmological}).  Similarly, we will relate the global and asymptotic behaviour of entire classes of EoS to the properties of a function which depends on second derivatives of the pressure $P(\epsilon)$.  An analogous study has been carried out in \cite{awad2013fixed} for the generic functional form $P=P(H)$.

\section{The dynamical system}
\label{sec:sys}

In order to compose well-defined dimensionless variables which are valid for both $k>0$ and $\displaystyle k\leq 0$ one can introduce the normalization $D^2=\displaystyle H^2+|k|/a^2$. Thus, we construct the new dimensionless variables as follows
\begin{align}
 \Omega_{\epsilon} = \frac{\epsilon}{3\,D^2}\quad &,\quad \Omega_H = \frac{H}{D} \, ,\label{eq:secvar1}\\
 \Omega_{P} =\frac{P}{D^2}\quad &,\quad \Omega_{\partial P} = \frac{\partial P}{\partial \epsilon}\, ,\label{eq:cpvar} \\
 \Gamma = \frac{\partial^2P}{\partial\epsilon^2}\epsilon \, ,
 \label{eq:secvar2}
\end{align}
where
\begin{equation}
D^2=H^2+\frac{|k|}{a^2}\, .
\end{equation}
Note that the EoS is defined by the parameter $\Gamma$.  Since we allow $\partial P/\partial \epsilon$ to be negative, we relax its interpretation as speed of sound and instead of denoting it as $c_s^2$ we have renamed it $\Omega_{\partial P}$.

In order to recast the evolution equations as an autonomous system, we take derivatives of the dimensionless variables with respect to the evolution parameter $\tau$, related to the cosmic time by $d\tau = D\, dt$.  This time parameter is well-defined throughout the whole cosmic evolution, in particular during possible recollapsing or bouncing scenarios where $H=0$.  The resulting system is given by
\begin{align}
 \Omega_{\epsilon}' &= -\Omega_H \left[\Omega_p+\Omega_{\epsilon} \left(3+2\left(\frac{\dot{H}}{D^2} + \Omega_H^2 - 1\right) \right) \right]\label{eq:omegaep} \, ,\\
 \Omega_H' &= \left( 1 - \Omega_H^2\right)\, \left( \frac{\dot{H}}{D^2} + \Omega_H^2\right) \, , \label{eq:omegahp}\\
 \Omega_P' &= -\Omega_H \left[ 3 \Omega_{\partial P}\left( \Omega_P+3 \Omega_{\epsilon} \right)  +2 \Omega_P\left( \frac{\dot{H}}{D^2} + \Omega_H^2 - 1\right)\right] \, ,
 \label{eq:omegap}\\
  \Omega_{\partial P}' & =-\Omega_H\left(\frac{\Omega_P}{\Omega_{\epsilon}}+3\right)\,\Gamma\label{eq:Csp} \, .
\end{align}

%\gl{Before submitting check whether we have used the equation below.}
%A useful relation is the time evolution of $D$ in terms of the dimensionless variables:
%\begin{equation}\label{ddot}
% \frac{\dot{D}}{D^2} = \Omega_H\, \left( \frac{\dot{H}}{D^2} + \Omega_H^2 - 1 \right)\, .
%\end{equation}
%Note that here $\Omega_P$ is not an independent dimensionless variable. One can always write $\Omega_P=f(\Omega_{\epsilon},\Omega_{\partial P})$. \textcolor{red}{GL:Do you mean in general or in special cases?} \textcolor{blue}{I think i should be possible to rewrite $\Omega_P=f(\Omega_{\epsilon},\Omega_{\partial P})$ or at least as a function of $\Omega_{\epsilon}$.}

\subsection{Positive curvature}

When $k>0$, the Friedmann equation can be expressed in terms of the variables Eqs.~\eqref{eq:secvar1}-\eqref{eq:secvar2} in the following form:
\begin{equation}
 \Omega_{\epsilon}=1\label{eq:fried_pos} \, .
 \end{equation}
From Raychaudhuri equation we get
\begin{equation} \label{eq:DeRaychPos}
\frac{\dot{H}}{D^2}=-\frac{1}{2}\left(\Omega_P+1\right)-\Omega_H^2 \, .
\end{equation}

\subsection{Non-positive curvature}\label{sec:nprych}

Applying the same definitions given by Eqs.~\eqref{eq:secvar1}-\eqref{eq:secvar2} to the case of non-positive spatial curvature $k\leq0$, one can reexpress the Friedmann constraint~\eqref{eq:fried} as
\begin{equation}\label{eq:fried_neg}
    \Omega_{\epsilon}= 2\, \Omega_H^2-1,
\end{equation}
and the Raychaudhuri equation~\eqref{eq:raych} as
\begin{equation}\label{eq:DeRaychNPos}
    \frac{\dot{H}}{D^2}=-\frac{1}{2}\left(\Omega_P+1\right)+\left(1-2\Omega_H^2\right).
\end{equation}
Since by definition $\Omega_H^2\le 1$ and by assumption $\epsilon \ge 0$, Eq.~\eqref{eq:fried_neg} implies that $0\le \Omega_\epsilon \le 1$ and $\frac{1}{2}\leq\Omega_H^2\leq1$.  However, as we will see, the system of evolution equations does not include automatically the requirement of positivity of energy and the trajectories might cross towards the negative energy regions: hence, we will have to select by hand the parts of variable space that we consider physically plausible.

\section{Critical points and their interpretation}\label{sec:CritPoi_ss}

%\begin{table*}
%  \caption{The critical elements of the system and their stability.  
%  }
%\begin{tabular}{c | c c c c | c c c c}\label{tab:CritPoints2}
% & $\Omega_{\epsilon}$ &  $\Omega_H$ & $\Omega_P$  &  Curvature & $q$ & stability\\
%\hline
% & & & &  & & & & \\
%  $A_+$ & $1$ & $1$ & $-3$  & flat & $-1$ &  source\\
%  $A_-$ & $1$ & $-1$ & $-3$   & flat & $-1$ &  sink\\
%  $B$ & $1$  & $0$ & $-1$   & positive & 0 & saddle\\
%\end{tabular}
%\end{table*}

The critical elements of the system are those values of the variables such that $\mathbf{\Omega}'=0$. Once the critical points are found, one can provide a cosmological interpretation in terms of the {\it deceleration parameter}
\begin{align}
 q &=-1-\frac{\dot{H}}{H^2}\nonumber\\
   &=-1-\Omega_H^{-2}\, \frac{\dot{H}}{D^2}\, ,
\end{align}
where we used the definition of $\Omega_H$.  From the Raychaudhuri Eq.~\eqref{eq:DeRaychPos} for positive curvature we see that, in order to have accelerated expansion, {\it i.e.} $q<0$, one needs $\Omega_P<-1$.  For the negative curvature case, by using Eq.~\eqref{eq:DeRaychNPos}, one has instead $\Omega_P<1-2\, \Omega_H^2=-\Omega_\epsilon$.  Thus, of course, having negative pressure is a necessary but not sufficient condition for accelerated expansion.

\subsection{Two de Sitter critical lines}\label{sec:CPA2}

The system presents two critical lines with a de Sitter behavior, with coordinates $\{\Omega_{\epsilon},\Omega_{H},\Omega_{P},\Omega_{\partial P}\}=\{1,\pm 1,-3, \forall \}$. Note that since $\Gamma$ has not yet been defined, these critical elements are independent of the EoS of the fluid. Specific choices of EoS can provide certain values for $\Omega_{\partial P}$, as we will show later on.  

Taking into account the definitions of $\Omega_{\epsilon}$ and $\Omega_{P}$, for both of these lines one could claim that $P=-\epsilon$, which would imply that $\Omega_{\partial P}=-1$. However, this is not the case, since at this point the EoS is still kept unspecified. Actually, as we will see in Sec.~\ref{sec:examples}, once an EoS is specified this critical point corresponds to the intersection between the function $P=P(\epsilon)$ and the $P=-\epsilon$ line.

The line with $\Omega_H=1$ (called $A_+$) has the typical cosmological constant behaviour given by $q=-1$.  The corresponding eigenvalues are
\begin{equation} \label{eq:eigap}
\{\lambda^{A_{+}}_{i} \}=\{-2 , 0 ,-3\left(1+\Omega_{\partial P}\right)  \}.
\end{equation}
 Thus, Eq.~\eqref{eq:eigap} implies that for $\Omega_{\partial P} < -1$ critical points along the line $A_+$ are saddles. If $\Omega_{\partial P} \geq -1$ the center manifold theorem does not provide the stability, thus we will discuss it through numerical examples for specific $\Gamma$.  

The line with $\Omega_{H}=-1$ (called $A_{-}$) describes an exponentially shrinking universe with $q=-1$. The eigenvalues in this case are
\begin{equation}\label{eq:eigam}
\{\lambda^{A_{-}}_{i} \}=\{2 , 0 ,3\left(1+\Omega_{\partial P}\right)  \},
\end{equation}
 Eq.~\eqref{eq:eigam} implies that $A_{-}$ is saddle for $\Omega_{\partial P} < -1$. Again, for $\Omega_{\partial P} \geq -1$ we will use numerical examples to discuss the stability for specific $\Gamma$.

\subsection{Static universe critical line}\label{sec:CPB2}

For positive curvature, the coordinates $\{\Omega_{\epsilon},\Omega_{H},\Omega_{P},\Omega_{\partial P}\}=\{1,0,-1,\forall \}$ correspond to a critical line (called $B$) describing a static universe, i.e $a = \textrm{const.}$. These points have eigenvalues
\begin{equation}\label{eq:eigb}
\{\lambda^{B}_{i} \}=\{0, -\sqrt{1+3 \Omega_{\partial P}} ,\sqrt{1+3 \Omega_{\partial P}} \}.
\end{equation}
 Regarding the stability, as long as $1+3 \Omega_{\partial P} > 0$, $B$ is saddle; for $1+3 \Omega_{\partial P} < 0$ it is center; for $\Omega_{\partial P}=-1/3$ the corresponding point is degenerate and all eigenvalues are zero, hence the center manifold theory cannot say anything about its stability: however, from a numerical inspection we find that in this case the point is marginally unstable.\footnote{A critical line (denoted by $\Bar{B}$) corresponding to a static universe exist also for the case of negative curvature. These points are located at $\{\Omega_{\epsilon},\Omega_{H},\Omega_{P},\Omega_{\partial P}\}=\{-1,0,1,\forall \}$. This location, however, lies at $\Omega_\epsilon<0$, which, as discussed in Sec.~\ref{sec:nprych}, is excluded from our study.}

\section{General features of $\Gamma$: invariant subsets and critical points}
\label{sec:roottrack}
Since we have assumed that $P=P(\epsilon)$, the definition of $\Gamma$ Eq.~\eqref{eq:secvar2} implies that $\Gamma=\Gamma(\epsilon)$. But the energy density is not a dimensionless parameter of the system and since $\Gamma$ should not depend on the geometry, the only valid option for its functional form is $\Gamma=\Gamma\left(\Omega_{\partial P},\frac{\Omega_P}{\Omega_\epsilon}\right)$. In some cases (see Appendix~\ref{sec:gammalinratio}) one can express $\frac{\Omega_P}{\Omega_\epsilon}$ as a function of $\Omega_{\partial P}$ and hence $\Gamma=\Gamma(\Omega_{\partial P})$ or vice versa. 

Eq.~\eqref{eq:Csp} of our dynamical system has been derived by combining the barotropicity of the effective fluid with the continuity equation \eqref{eq:state}, namely
\begin{align}
 \dot{\Omega}_{\partial P} &= \frac{\partial \Omega_{\partial P}}{\partial \epsilon}\, \dot{\epsilon}\nonumber\\
 &= -3\, H\, \left( 1+\frac{P}{\epsilon} \right)\, \left[\frac{\partial \Omega_{\partial P}}{\partial \epsilon}\, \epsilon \right]\, ,\label{eq:cs2dot}
\end{align}
where the last square bracket defines the parameter $\Gamma$.  Eq.~\eqref{eq:cs2dot} is independent of our choice of dimensionless variables and it clearly indicates that any root of $\Gamma$ will be a stationary point in time for $\Omega_{\partial P}$: this can happen either when $\epsilon=0$ or whenever $\Omega_{\partial P}$ has an extremum with respect to the energy density, {\it i.e.} whenever $P(\epsilon)$ has an inflection point.  Physically we expect $\epsilon=0$ only asymptotically, either past or future depending whether the model is collapsing or expanding; the second case instead can happen for some finite energy density and at a finite time. Hence, any inflection point of the EoS for $\epsilon>0$ will create an invariant subset in the parameter space of the dynamical system in the case that $\Gamma=\Gamma(\Omega_{\partial P})$.

By choosing appropriately the form of the function $\Gamma$ we can impose physically meaningful constraints on the evolution of the system.  For instance, being $\Omega_{\partial P}$ a dynamical variable, one could require the causality condition $\Omega_{\partial P}\leq1$ by imposing that $\Gamma(\Omega_{\partial P}=1)=0$ (see sec.~\ref{sec:causal}).  Such condition cannot be imposed at the level of equation of state, since this  would not be part of the evolution equations of the system and it wouldn't stop the dynamical system from crossing the value $\Omega_{\partial P}=1$.  Something analogous can happen for any other condition which is not imposed at the level of evolution equations (see {\it e.g.} Sec.~\ref{sec:nproots}).

Critical lines $A_\pm$ and $B$ are independent of the EoS since they exist for any $\displaystyle\Omega_{\partial P}$. However, once the function $\Gamma$ is chosen, its roots $\displaystyle \Tilde{\Omega}_{\partial P}$ introduce invariant subsets lying on $\{\Omega_{H},\Omega_{P}\}$ planes and critical points lying in these planes\footnote{It is not necessary to analyse the roots of $\Gamma$ with respect to the combination $\Omega_P/\Omega_\epsilon$, because having a constant ratio $\Omega_P/\Omega_\epsilon$ is equivalent to having a constant $\Omega_{\partial P}$ and hence the critical elements in the two cases can be related to each other.}. These critical points are located 
%\mk{saying inside the invariant subset is true ? } at  $\displaystyle \{\Omega_{\epsilon},\Omega_{H},\Omega_{P},\Omega_{\partial P}\}=\{1,\pm 1,3\, \Tilde{\Omega}_{\partial P},\Tilde{\Omega}_{\partial P}\}$.
at $\displaystyle \{\Omega_{H},\Omega_{P}\}=\{\pm 1,3\, \Tilde{\Omega}_{\partial P}\}$, hence for each root of $\Gamma$ there will be a pair of critical points $C_\pm$. Moreover, any new invariant subset might intersect the critical lines: we denote the resulting critical points with the same name as the respective critical lines throughout the text.

The scale factor in $C_+$ grows as $\displaystyle a \sim (t-t_0)^{\frac{2}{3\,(\Tilde{\Omega}_{\partial P}+1)}} $, while in $C_-$ the scale factor decreases as $\displaystyle a \sim (t_0-t)^{\frac{2}{3\,(\Tilde{\Omega}_{\partial P}+1)}} $. Moreover, the deceleration parameter at these points is $  q=\frac{1}{2}(3\,\Tilde{\Omega}_{\partial P}+1)$. Therefore, according to this parameter these critical points describe an accelerated phase if $ \Tilde{\Omega}_{\partial P}< -\frac{1}{3}$ and a decelerated phase if $ \Tilde{\Omega}_{\partial P}> -\frac{1}{3}$.

On the invariant subset $\{\Omega_H,\Omega_P\}$ point $C_+$ has  eigenvalues 
\begin{equation}
    \{\lambda^{C_+}_{i} \}=\{3\,(1+\Tilde{\Omega}_{\partial P}),1+3\,\Tilde{\Omega}_{\partial P} \},\label{eq:eigcp}
\end{equation}{}
and $C_-$ has 
\begin{equation}
    \{\lambda^{C_-}_{i} \}=\{-3\,(1+\Tilde{\Omega}_{\partial P}),-(1+3\,\Tilde{\Omega}_{\partial P} )\}.\label{eq:eigcm}
\end{equation}
From Eqs.~\eqref{eq:eigcp} and~\eqref{eq:eigcm} we see that for $-\frac{1}{3}<\Tilde{\Omega}_{\partial P}$ point $C_+$ ($C_-$) represent a source (sink). For the case $-1< \Tilde{\Omega}_{\partial P}<- \frac{1}{3}$ instead $C_\pm$ are saddle. Finally, for $\Tilde{\Omega}_{\partial P}<-1$ point $C_+$ ($C_-$) is a sink (source).
 
Since the stability of $A_\pm$, $B$ (see Sec.~\ref{sec:CritPoi_ss}) and $C_\pm$ depends on the value of $\Tilde{\Omega}_{\partial P}$, we split our analysis into the following three ranges
\begin{itemize}
    \item $-\frac{1}{3}<\Tilde{\Omega}_{\partial P}$,
    \item $-1< \Tilde{\Omega}_{\partial P}<- \frac{1}{3}$,
    \item $\Tilde{\Omega}_{\partial P}<-1$.
\end{itemize}
The behaviour of these cases will be depicted for positive and non-positive curvatures, choosing one representative value of $\Tilde{\Omega}_{\partial P}$ for each range, noting that the topology of the trajectories is independent of the specific value inside each range. Assuming further that the function $\Gamma$ has only one root, there is only one pair of critical points $C_\pm$. 

\subsection{Positive curvature}
\label{sec:rootpos}

\begin{figure}[htp]
\begin{center}
\captionsetup[subfloat]{position=top} 
 {\subfloat[$\Tilde{\Omega}_{\partial P}=1$]{\includegraphics[width=0.38\textwidth]{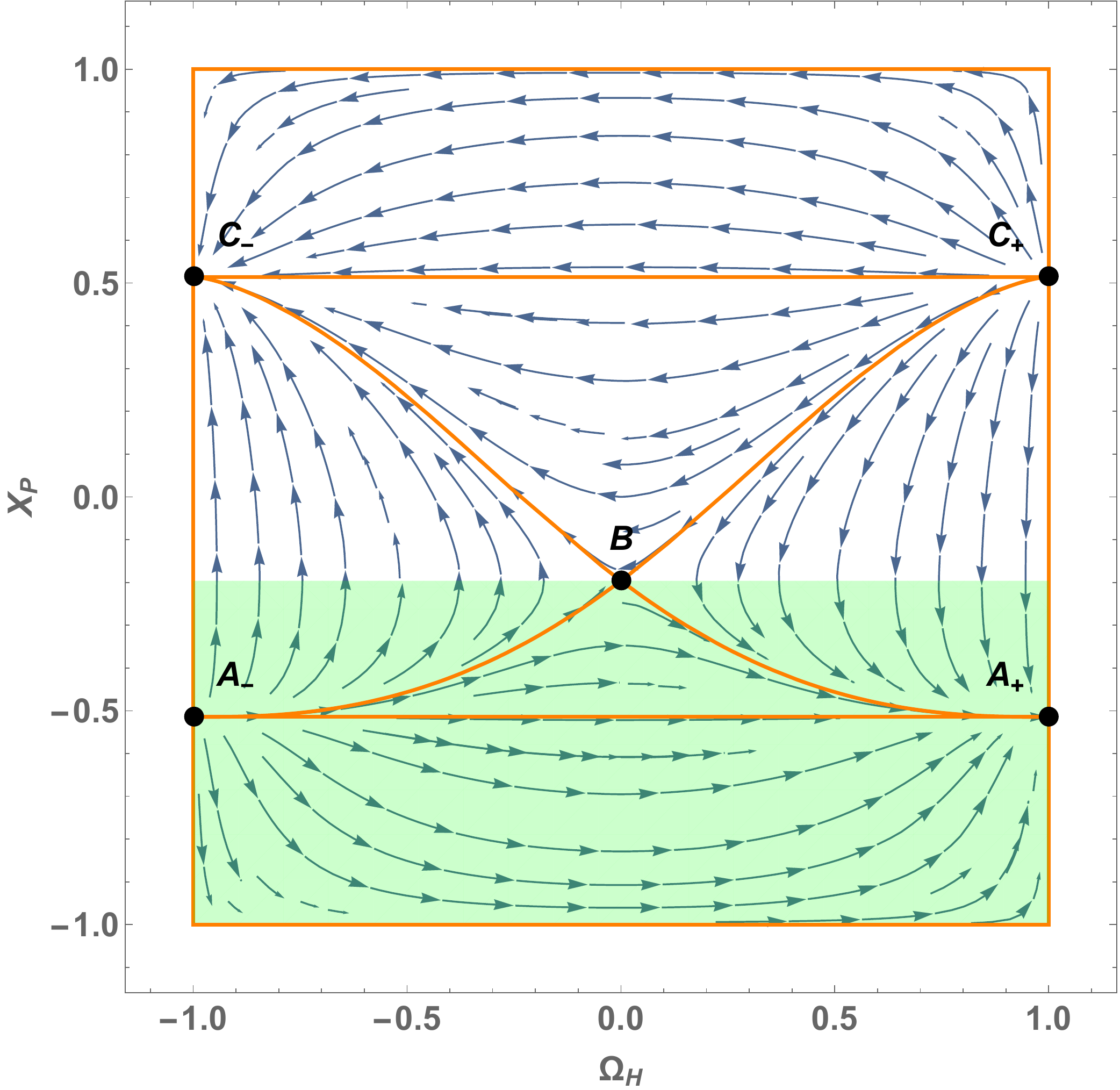}\label{fig:gen1}}}
{  \subfloat[$\Tilde{\Omega}_{\partial P}=-\frac{1}{2}$]{\includegraphics[width=0.38\textwidth]{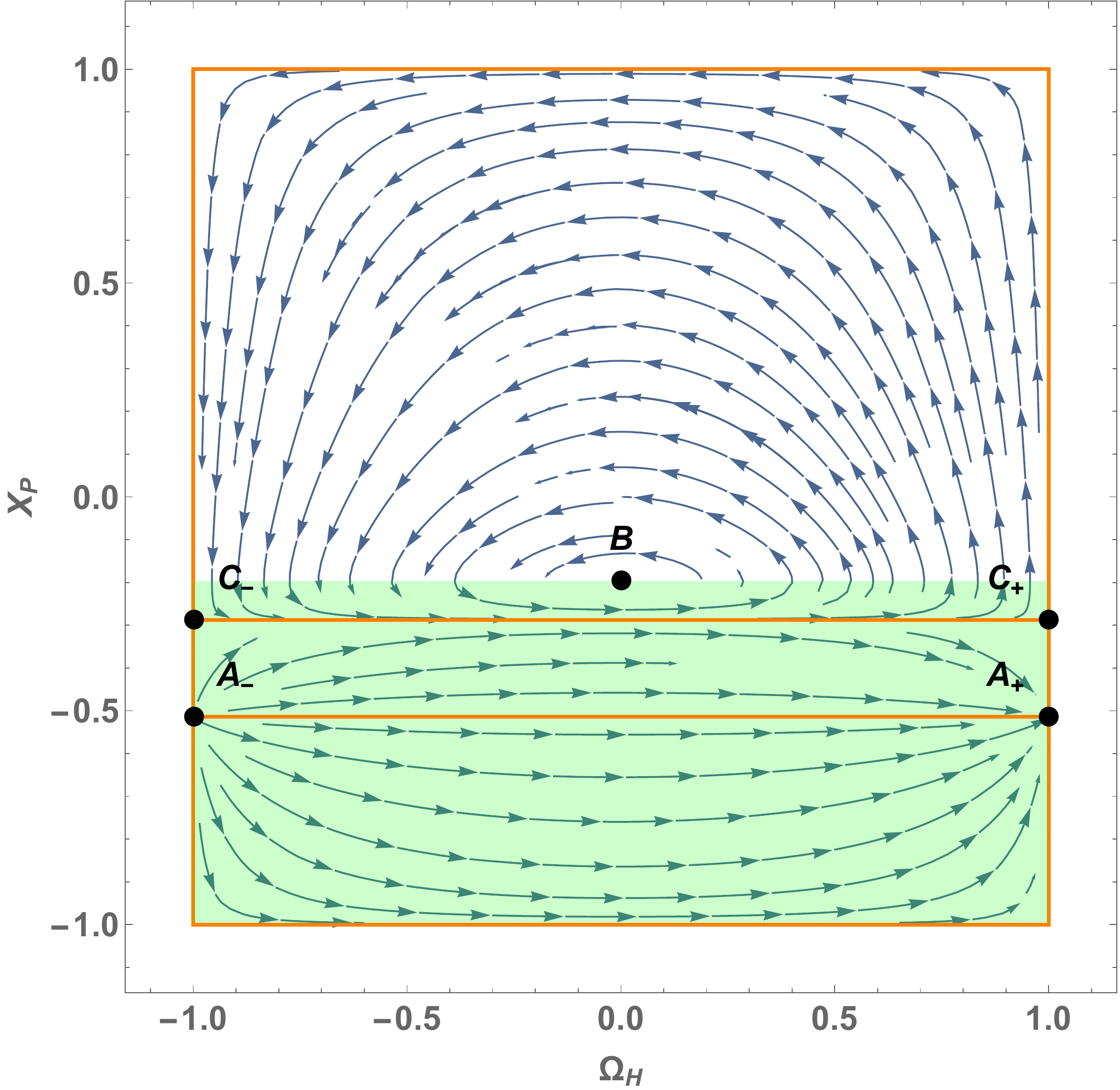}\label{fig:gen2}}}
{  \subfloat[$\Tilde{\Omega}_{\partial P}=-2$]{\includegraphics[width=0.38\textwidth]{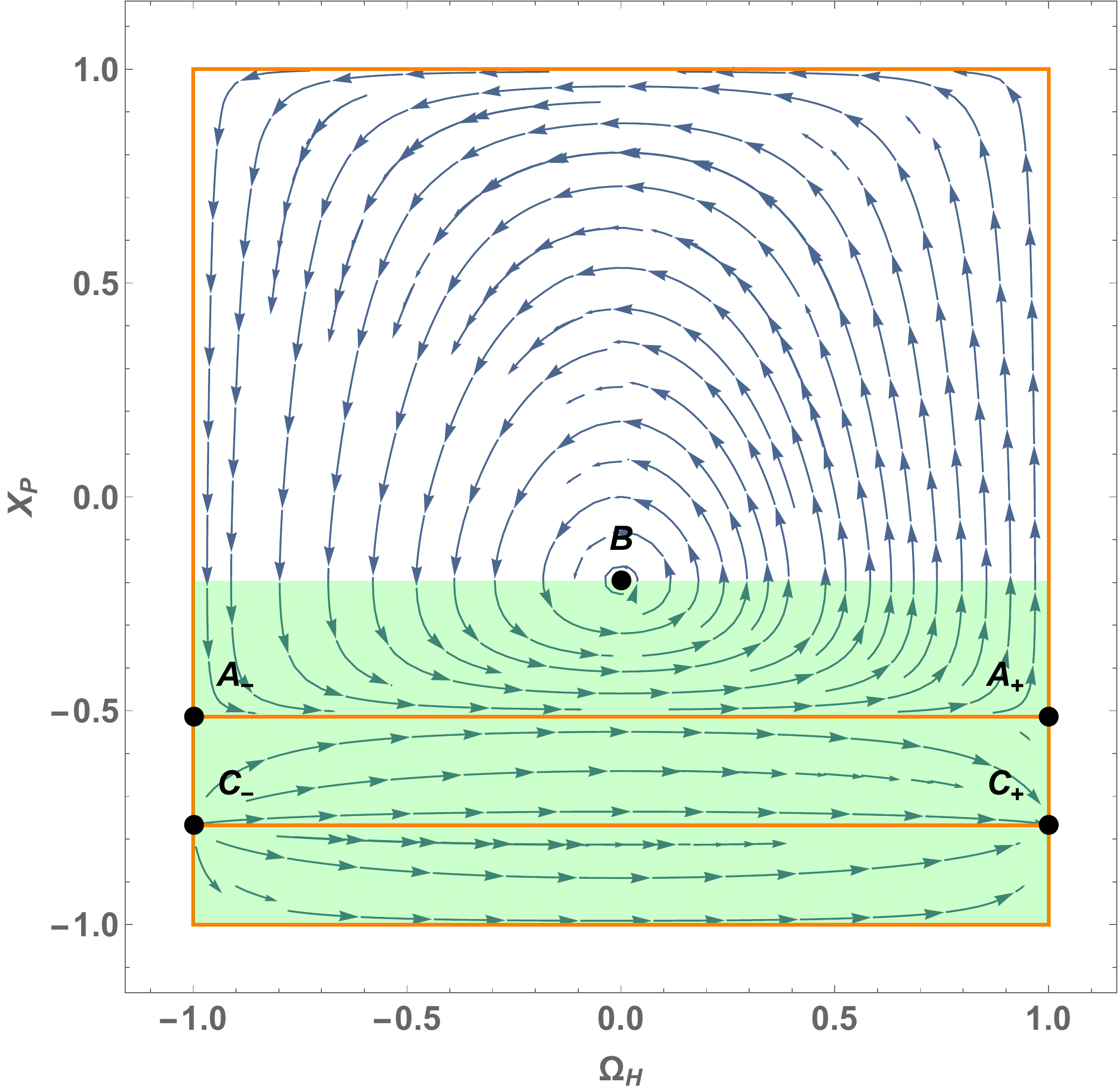}\label{fig:gen3}}}   
\end{center}
    \caption{Invariant subsets for positive spatial curvature and $\zeta=0.2$ plotted for three representative values of $\Tilde{\Omega}_{\partial P}$ in the ranges given in Sec.~\ref{sec:rootpos}. The orange thick lines are the separatrices of the system and the green shaded regions denote the part of the variable space where the universe is accelerating.}\label{fig:genp}
\end{figure}

For positive curvature the system has invariant subsets located at $\Omega_P=-3$ and $\Omega_P=3\,\Tilde{\Omega}_{\partial P}$. In order to study the behaviour at $\Omega_P=\pm \infty$ one can compactify $\Omega_P$ by using the transformation
\begin{equation}\label{eq:xp}
    X_P=\frac{\zeta\Omega_P}{\sqrt{1+\zeta^2\Omega_P^2}} \in [-1,1],
\end{equation}
where $\zeta>0$ is just a constant rescaling parameter. The evolution equation for this variable is 
\begin{align}\label{eq:xpp}
    X_P' &= \frac{\Omega_H}{\zeta} \sqrt{1-X_P^2}\nonumber\\
  &  \left( X_P+3\,\zeta\,\sqrt{1-X_P^2}\right)\,\left(X_P-3\,\zeta\,\Omega_{\partial P}\,\sqrt{1-X_P^2}\right),
\end{align}
which along with the Eq.~\eqref{eq:omegahp} defines the compactified system.

\paragraph{Invariant subsets for $-\frac{1}{3}<\Tilde{\Omega}_{\partial P}$.}
The corresponding invariant subsets divide the  variable space into three disjoint regions. The portrait of the variable space is depicted in Fig.\ref{fig:gen1} where the value $\Tilde{\Omega}_{\partial P}=1$ has been chosen. The sources of the system are $C_+$ and $A_-$, $B$ is a saddle point, and $A_+$ and $C_-$ are sinks.

\begin{itemize}
    \item The region $3\,\Tilde{\Omega}_{\partial P}\leq \Omega_P$ describes recollapsing models starting from expanding $C_+$ and going towards contracting $C_-$.
    \item The region bounded between $-3\leq \Omega_P\leq 3\,\Tilde{\Omega}_{\partial P}$ is divided into four subregions by separatrices. The separatrices  meet at the static universe point $B$. The right subregion is characterized by trajectories starting from decelerating expansion in $C_+$ and going towards the de Sitter point $A_+$. The left subregion describes cosmologies which start from accelerated Anti-de Sitter $A_-$ and end their collapse decelerating at $C_-$. The upper subregion describes recollapsing senarios starting from expanding $C_+$ and ending at the contracting $C_-$. The lower subregion describes bouncing models starting from the Anti-de Sitter $A_-$ and ending at the expanding de Sitter $A_+$.
    \item The last region lying in $\Omega_P\leq -3$ also describes bouncing models starting from the Anti-de Sitter $A_-$ and going to the expanding de Sitter $A_+$.
\end{itemize}

\paragraph{Invariant subsets for $-1< \Tilde{\Omega}_{\partial P}<- \frac{1}{3}$.} In this range,  points $A_-$ and $A_+$ still behave as source and sink respectively. Points $C_\pm$ become saddle points, while point $B$ becomes a center. The variable space portrait is illustrated in Fig.~\ref{fig:gen2} for $\Tilde{\Omega}_{\partial P}=-\frac{1}{2}$. In this range the full invariant subset is divided into the three regions same as in the previous case. 
\begin{itemize}
    \item The region in the range  $3\,\Tilde{\Omega}_{\partial P}\leq \Omega_P$ is dominated by the presence of the center $B$. The trajectories in this region describe cyclic models which go through alternating accelerated and decelerated phases.
   \item The bounded region $-3\leq \Omega_P\leq 3\,\Tilde{\Omega}_{\partial P}$ describes bouncing universes starting from Anti-de Sitter $A_-$ and going to expanding de Sitter $A_+$. 
   \item The region for the case $\Omega_P\leq -3$ also represents bouncing universes starting from Anti-de Sitter $A_-$ and going to the expanding de Sitter $A_+$.
\end{itemize}

\paragraph{Invariant subsets for $\Tilde{\Omega}_{\partial P}<-1$.} In this range, points $A_\pm$ become saddle points. Point $B$ still describes center, while point $C_-$ is a source and $C_+$ is a sink. The variable space dynamic for $\Tilde{\Omega}_{\partial P}=-2$ is depicted in Fig.~\ref{fig:gen3}. Similar to the previous cases the variable space is divided into  three independent regions. These three regions are topologically the same as in the previous case. The only differences are that $A_\pm$ and $C_\pm$ have swapped stability properties and $C_\pm$ are located at lower values of $\Omega_P$ than $A_\pm$.

\subsection{Non-positive curvature}\label{sec:nproots}

\begin{figure}[htp]
    \centering
    \captionsetup[subfloat]{position=top} 
{\subfloat[$\Tilde{\Omega}_{\partial P}=1$]{\includegraphics[width=0.38\textwidth]{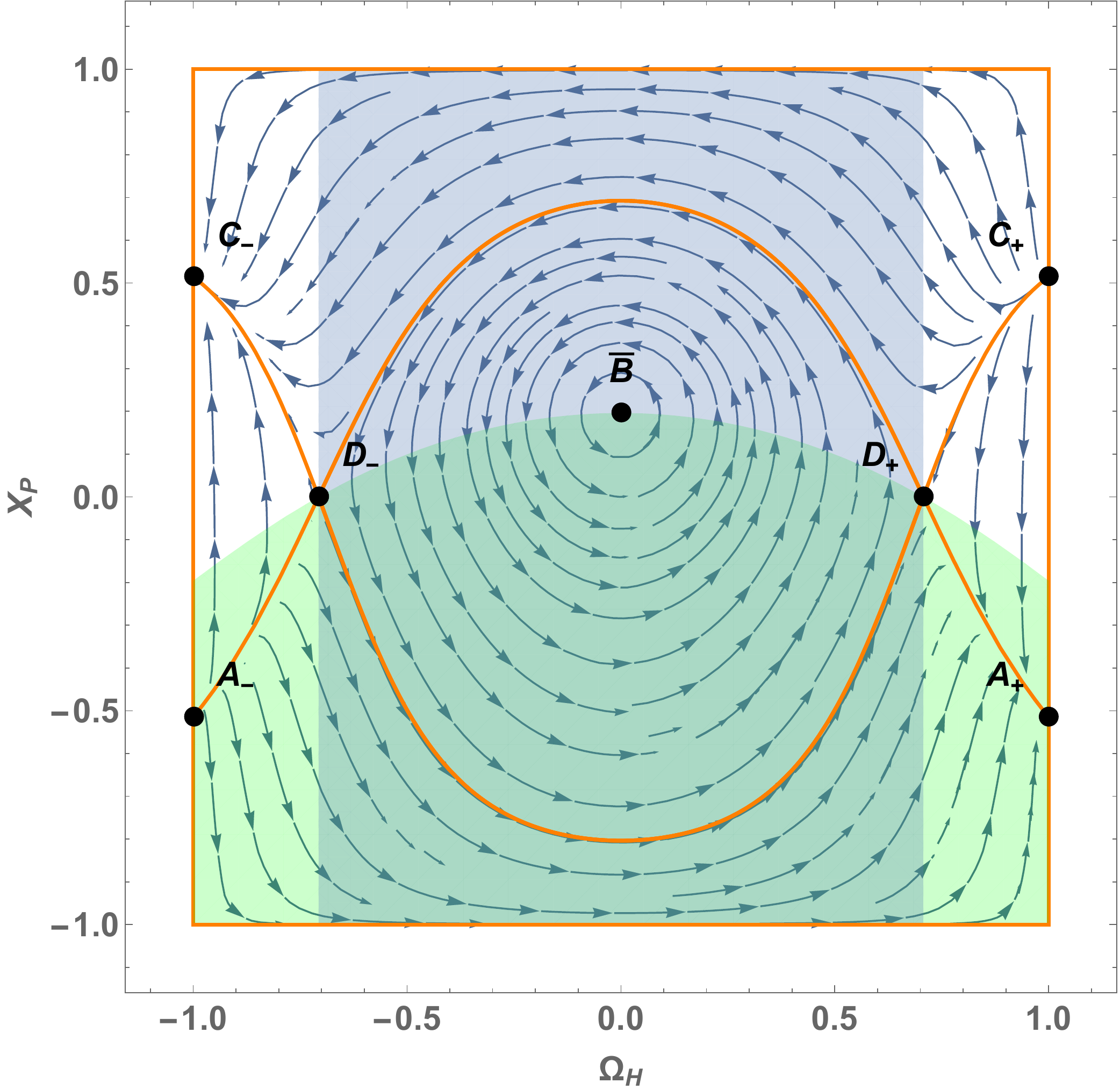}\label{fig:genn1}}}
{  \subfloat[$\Tilde{\Omega}_{\partial P}=-\frac{1}{2}$]{\includegraphics[width=0.38\textwidth]{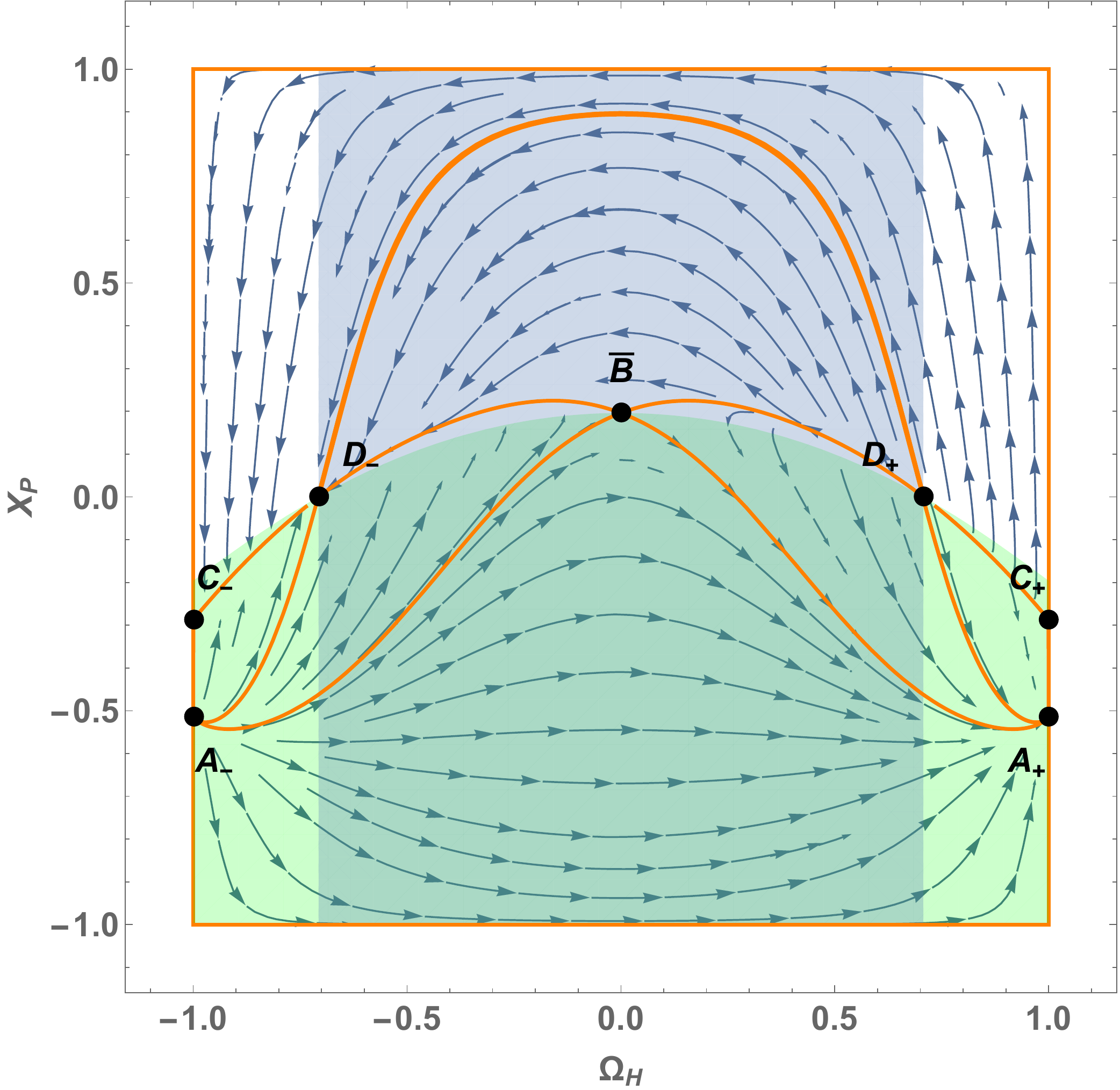}\label{fig:genn2}}}
{  \subfloat[$\Tilde{\Omega}_{\partial P}=-2$]{\includegraphics[width=0.38\textwidth]{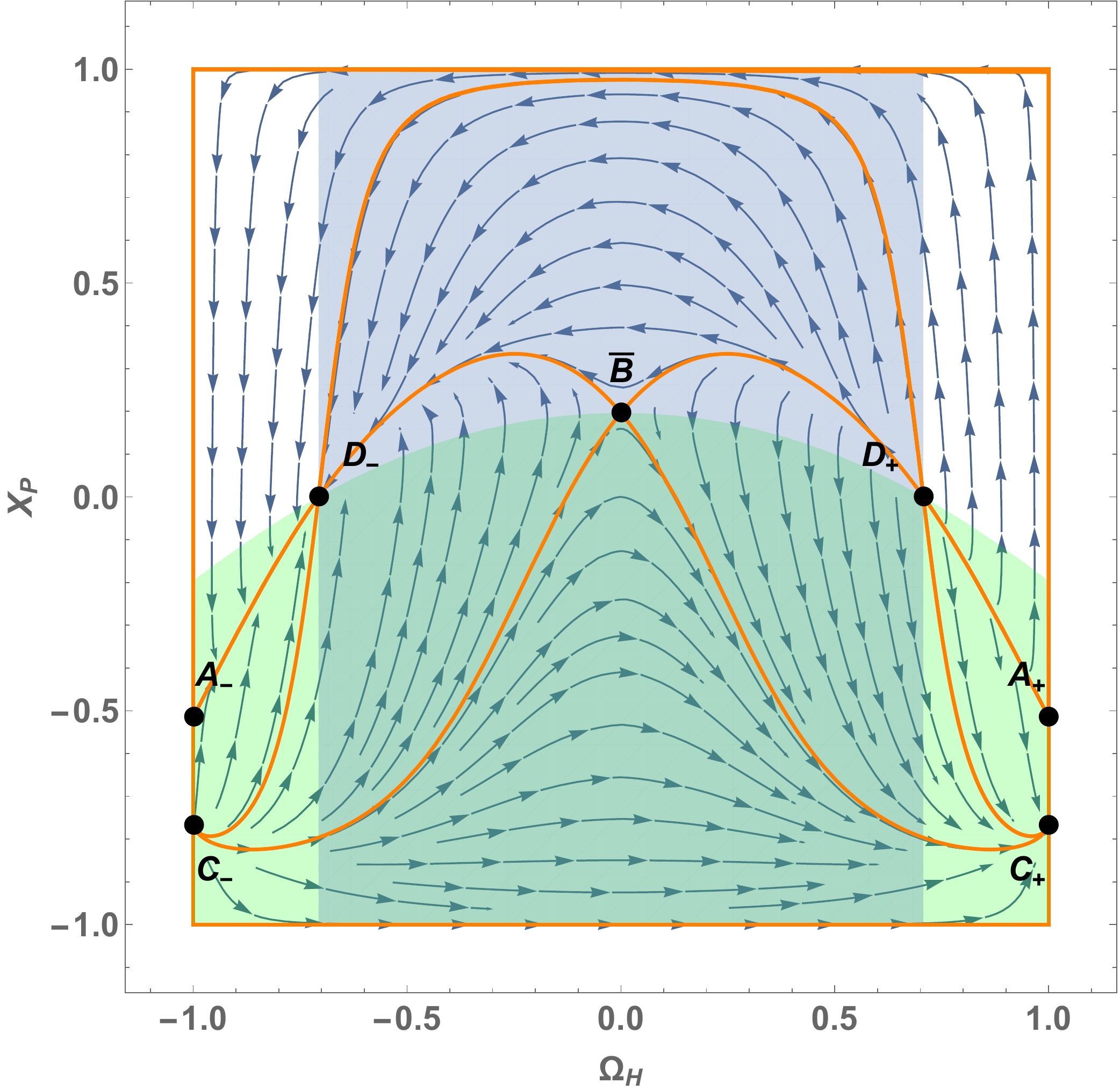}\label{fig:genn3}}}
    \caption{Invariant subsets for negative spatial curvature and $\zeta=0.2$ plotted for three representative values of $\Tilde{\Omega}_{\partial P}$ in the ranges given in Sec.~\ref{sec:nproots}. The orange thick lines are the separatrices. The  blue shaded areas are the regions excluded by our assumption that $\Omega_\epsilon>0$. The green shaded region are the part of the variable space where we have accelerating universe.
    }\label{fig:genn}
\end{figure}

There are additional critical points for the non-positive curvature once we consider $\Gamma( \Tilde{\Omega}_{\partial P})=0$. These critical points are located at $\{\Omega_{H},\Omega_{P}\}=\{\pm \frac{1}{\sqrt{2}},0\}$ and describe the Milne universe. At these points the deceleration parameter is $q=0$ and the scale factor evolves as $a=\pm \mid k \mid (t+c_1)$ for $\Omega_H=\pm\,\frac{1}{\sqrt{2}}$ respectively.

A critical point with $\Omega_H=\frac{1}{\sqrt{2}}$, which we denote as $D_{+}$, has the eigenvalues 
\begin{equation}
    \{\lambda^{D_+}_{i} \}=\{\sqrt{2}, -\frac{\sqrt{2}}{2}\,\left(1+3\,\Tilde{\Omega}_{\partial P}\right) \},\label{eq:eigdp}
\end{equation}
in the invariant subset $\lbrace \Omega_H,\Omega_P\rbrace$. When $\Omega_H=-\frac{1}{\sqrt{2}}$, the critical point is denoted as $D_-$ and has the eigenvalues
\begin{equation}
    \{\lambda^{D_-}_{i} \}=\{-\sqrt{2}, \frac{\sqrt{2}}{2}\,\left(1+3\,\Tilde{\Omega}_{\partial P}\right) \}.\label{eq:eigdm}
\end{equation}
Eqs.~\eqref{eq:eigdp} and~\eqref{eq:eigdm} imply that for $-\frac{1}{3}<\Tilde{\Omega}_{\partial P}$ the critical points $D_\pm$ are saddles, while for $-\frac{1}{3}>\Tilde{\Omega}_{\partial P}$, $D_+$ is a source and $D_-$ is a sink. 

As in the positive curvature case, one can compactify the variable space by applying the transformation~\eqref{eq:xp} to obtain
\begin{align}\label{eq:xppn}
      X_P' &= \frac{\Omega_H}{\zeta} \sqrt{1-X_P^2}\,( 9\,\zeta^2\,\Omega_{\partial P}\,(1-2\,\Omega_H^2)\,(1-X_P^2)+\nonumber\\
  &  \zeta\, X_P\,\sqrt{1-X_P^2}\,(1-3\,\Omega_{\partial P}+2\,\Omega_H^2)+X_P^2)).
\end{align}

Our assumption of non-negative energy density causes the Friedman constraint to limit the physically admissible values of $\Omega_H$ as discussed in Sec.~\ref{sec:nprych}. This assumption is not imposed at the level of the evolution equations and hence some trajectories might cross to the forbidden area. The only trajectories lying entirely in the physical region are confined between the separatrices connecting the points $\lbrace C_{+},D_{+},A_{+}\rbrace$ for $\Omega_H>0$ and $\lbrace C_{-},D_{-},A_{-}\rbrace$ for $\Omega_H<0$, as can be seen in Fig.~\ref{fig:genn}. In the following we will focus our discussion only in these physical regions.

\paragraph{Invariant subsets for $-\frac{1}{3}<\Tilde{\Omega}_{\partial P}$.}
The points $C_+$ and $A_-$ are sources, $D_\pm$ are saddle points while the points $C_-$ and $A_+$ are sinks. Fig~\ref{fig:genn1} shows the variable space dynamics for $\Tilde{\Omega}_{\partial P}=1$. For positive $\Omega_H$, the trajectories start from $C_+$ and go towards the expanding de Sitter $A_+$. On the other hand, for negative $\Omega_H$, the trajectories begin from the Anti-de Sitter universe $A_-$ and go towards the contracting $C_-$. In both cases the trajectories can pass close to the saddle points $D_+$ and $D_-$ respectively.
\paragraph{Invariant subsets for $-1<\Tilde{\Omega}_{\partial P}<-\frac{1}{3}$.}
The variable space is plotted in Fig.~\ref{fig:genn2} for $\Tilde{\Omega}_{\partial P}=-\frac{1}{2}$. In this case points $D_+$ and $A_-$ represent sources, $C_\pm$ become saddle points and the sinks are $A_+$ and $D_-$. For $\Omega_H>0$ the trajectories start from expanding Milne universe $D_+$ and going towards the expanding de Sitter $A_+$. On the other hand, for $\Omega_H<0$, we see that the past attractor is now the Anti-de Sitter $A_-$ and the trajectories move towards the collapsing Milne universe. In both cases some trajectories may approach transiently $C_+$ and $C_-$ respectively. 
\paragraph{Invariant subsets for $\Tilde{\Omega}_{\partial P}<-1$.}
The variable space portrait for this case is illustrated in Fig.~\ref{fig:genn3} where $\Tilde{\Omega}_{\partial P}=-2$ is chosen. In contrast to the other cases, here $A_\pm$ become saddle points. Points $D_+$ and $C_-$ represent sources and $C_+$ and $D_-$ are sinks. For $\Omega_H>0$ the trajectories start from expanding Milne universe and go towards the late attractor $C_+$. For $\Omega_H<0$, the trajectories emerge from the past attractor $C_-$ and end up at the contracting Milne universe $D_-$. In both cases, there are some trajectories passing transiently trough $A_+$ and $A_-$ respectively.

\section{Examples}
\label{sec:examples}

In this section we close the system of equations following two approaches: first by choosing a specific form of the function $\Gamma$, and then by imposing instead a form of EoS from which $\Gamma$ can be derived.  As we will see, both cases will have the simplest functional form of $\Gamma$, that is linear in $\Omega_{\partial P}$, i.e. 
\begin{equation}\label{eq:linearG}
    \Gamma=\alpha\, \Omega_{\partial P}+\beta,
\end{equation}
where $\alpha$ and $\beta$ are free parameters.  One can find the functional form of the EoS by integrating Eq.~\eqref{eq:secvar2}. Namely, when $0\neq \alpha \neq -1$
\begin{equation} \label{eq:linp}
    P=\frac{\alpha \Omega_{\partial P\star}+\beta}{\alpha\,(1+\alpha)\,\epsilon_\star^{\alpha}} \epsilon^{1+\alpha}-\frac{\beta}{\alpha} \epsilon+P_\star\, ,
\end{equation}
and 
\begin{equation} \label{eq:lincs}
    \Omega_{\partial P}=\frac{\partial P}{\partial \epsilon}= \frac{\alpha \Omega_{\partial P\star}+\beta}{\alpha\,\epsilon_\star^{\alpha}} \epsilon^{\alpha}-\frac{\beta}{\alpha}\, ,
\end{equation}
where $P_\star,~\Omega_{\partial P\star},~\epsilon_\star$ are EoS integration constants.\footnote{The two special cases $\alpha=0$ and $\alpha=-1$  will not be discussed, since in the former case the EoS can violate causality, while in the latter case the pressure diverges as $\epsilon\rightarrow0$.} Depending on the free parameters, the first term in eq.~\eqref{eq:linp} can represent a generalized Chaplygin gas \cite{bento2002generalized}, while the second term includes a typical linear EoS; however in general both terms can describe more exotic fluids.

We can rewrite Eq.~\eqref{eq:linp} as
\begin{equation} \label{eq:genomegp}
    \Omega_P=\frac{3}{1+\alpha}\, \Omega_{\epsilon}\left( \Omega_{\partial P}-\beta\right)+\Omega_{P_\star}\, ,
\end{equation}
where $\displaystyle\Omega_{P_\star}=\frac{P_\star}{D^2}$. Eq.~\eqref{eq:genomegp} is a constraint between $\displaystyle\Omega_{P_\star}$ and the dynamical variables of our system and it implies that, even though this quantity appears as a new variable, we can retreat it during the evolution.

\subsection{$\beta=-\alpha$: causality condition}\label{sec:causal}
In light of the discussion of Sec.~\ref{sec:roottrack}, we impose the causality condition to this model by choosing $\beta=-\alpha$, which implies $\Gamma =\alpha(\Omega_{\partial P}-1$).  The EoS~\eqref{eq:linp} in this case reduces to 
\begin{align}\label{eq:linpr}
       P=\frac{\Omega_{\partial P\star}-1}{(1+\alpha)\,\epsilon_\star^\alpha} \epsilon^{1+\alpha}+ \epsilon+P_\star\, .
\end{align}
Note that this represents a combination of a stiff EoS and an exotic fluid.

By demanding further that when the energy density tends to zero, the pressure does not diverge, Eq.~\eqref{eq:linpr} implies that $\alpha > -1$.\footnote{Note that, since we have assumed causality, it holds that $1-\Omega_{\partial P\star}\ge 0$. Thus, $\displaystyle\frac{\Omega_{\partial P\star}-1}{(1+\alpha)\,\epsilon_\star^\alpha}\leq 0$.}

The EoS~\eqref{eq:linpr} has an extremum at 
\begin{align}
    \epsilon_e=\frac{\epsilon_\star}{(1-\Omega_{\partial P\star})^{1/\alpha}}\, ,
\end{align}
which is maximum if $\alpha>0$ and minimum if $\alpha<0$. If one ignores an early inflationary epoch, then the pressure of the fluid should be positive for large energy densities. For low energy densities, to reproduce the effect of dark energy, one would expect negative pressure. Thus, the EoS we want has a minimum, i.e. $-1<\alpha<0$. 

For $P=P_\star$ Eq.~\eqref{eq:linpr} implies that either $$\Omega_{\partial P\star}=-\alpha$$ or $\epsilon_\star=0$. The latter leads to the trivial EoS of the stiff fluid, since $-1<\alpha<0$. Thus, we choose the former. If further we make the reasonable demand that the pressure tends to zero along with the energy density, then $${P_\star}=0\, ,$$ which brings the constraint~\eqref{eq:genomegp} to
\begin{align}\label{eq:genomegpr}
    \Omega_P=\frac{3}{1+\alpha}\, \Omega_{\epsilon}\left( \Omega_{\partial P}+\alpha\right)\, .
\end{align}
This, combined with the Friedmann constraint, reduces the system to two dimensions. Namely, due to the Friedmann constraints~\eqref{eq:fried_pos} or \eqref{eq:fried_neg} one can disregard the $\Omega_\epsilon$ evolution eq.~\eqref{eq:omegaep}, and due to the constraint~\eqref{eq:genomegpr} we can ignore the $\Omega_P$ evolution equation~\eqref{eq:omegap}. Thus, the remaining dynamical variables are $\Omega_{\partial P}$ and $\Omega_H$.  Note that constraint~\eqref{eq:genomegpr} introduces an invariant subset at $\Omega_{\partial P}=-(1+2\alpha)$.  As we want to preserve causality and also allow for positive $\Omega_{\partial P}$, we will focus our analysis in the compact region $\Omega_{\partial P} \in \left[ -(1+2\alpha) , 1 \right]$.

The critical lines $A_{\pm}$ intersect the new invariant subset at 
\begin{align}
        \lbrace \Omega_H,\Omega_{\partial P}\rbrace = \lbrace \pm 1,-(1+2\alpha)\rbrace \, ,
\end{align}
while the critical line $B$ intersects it at 
\begin{align}
        \lbrace \Omega_H,\Omega_{\partial P}\rbrace = \lbrace 0,-\frac{1+4\alpha}{3}\rbrace \, .
\end{align}
The critical points $D_\pm$ lie at
 \begin{align}
        \lbrace \Omega_H,\Omega_{\partial P}\rbrace = \lbrace \pm \frac{1}{\sqrt{2}},1 \rbrace \, .
\end{align} 
Given the chosen form of $\Gamma$ and the assumptions on the parameters made above, for the negative curvature we get a new pair of Milne-like critical points $E_\pm$ at
\begin{align}
        \lbrace \Omega_H,\Omega_{\partial P}\rbrace = \lbrace \pm \frac{1}{\sqrt{2}},-(1+2\,\alpha) \rbrace \, .
\end{align} 
In the invariant subset $\lbrace \Omega_H,\Omega _{\partial P}\rbrace$ critical point $E_+$ has eigenvalues 
\begin{equation}
    \lbrace \lambda_i^{E_+}\rbrace=\lbrace \sqrt{2} , 3\,\sqrt{2}\,\alpha \rbrace,
\end{equation}
while critical point $E_-$ has eigenvalues 
\begin{equation}
    \lbrace \lambda_i^{E_+}\rbrace=\lbrace \sqrt{2} , 3\,\sqrt{2}\,\alpha \rbrace,
\end{equation}
which shows that points $E_\pm$ represent saddle points in the range $-1<\alpha<0$.
    The critical points $C_\pm$ lie at
\begin{align}
        \lbrace \Omega_H,\Omega_{\partial P}\rbrace = \lbrace \pm 1,1\rbrace \,.
\end{align}
Points $C_\pm$ describe cases of stiff matter dominated universe, in which the scale factor evolves as $a \sim t^{\frac{1}{3}}$ and the cosmological parameter $q=2$.
Point $C_+$ ($\Omega_H=1$) has eigenvalues 
\begin{equation} \label{eigcp}
\{\lambda^{C_+}_{i} \}=\{4,\,-6\, \alpha   \} \, ,
\end{equation}
while $C_-$ ($\Omega_H=-1$) has eigenvalues 
\begin{equation} \label{eigcm}
\{\lambda^{C_-}_{i} \}=\{-4,\,6\, \alpha   \} \, .
\end{equation}
Note that points $C_{\pm}$ lie on the invariant subset $\Omega_{\partial P}=1$.
All the aforementioned critical points along with their stability are summarized in Table~\ref{tb:CrPlin}.

\begin{figure}[ht]
    \centering
      {\includegraphics[width=0.45\textwidth]{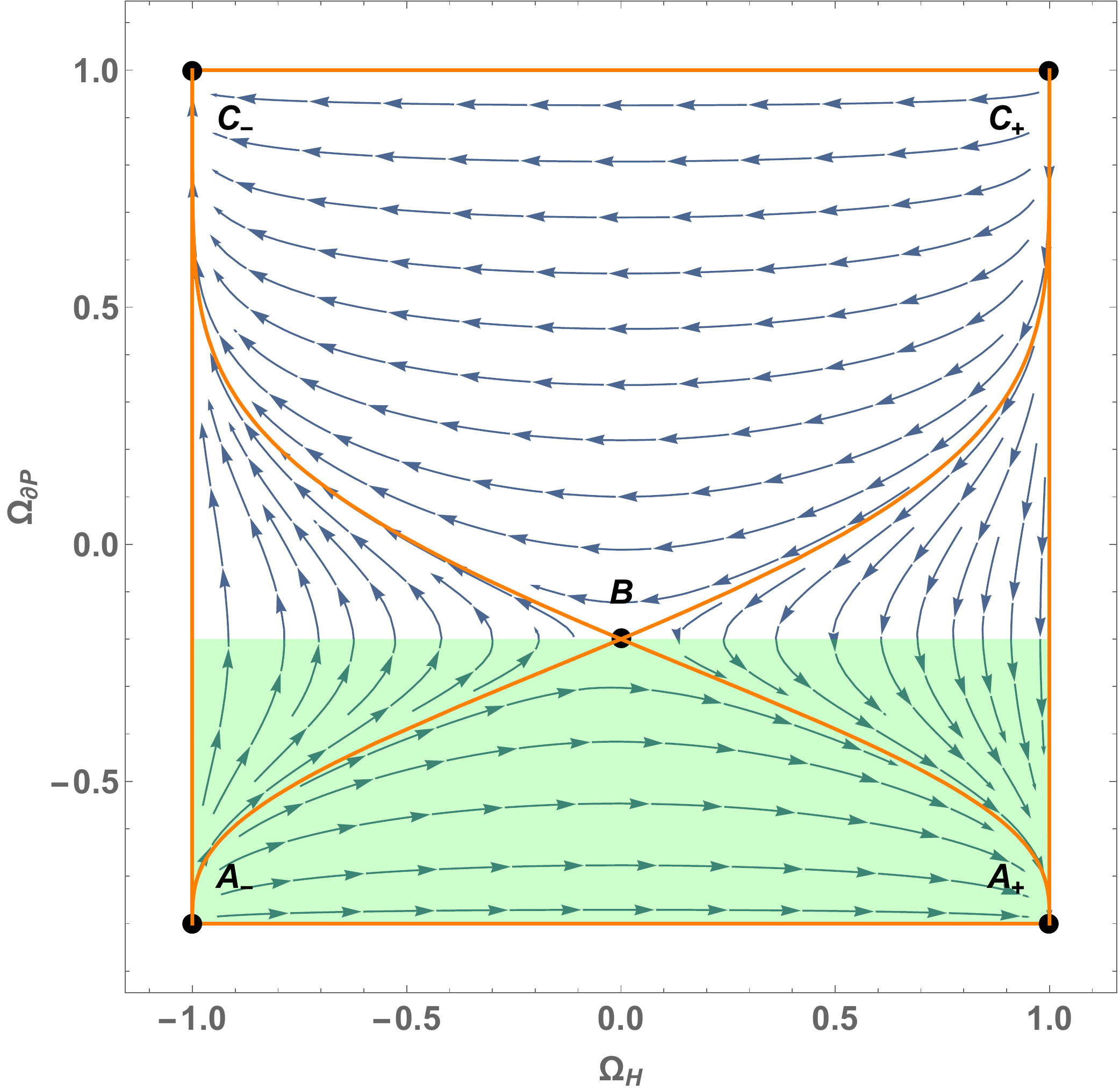}}
    \caption{Invariant subset of the system studied in Sec.~\ref{sec:causal} for non-negative curvature with  $\alpha=-0.1$ . The green shaded area denotes the phase of accelerated expansion $q<0$ and the orange thick lines indicate the separatrices.}\label{fig:invcsh}
\end{figure}

\begin{figure}[ht]
    \centering
      {\includegraphics[width=0.45\textwidth]{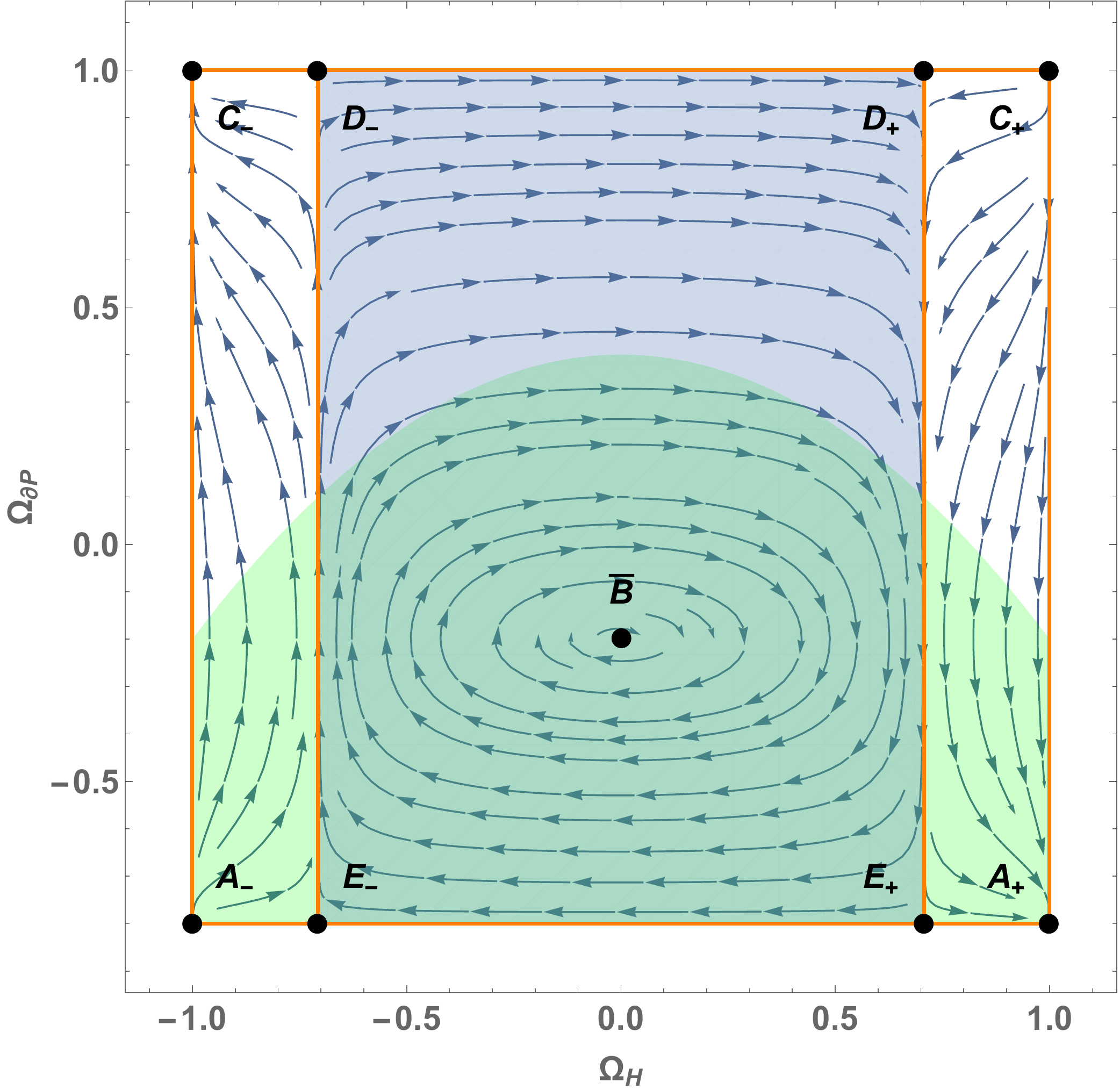}}
    \caption{Invariant subset of the system studied in Sec.~\ref{sec:causal} with $\alpha=-0.1$ for non-positive curvature. The  blue shaded area is the region excluded by our assumption that $\Omega_\epsilon>0$. The green shaded region is the part of the variable space where we have accelerating universe.}\label{fig:invcshnk}
\end{figure}

In Figs.~\ref{fig:invcsh} and~\ref{fig:invcshnk} we show these critical points for the cases $k \geq 0$ and $k \leq 0$ respectively where the free parameter $\alpha= -0.1$. The variable space for $k \geq 0$, i.e. Fig.~\ref{fig:invcsh}, is divided into four subregions from the respective separatrices. All four can transiently pass through a static phase in case they approach point $B$.  The right subregion starts from the stiff matter era $C_+$ expanding exponentially towards the de Sitter point $A_+$. The left subregion describes cosmologies starting from the contracting Anti-de Sitter point $A_-$ and collapsing to the future stiff matter attractor $C_-$. The upper subregion describes recollapsing scenarios starting from the expanding stiff point  $C_+$ and ending at the stiff point $C_-$. The lower subregion is describing bouncing models from the contracting de sitter $A_{-}$, to the expanding de Sitter $A_+$. On the other hand, the variable space for $k\leq 0$ is divided into two subregions since $\frac{1}{2}\leq \Omega_H^2\leq 1$. The right subregion describes models starting from the stiff matter source $C_+$, which expand towards the de Sitter point attractor $A_+$. Whereas, the left subregion describes scenarios starting from contracting de Sitter, point $A_-$, and eventually collapsing to the stiff mater point attractor $C_-$. Note that the variable spaces depicted in Figs.~\ref{fig:invcsh} and~\ref{fig:invcshnk} depend only on the free parameter $\alpha$ and they just rescale accordingly. Namely, changing the value of $\alpha$ in the interval $-1 < \alpha < 0$, changes the coordinate $\Omega_{\partial P}$ of points $A_\pm$ in the interval $[-1,1]$ (alongside with the invariant subset $\Omega_{\partial P}=-(1+2\alpha)$) and $B$ in the interval $[-1/3,1]$.

\begin{table}
  \caption{The critical points of the system described in Sec.~\ref{sec:causal}  on the $\{\Omega_H,~\Omega_{\partial P}\}$ plane and their stability for $-1<\alpha<0$ and non-negative curvature.}
\begin{tabular}{c | c c c c}\label{tab:CritPoints2}
 point &  $\Omega_H$ & $\Omega_{\partial P}$ & stability & curvature\\
\hline
  $A_+$ & $1$ & $-(1+2\alpha)$  &  sink & flat \\
  $A_-$ & $-1$ & $-(1+2\alpha)$ &  source & flat \\
  $B$   & $0$ &  $\displaystyle -\frac{1+4\alpha}{3}$  &  saddle & positive\\
  $C_+$ & $1$ & $1$ &  source & flat \\
  $C_-$ & $-1$ & $1$ &  sink & flat \\
  $D_+$ & $\frac{1}{\sqrt{2}}$ & $1$ & saddle & negative\\
  $D_-$ & $-\frac{1}{\sqrt{2}}$ & $1$ & saddle & negative\\
  $E_+$ & $\frac{1}{\sqrt{2}}$ & $-(1+2\alpha)$ & saddle & negative\\
  $E_-$ & $-\frac{1}{\sqrt{2}}$ & $-(1+2\alpha)$ & saddle & negative\\
\end{tabular}\label{tb:CrPlin}
\end{table}

 The setup $-1<\alpha<0$ has also the following consequences:
\begin{itemize}
    \item For the positive curvature, by combining the limits imposed on $\Omega_{\partial P}$ by the invariant subsets, the Friedmann constraint~\eqref{eq:fried_pos} and the constraint~\eqref{eq:genomegpr}, we have 
    \begin{equation}
     -3\leq \Omega_P\leq 3\ .
    \end{equation}
    \item For the negative curvature, Friedmann constraint together with constraint~\eqref{eq:genomegpr} and the limits imposed on $\Omega_{\partial P}$ by the invariant subsets, lead to
    \begin{equation}
    \left\{
    \begin{array}{c}
     -3\, \Omega_\epsilon \leq \Omega_P \leq 3\, \Omega_\epsilon\\
     0\leq \Omega_\epsilon\leq 1
     \end{array}
     \right.
    \end{equation}
    
\end{itemize}

\begin{figure}[ht]
    \centering
      {\includegraphics[width=0.48\textwidth]{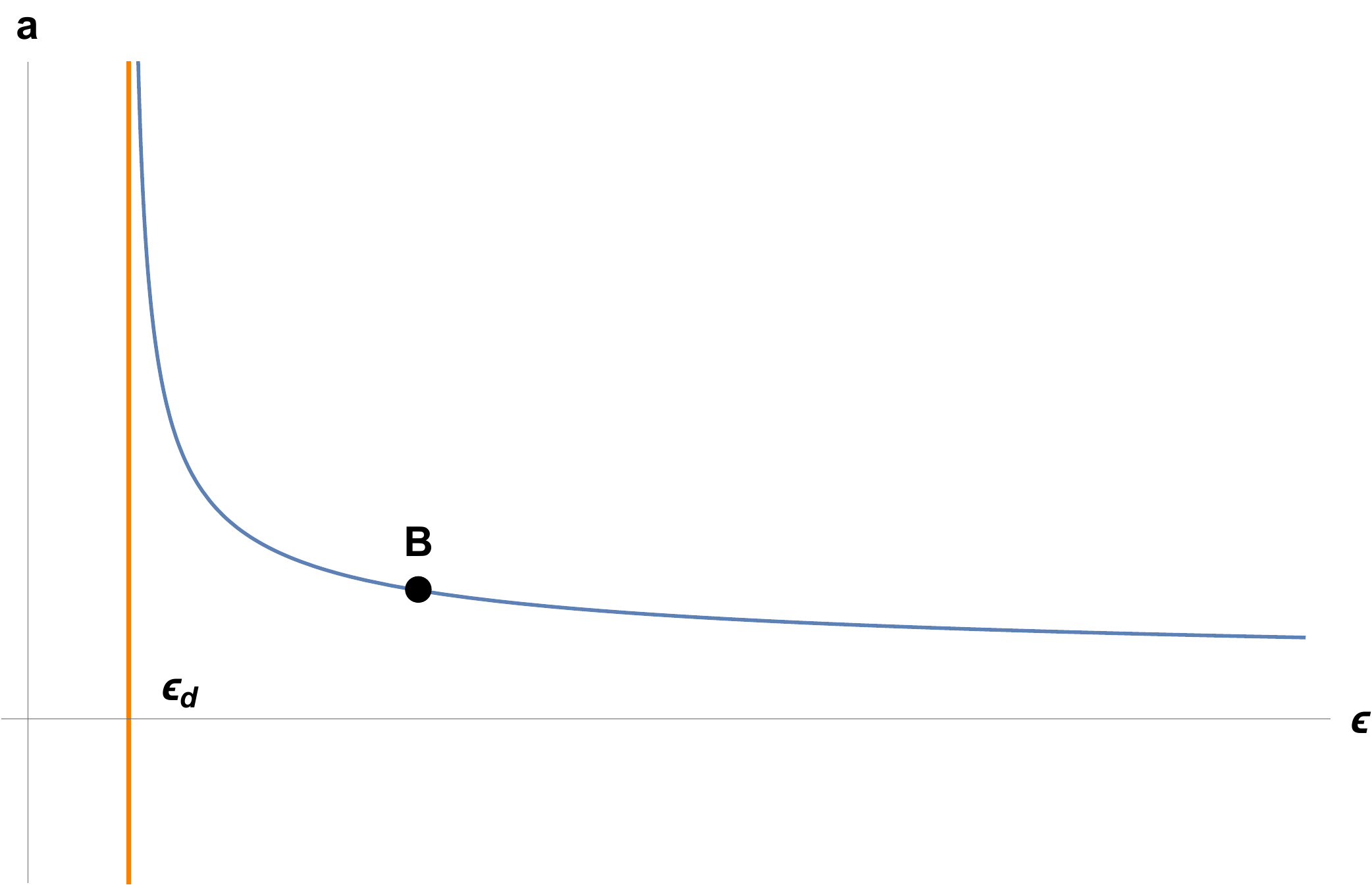}}
    \caption{Behaviour of the scale factor as a function of energy density as given by Eq.~\eqref{eq:scalefactor}. The orange line is the value $\epsilon=2^{-\frac{1}{|\alpha|}}\epsilon_\star$ where the scale factor diverges. }\label{fig:scalefactor}
\end{figure}

By substituting Eq.~\eqref{eq:linpr} with $P_\star=0$ into the continuity equation~\eqref{eq:state} we can calculate the scale factor in terms of $\epsilon$ as follows (taking into account that $-1<\alpha<0$)
\begin{align}\label{eq:scalefactor}
a=a_0 \left(\frac{2-\left(\frac{\epsilon_\star} {\epsilon_0}\right)^{|\alpha|}}{2-\left(\frac{\epsilon_\star} {\epsilon}\right)^{|\alpha|}}\right)^\frac{1}{6|\alpha|}\left(\frac{\epsilon_0}{\epsilon}\right)^{1/6}\, .
\end{align}
From Eq.~\eqref{eq:scalefactor} and Eq.~\eqref{eq:lincs} we can explain the behaviour of the scale factor in the different subregions of Fig.~\ref{fig:invcsh} and Fig.~\ref{fig:invcshnk}, with the aid of Fig.~\ref{fig:scalefactor}. Points $C_\pm$ correspond to $\epsilon\rightarrow\infty$ where the scale factor $a\rightarrow 0$. On the other hand, $A_\pm$ are points in which $\displaystyle~\epsilon~\rightarrow~2^{-\frac{1}{|\alpha|}}\epsilon_\star \equiv \epsilon_d$ and the scale factor diverges. The latter actually happens when the EoS intersects $P=-\epsilon$. Point $B$ corresponds to $\displaystyle\epsilon_B=\left(3/4\right)^\frac{1}{|\alpha|}\,\epsilon_\star$ which has a finite scale factor value. In the right subregion of Fig.~\ref{fig:invcsh}  the scale factor evolves from $a=0$ at $\epsilon=\infty$ to the point $\epsilon=2^{-\frac{1}{|\alpha|}}\epsilon_\star$ where the scale factor diverges. The left subregion has the opposite behaviour, namely the scale factor starts from infinite value and decreases to zero. In the upper subregion the scale factor starts from zero, increases and then decreases to zero again. The  maximum value it can attain  is $\epsilon_\text{max}>\epsilon_B$. In the lower subregion, the scale factor starts from infinity, decreases and then increases again to infinity. The minimum value it can attain is $\epsilon_\text{min}<\epsilon_B$.  The behaviour of the scale factor in the left and right subregions of Fig.~\ref{fig:invcshnk} are analogous to the behaviour in the left and right subregions of Fig.~\ref{fig:invcsh} respectively.

When $\alpha > -\frac{1}{2}$, then $\epsilon_d< \epsilon_e$ and  $\Omega_{\partial P}$ can attain negative values. For $\alpha=-1/2$ the energy density of the scale factor divergence coincides with the EoS's minimum, i.e $\epsilon_d=\epsilon_e$ and this happens when $\Omega_{\partial P}=0$. For $\alpha <-\frac{1}{2}$, $\Omega_{\partial P}>0$.

\subsection{The quadratic EoS}
\label{sec:quadeos}

\begin{figure*}[htp]
    \centering
\includegraphics[width=0.43\textwidth]{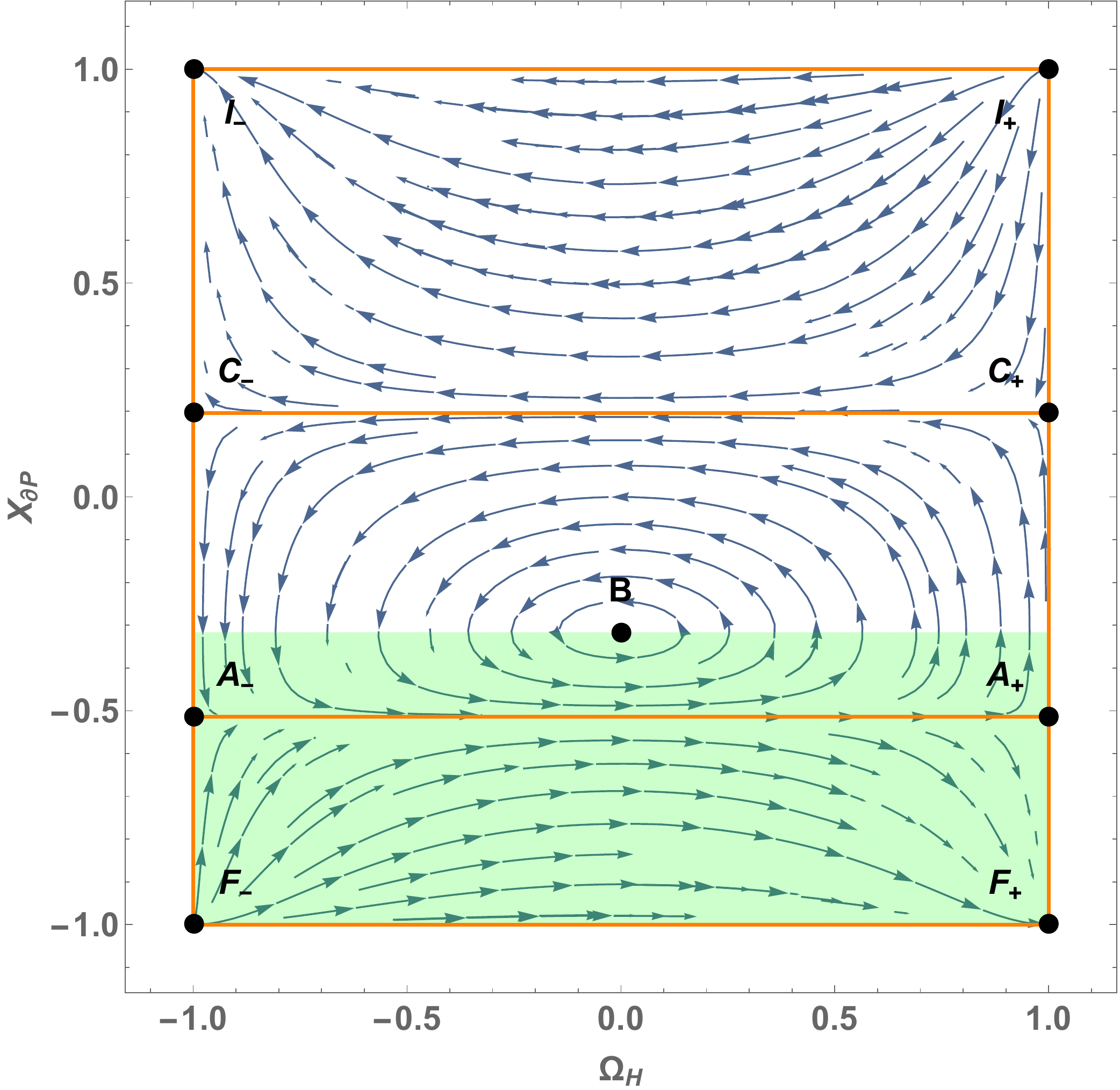}
\includegraphics[width=0.43\textwidth]{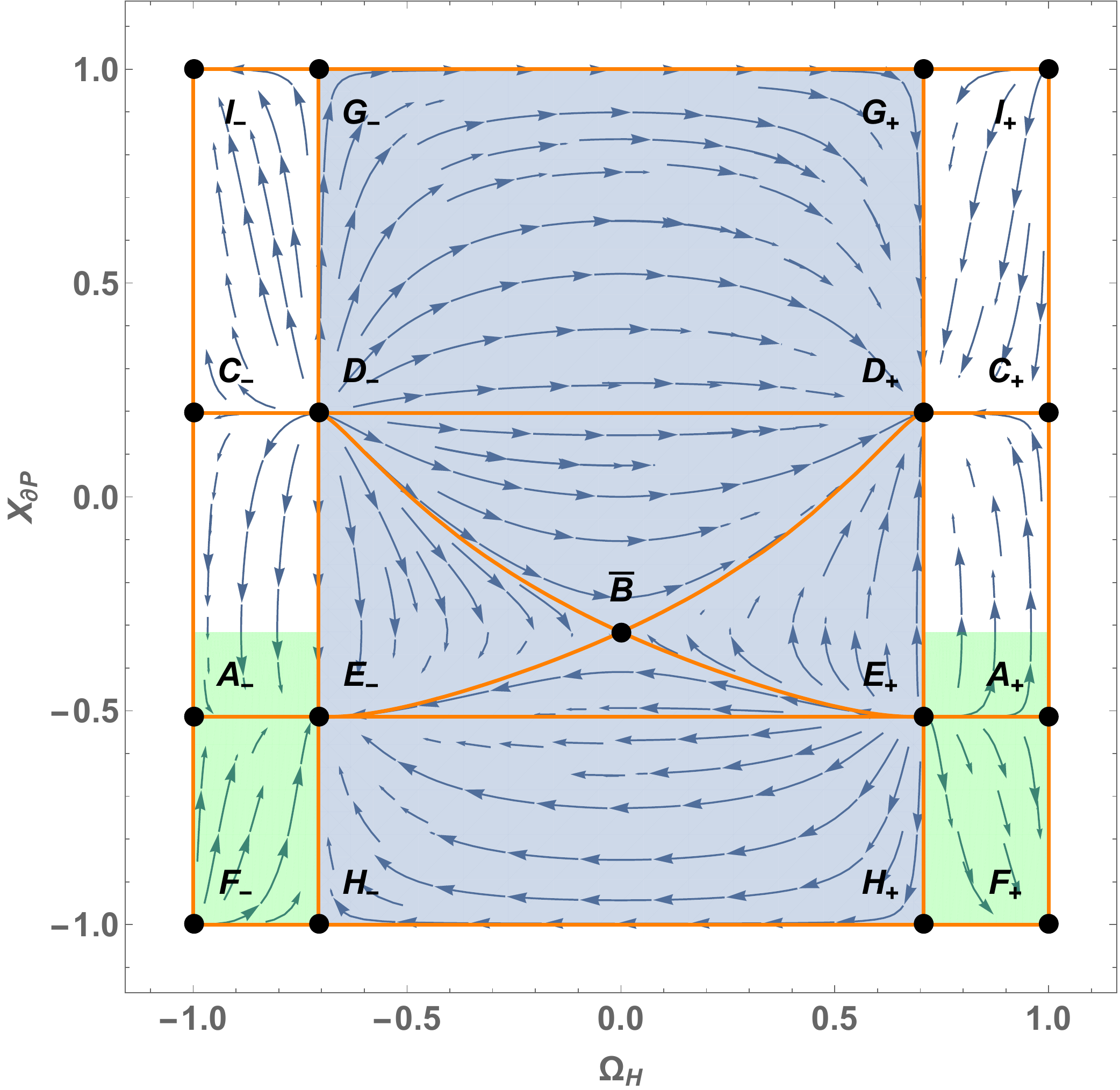}
    \caption{Invariant subsets $\lbrace \Omega_H,X_{\partial P}\rbrace$ for the quadratic EoS with $P_\star=0$ in the case of positive spatial curvature (left panel) and negative spatial curvature (right panel) when $\zeta=0.2$ and $\sigma=1$. The orange thick lines are the separatrices of the system and the green shaded regions denote the part of the variable space where the universe is accelerating.}\label{fig:quadpz1}
\end{figure*}

\begin{figure*}[htp]
    \centering
\includegraphics[width=0.43\textwidth]{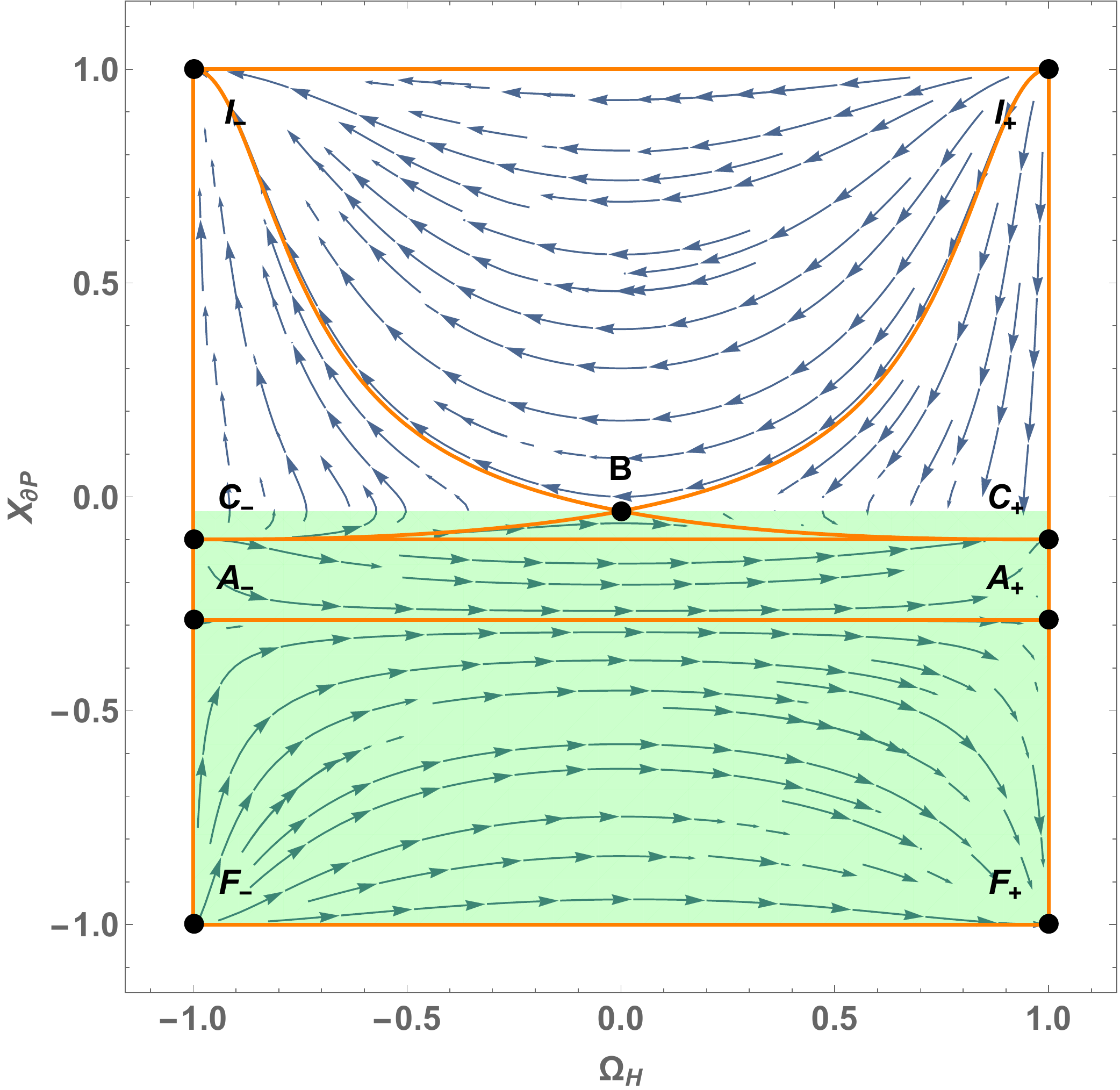}
\includegraphics[width=0.43\textwidth]{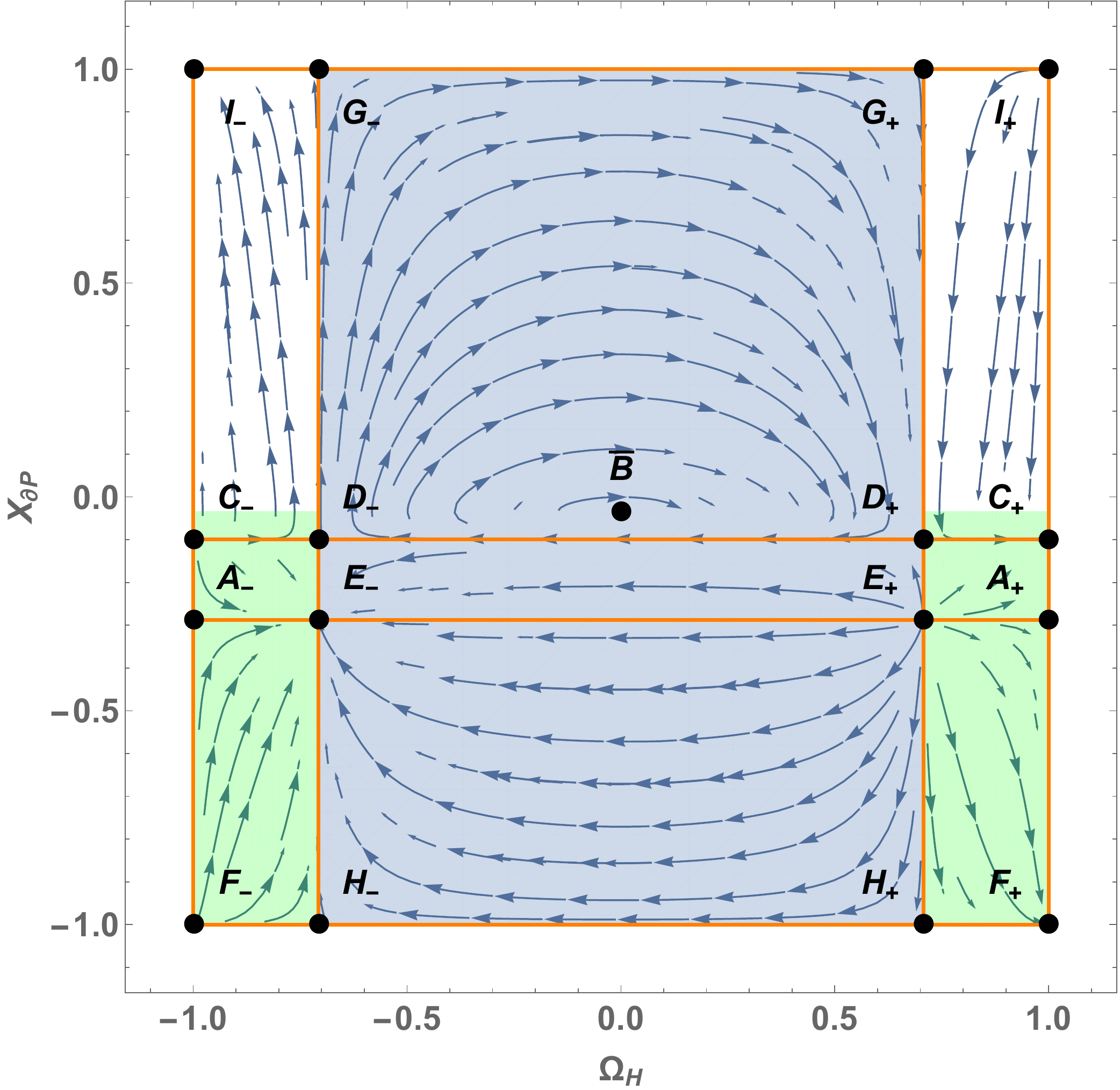}
    \caption{Invariant subsets $\lbrace \Omega_H,X_{\partial P}\rbrace$ for the quadratic EoS with $P_\star=0$ in the case of positive spatial curvature (left panel) and negative spatial curvature (right panel) when $\zeta=0.2$ and $\sigma=-0.5$. The orange thick lines are the separatrices of the system and the green shaded regions denote the part of the variable space where the universe is accelerating.}\label{fig:quadpz-5}
\end{figure*}

\begin{figure*}[htp]
    \centering
\includegraphics[width=0.43\textwidth]{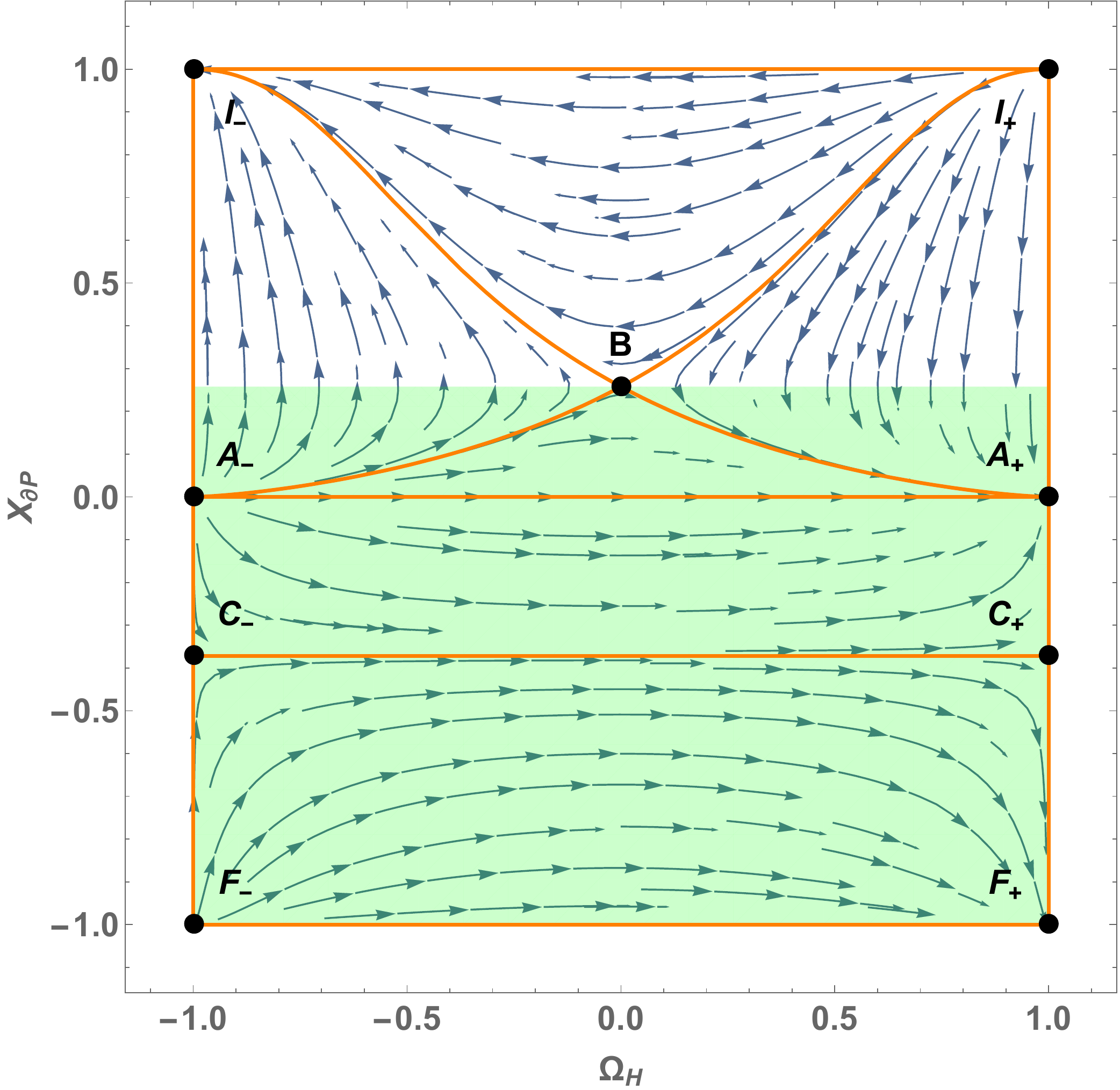}
\includegraphics[width=0.43\textwidth]{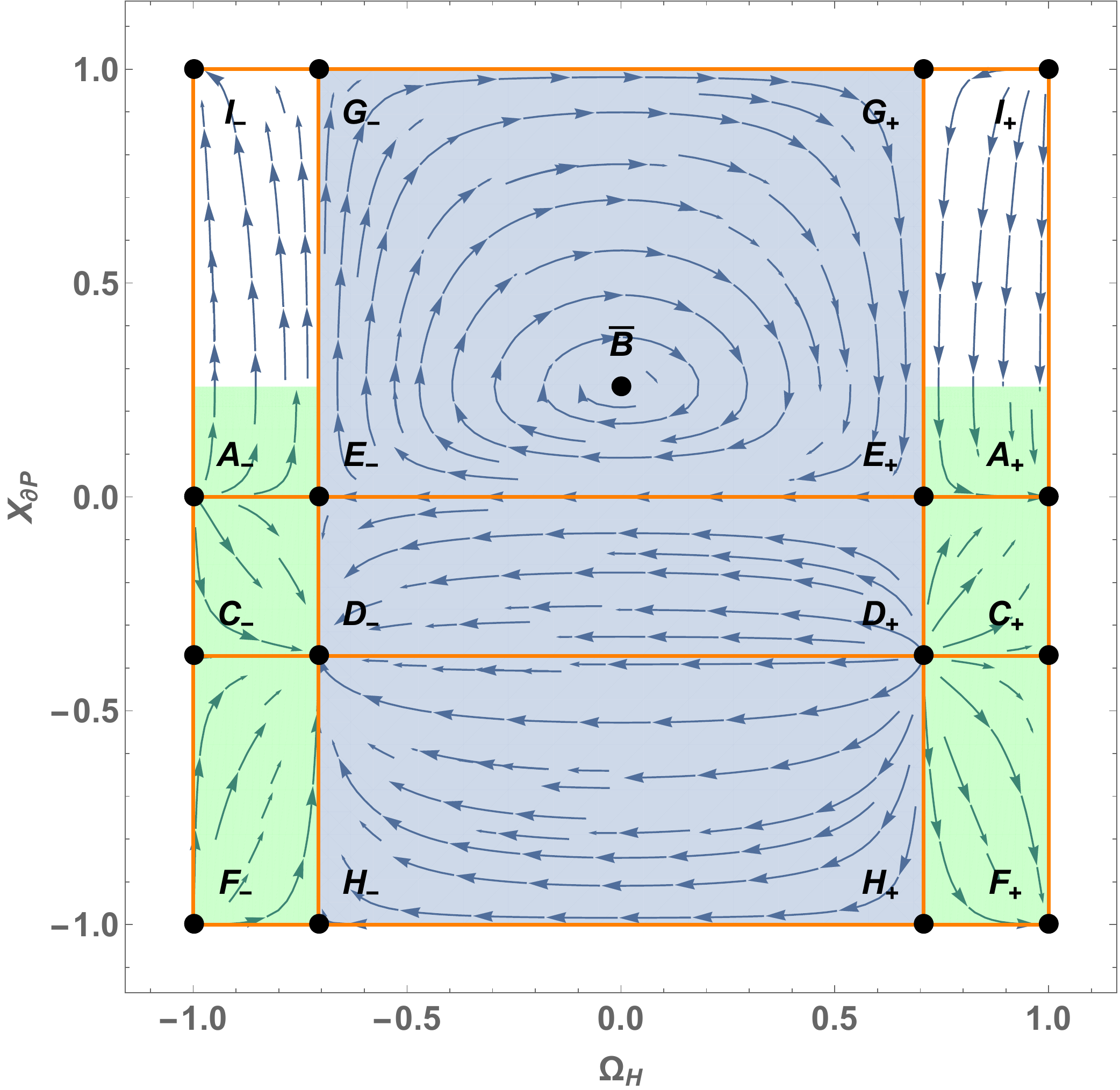}
    \caption{Invariant subsets $\lbrace \Omega_H,X_{\partial P}\rbrace$ for the quadratic EoS with $P_\star=0$ in the case of positive spatial curvature (left panel) and negative spatial curvature (right panel) when $\zeta=0.2$ and $\sigma=-2$. The orange thick lines are the separatrices of the system and the green shaded regions denote the part of the variable space where the universe is accelerating.}\label{fig:quadpz-2}
\end{figure*}
In \cite{quadratic,quadratic2} the quadratic EoS
\begin{equation}\label{eq:quadraticeos}
    P=\frac{\delta}{\epsilon_c}\,\epsilon^2+\sigma\,\epsilon+P_\star ,
\end{equation}
was studied thoroughly. Here we compare their results with our formalism, by adopting the reasoning in \cite{quadratic} and viewing the EoS~\eqref{eq:quadraticeos} as a Taylor expansion of an unknown barotropic EoS  around $\epsilon=0$ without necessarily demanding that $P_\star=0$.  From Eq.~\eqref{eq:quadraticeos} we can write
\begin{equation} \label{eq:quadodp}
    \Omega_{\partial P}= 2 \frac{\delta}{\epsilon_c}\,\epsilon+\sigma,
\end{equation}
\begin{equation}\label{eq:quadomegap}
    \Omega_P=\frac{3}{2}\,\Omega_\epsilon\,\left(\Omega_{\partial P}+\sigma\right)+\Omega_{P_\star}
\end{equation}
and derive the $\Gamma$ function in our variables as follows
\begin{equation}\label{eq:gammaquad}
    \Gamma=\Omega_{\partial P}-\sigma,
\end{equation}
which shows that the quadratic EoS brings a $\Gamma$ which is linear in $\Omega_{\partial P}$. In order to analyze the behaviour of the quadratic EoS we split it into the three cases as done in \cite{quadratic}.

\paragraph{$\delta=0$ the linear EoS.} In this case, since $\displaystyle\Omega_{\partial P}=\sigma$ is a constant, the system is similar to the cases analyzed in Sec.~\ref{sec:roottrack}.

\paragraph{$P_\star=0$.} This amounts to assuming that the pressure tends to zero along with the energy density. Thus, Eq.~\eqref{eq:quadomegap} provides the constraint 
\begin{equation}
    \Omega_P=\frac{3}{2}\,\Omega_\epsilon\,\left(\Omega_{\partial P}+\sigma\right).
\end{equation}
Similarly to the example given in Sec.~\ref{sec:causal} this constraint reduces the system to two dimensions. Therefore, the critical points $A_\pm$ lie at
\begin{equation}
    \lbrace \Omega_H,\Omega_{\partial P}\rbrace=\lbrace \pm1,-(2+\sigma)\rbrace.
\end{equation}
The critical point $B$ is located at
\begin{equation}
    \lbrace \Omega_H,\Omega_{\partial P}\rbrace=\lbrace 0,-\left(\frac{2}{3}+\sigma\right)\rbrace.
\end{equation}
Critical points $C_\pm$ are located at
\begin{equation}
    \lbrace \Omega_H,\Omega_{\partial P}\rbrace=\lbrace \pm 1,\sigma\rbrace.
\end{equation}
Critical lines $D_\pm$ are located at
\begin{equation}
    \lbrace \Omega_H,\Omega_{\partial P}\rbrace=\lbrace \pm\frac{1}{\sqrt{2}},\sigma\rbrace.
\end{equation}
There is also a pair of Milne-like critical points $E_\pm$ for negative curvature at
\begin{equation}
    \lbrace\Omega_H,\Omega_{\partial P}\rbrace=\lbrace\pm\frac{1}{\sqrt{2}},-(2+\sigma)\rbrace.
\end{equation}
As in Sec.~\ref{sec:roottrack} in order to compactify the variable space we use the transformation
\begin{equation}\label{eq:xdp}
    X_{\partial P}=\frac{\zeta\Omega_{\partial P}}{\sqrt{1+\zeta^2\Omega_{\partial P}^2}} \in [-1,1].
\end{equation}
The evolution equation for $X_{\partial P}$ becomes
\begin{align}
&X_{\partial P}^{'}= \frac{3\,\Omega_H}{2\,\xi}\,\sqrt{1-X_{\partial P}^2}\times \nonumber \\
    &\left(\sigma\,\xi^2\,(\sigma+2)\,(1-X_{\partial P}^2)+X_{\partial P}\,\left(1+2\,\xi\,\,\sqrt{1-X_{\partial P}^2}\right) \right)
    \end{align}
and it holds for both curvatures.

Once we have compactified the variable space two more pairs of critical points appear for the flat case, while two Milne-like pairs of critical points appear for the negative curvature. In particular, $I_\pm$ appear at $\lbrace \Omega_H,\Omega_{\partial P}\rbrace=\lbrace \pm 1 ,+\infty \rbrace$, $F_\pm$ appear at $\lbrace \Omega_H,\Omega_{\partial P}\rbrace=\lbrace \pm 1,-\infty\rbrace$, $G_\pm$ appear at $\lbrace \Omega_H,\Omega_{\partial P}\rbrace=\lbrace \pm \frac{1}{\sqrt{2}} ,+\infty \rbrace$, while $H_\pm$ appear at $\lbrace \Omega_H,\Omega_{\partial P}\rbrace=\lbrace \pm \frac{1}{\sqrt{2}},-\infty\rbrace$.   

In Figs.~\ref{fig:quadpz1}-~\ref{fig:quadpz-2} the variable spaces $\lbrace \Omega_H,X_{\partial P}\rbrace$ for different ranges of $\sigma$ are plotted. These variable spaces are divided into two main subregions depending on the sign of $\delta$: the subregions above the separatrices connecting points $C_\pm$ correspond to an EoS with $\delta=+1$ and the rest of variable space describes the case $\delta=-1$.

For comparing our analysis to the one in \cite{quadratic} let's denote $\Tilde{X}_{\partial P}=X_{\partial P}\left(\Omega_{\partial P}=\sigma\right)$  by using Eq.~\eqref{eq:xdp}. Figs.~4 and~7 of \cite{quadratic} correspond to $X_{\partial P}>\Tilde{X}_{\partial P}$ and $X_{\partial P}<\Tilde{X}_{\partial P}$ in our Fig.~\ref{fig:quadpz1} respectively. Figs.~3 and 6 of \cite{quadratic} correspond to $X_{\partial P}>\Tilde{X}_{\partial P}$ and $X_{\partial P}<\Tilde{X}_{\partial P}$ in our Fig.~\ref{fig:quadpz-5} respectively. Figs.1 and 5 of \cite{quadratic}  correspond to $X_{\partial P}>\Tilde{X}_{\partial P}$ and $X_{\partial P}<\Tilde{X}_{\partial P}$ in our Fig.~\ref{fig:quadpz-2} respectively.  When $X_{\partial P}=\Tilde{X}_{\partial P}$, Eq.~\eqref{eq:quadodp} implies either $\epsilon=0$ or $\delta=0$.  However, due to the Friedmann constraint $\epsilon$ is in general different from zero; hence we have $\delta=0$, which is case {\it a.} above.

The points $A_\pm$, $B$, $I_\pm$ and $F_\pm$ were identified also in \cite{quadratic}.  Additionally, we find a pair of fluid-dominated models ($C_\pm$) and two pairs of Milne-like solutions for negative curvature ($D_\pm$ and $E_\pm$).  Moreover, through compactification we are able to identify the critical points $G_\pm$ and $H_\pm$ at infinity. On the other hand, contrary to \cite{quadratic}, in our setting we cannot identify a critical element corresponding to Minkowski spactime.

\begin{table}[]
\caption{Quadratic EoS: number of critical points appearing on the invariant subsets for different ranges of the parameters with $P_\star=0$. $C_\pm$ exist for any parameter ranges and curvature, while points $D_\pm$ exist for any parameter ranges for negative curvature. The points at infinity are not included.}
\label{tab:p0tab}
\begin{tabular}{|l|c|c|c|c|c|c|c|}
\hline
\multicolumn{1}{|c|}{$\delta$} & $\sigma$                                      & $A_+$ & $A_-$ & $B$ & $E_+$ & $E_-$ & Figure                                  \\ \hline
                               & $-\frac{1}{3}<\sigma$                         & 0     & 0     & 0   & 0     & 0     & Fig.\ref{fig:quadpz1}  \\ \cline{2-8} 
\multicolumn{1}{|c|}{$+1$}     & \multicolumn{1}{l|}{$-1<\sigma<-\frac{1}{3}$} & 0     & 0     & 1   & 0     & 0     & Fig.\ref{fig:quadpz-5} \\ \cline{2-8} 
                               & $\sigma<-1$                                   & 1     & 1     & 1   & 1     & 1     & Fig.\ref{fig:quadpz-2} \\ \hline
                               & $-\frac{1}{3}<\sigma$                         & 1     & 1     & 1   & 1     & 1     & Fig.\ref{fig:quadpz1}  \\ \cline{2-8} 
\multicolumn{1}{|c|}{$-1$}     & $-1<\sigma<-\frac{1}{3}$                      & 1     & 1     & 0   & 1     & 1     & Fig.\ref{fig:quadpz-5} \\ \cline{2-8} 
                               & $\sigma<-1$                                   & 0     & 0     & 0   & 0     & 0     & Fig.\ref{fig:quadpz-2} \\ \hline
\end{tabular}
\end{table}
\paragraph{Generic quadratic EoS.} We can write $P_\star$ as
\begin{equation}\label{eq:psquad}
    P_\star=\Delta\frac{\epsilon_c}{4\,\delta}\,(\sigma-\xi)^2,
\end{equation}
where $\xi\in\lbrace-\infty,+\infty\rbrace$ and $\Delta=-\delta\,\text{sgn}(\sigma-\xi)$. Writing it in terms of our dimensionless variables by using Eq.~\eqref{eq:quadodp} we get
\begin{equation}\label{eq:quadops}
    \Omega_{P_\star}=\frac{3}{2}\,\Delta\,\Omega_\epsilon\,\frac{(\sigma-\xi)^2}{\Omega_{\partial P}-\sigma}.
\end{equation}
By combining Eqs.~\eqref{eq:psquad} and \eqref{eq:quadops} along with our assumption $\Omega_{\epsilon}>0$, we get the following constraints on the allowed values of $\Omega_{\partial P}$ for $\sigma\neq\xi$:
\begin{align}
 &\text{if}\quad \delta>0\quad \Rightarrow\quad \Omega_{\partial P}>\sigma\\
 &\text{if}\quad \delta<0\quad \Rightarrow\quad \Omega_{\partial P}<\sigma
\end{align}
In the case $\sigma=\xi$, then $P_*=0$ and we reduce to the previous case.

By combining Eqs.~\eqref{eq:quadops} and \eqref{eq:quadomegap} we get the constraint
\begin{equation}\label{eq:quadconst}
    \Omega_P=\frac{3}{2}\,\frac{\Omega_\epsilon}{\Omega_{\partial P}-\sigma}\,\left(\Omega_{\partial P}^2-\sigma^2+\Delta\,(\sigma-\xi)^2\right),
\end{equation}
which together with Friedmann constraints \eqref{eq:fried_pos} or \eqref{eq:fried_neg} reduce the system to two dimensions, namely the remaining dynamical variables are $\Omega_{\partial P}$ and $\Omega_H$.  

Note that for $\Omega_{\partial P}=\sigma$, Eq.~\eqref{eq:quadops} becomes singular: however, this singularity does not affect the evolution equation~\eqref{eq:Csp}, since the denominator is cancelled by the $\Gamma$ given by Eq.~\eqref{eq:gammaquad}.  Actually, $\Omega_{\partial P}= \sigma$ is the intersection line between the plane $\{ \Omega_{\partial P} , \Omega_H \}$ and the case {\it a.} above.

The critical points discussed in Secs.~\ref{sec:CritPoi_ss} and~\ref{sec:roottrack} are now the following:
\begin{itemize}
    \item the critical points $A_\pm$ are located at 
        \begin{equation}
        \lbrace \Omega_H,\Omega_{\partial P}\rbrace=\lbrace \pm1,-1\pm\sqrt{1-\sigma\,(2-\sigma)-\Delta\,(\sigma-\xi)^2}\,\rbrace;
    \end{equation}
    \item the critical point $B$ is located at
        \begin{equation}
        \lbrace \Omega_H,\Omega_{\partial P}\rbrace=\lbrace 0,-\frac{1}{3}\pm\sqrt{\frac{1}{9}+\sigma\,(\frac{2}{3}+\sigma)-\Delta\,(\sigma-\xi)^2}\,\rbrace;
    \end{equation} 
    \item the critical points $C_\pm$ are located at
    \begin{equation}
        \lbrace \Omega_H,\Omega_{\partial P}\rbrace=\lbrace \pm1,\sigma\rbrace;
    \end{equation}
   \item the critical points $D_\pm$ are located at
    \begin{equation}
        \lbrace \Omega_H,\Omega_{\partial P}\rbrace=\lbrace \pm\frac{1}{\sqrt{2}},\sigma\rbrace;
    \end{equation}
\end{itemize}
The additional Milne-like critical points $E_\pm$ are located at
    \begin{equation}
        \lbrace \Omega_H,\Omega_{\partial P}\rbrace=\lbrace \pm\frac{1}{\sqrt{2}},-1\pm\sqrt{1-\sigma\,(2-\sigma)-\Delta\,(\sigma-\xi)^2}\,\rbrace.
    \end{equation}

In appendix B we show representative cases which are summarized in Table 3.  The comparison of the number of critical points  between our study and the analysis of \cite{quadratic} for this generic case follows the same lines as in paragraph {\it b}.

\begin{table*}[]
\caption{Quadratic EoS: number of critical points appearing on the invariant subsets for different ranges of the parameters with $P_\star\neq 0$. $C_\pm$ exist for any parameter ranges and curvature, while points $D_\pm$ exist for any parameter ranges for negative curvature. The points at infinity are not included.}
\label{tab:gentab}
\begin{tabular}{|l|l|c|c|c|c|c|c|c|c|c|}
\hline
\multicolumn{1}{|c|}{$\delta$} & \multicolumn{1}{c|}{$\sigma$}                 & $P_\star$                                                                                           & $\xi$                     & $\Omega_{P_\star}$                                                                                                                                                               & $A_+$ & $A_-$ & $B$ & $E_+$ & $E_-$ & Figure                              \\ \hline
\multicolumn{1}{|c|}{}         &                                               & $ \frac{\epsilon_c (1+3 \sigma)^2}{36\, \delta}<P_\star$                                            & $-\frac{1}{3}<\xi$        & $\frac{1}{6} \Omega_\epsilon \frac{(1+3\,\sigma)^2}{\Omega_{\partial P}-\sigma}<\Omega_{P_\star}$                                                                                & 0     & 0     & 0   & 0     & 0     & Fig.\ref{fig:pD.1} \\ \cline{3-11} 
\multicolumn{1}{|c|}{}         &                                               & $P_\star=\frac{\epsilon_c (1+3 \sigma)^2}{36\, \delta}$                                             & $\xi=-\frac{1}{3}$        & $\Omega_{P_\star}= \frac{1}{6} \Omega_\epsilon \frac{(1+3\,\sigma)^2}{\Omega_{\partial P}-\sigma}$                                                                               & 0     & 0     & 1   & 0     & 0     & Fig.\ref{fig:pD.2} \\ \cline{3-11} 
                               & \multicolumn{1}{c|}{$\sigma<-1$}              & $\frac{\epsilon_c (1+ \sigma)^2}{4\, \delta}<P_\star<\frac{\epsilon_c (1+3 \sigma)^2}{36\, \delta}$ & $-1<\xi<-\frac{1}{3}$     & $\frac{3}{2} \Omega_\epsilon \frac{(1+\sigma)^2}{\Omega_{\partial P}-\sigma}  <\Omega_{P_\star}< \frac{1}{6} \Omega_\epsilon \frac{(1+3\,\sigma)^2}{\Omega_{\partial P}-\sigma}$ & 0     & 0     & 2   & 0     & 0     & Fig.\ref{fig:pD.3} \\ \cline{3-11} 
\multicolumn{1}{|c|}{}         &                                               & $P_\star=\frac{\epsilon_c (1+ \sigma)^2}{4\, \delta}$                                               & $\xi=-1 $                 & $\Omega_{P_\star}=\frac{3}{2} \Omega_\epsilon \frac{(1+\sigma)^2}{\Omega_{\partial P}-\sigma}$                                                                                   & 1     & 1     & 2   & 1     & 1     & Fig.\ref{fig:pD.4} \\ \cline{3-11} 
                               &                                               & $0<P_\star<\frac{\epsilon_c (1+ \sigma)^2}{4\, \delta}$                                             & $\sigma<\xi<-1 $          & $0<\Omega_{P_\star}<\frac{3}{2} \Omega_\epsilon \frac{(1+\sigma)^2}{\Omega_{\partial P}-\sigma}$                                                                                 & 2     & 2     & 2   & 2     & 2     & Fig.\ref{fig:pD.5} \\ \cline{3-11} 
\multicolumn{1}{|c|}{+1}       &                                               & $ P_\star<0$                                                                                        & $\xi<\sigma$              & $\Omega_{P_\star}<0$                                                                                                                                                             & 1     & 1     & 1   & 1     & 1     & Fig.\ref{fig:pD.6} \\ \cline{2-11} 
                               &                                               & $ \frac{\epsilon_c (1+3 \sigma)^2}{36\, \delta}<P_\star$                                            & $-\frac{1}{3}<\xi$        & $\frac{1}{6} \Omega_\epsilon \frac{(1+3\,\sigma)^2}{\Omega_{\partial P}-\sigma}<\Omega_{P_\star}$                                                                                & 0     & 0     & 0   & 0     & 0     & Fig.\ref{fig:pD.1} \\ \cline{3-11} 
                               & $-1<\sigma<-\frac{1}{3}$                      & $P_\star=\frac{\epsilon_c (1+3 \sigma)^2}{36\, \delta}$                                             & $\xi=-\frac{1}{3}$        & $\Omega_{P_\star}= \frac{1}{6} \Omega_\epsilon \frac{(1+3\,\sigma)^2}{\Omega_{\partial P}-\sigma}$                                                                               & 0     & 0     & 1   & 0     & 0     & Fig.\ref{fig:pD.2} \\ \cline{3-11} 
                               &                                               & $0<P_\star<\frac{\epsilon_c (1+3 \sigma)^2}{36\, \delta}$                                           & $\sigma<\xi<-\frac{1}{3}$ & $0 <\Omega_{P_\star}< \frac{1}{6} \Omega_\epsilon \frac{(1+3\,\sigma)^2}{\Omega_{\partial P}-\sigma}$                                                                            & 0     & 0     & 2   & 0     & 0     & Fig.\ref{fig:pD.3} \\ \cline{3-11} 
                               &                                               & $ P_\star<0$                                                                                        & $\xi<\sigma$              & $\Omega_{P_\star}<0$                                                                                                                                                             & 1     & 1     & 1   & 1     & 1     & Fig.\ref{fig:pD.6} \\ \cline{2-11} 
\multicolumn{1}{|c|}{}         & \multicolumn{1}{c|}{$-\frac{1}{3}<\sigma$}    & $0< P_\star$                                                                                        & $\sigma<\xi$              & $0<\Omega_{P_\star}$                                                                                                                                                             & 0     & 0     & 0   & 0     & 0     & Fig.\ref{fig:pD.1} \\ \cline{3-11} 
                               &                                               & $ P_\star<0$                                                                                        & $\xi<\sigma$              & $\Omega_{P_\star}<0$                                                                                                                                                             & 1     & 1     & 1   & 1     & 1     & Fig.\ref{fig:pD.6} \\ \hline
\multicolumn{1}{|c|}{}         & \multicolumn{1}{c|}{$\sigma<-1$}              & $0< P_\star$                                                                                        & $\sigma<\xi$              & $0<\Omega_{P_\star}$                                                                                                                                                             & 1     & 1     & 1   & 1     & 1     & Fig.\ref{fig:nD.1} \\ \cline{3-11} 
                               &                                               & $ P_\star<0$                                                                                        & $\xi<\sigma$              & $\Omega_{P_\star}<0$                                                                                                                                                             & 0     & 0     & 0   & 0     & 0     & Fig.\ref{fig:nD.6} \\ \cline{2-11} 
                               &                                               & $0<P_\star$                                                                                         & $\sigma<\xi$              & $0<\Omega_{P_\star}$                                                                                                                                                             & 1     & 1     & 1   & 1     & 1     & Fig.\ref{fig:nD.1} \\ \cline{3-11} 
                               & \multicolumn{1}{c|}{$-1<\sigma<-\frac{1}{3}$} & $\frac{\epsilon_c (1+ \sigma)^2}{4\, \delta}<P_\star<0$                                             & $-1<\xi<\sigma$           & $\frac{3}{2} \Omega_\epsilon \frac{(1+\,\sigma)^2}{\Omega_{\partial P}-\sigma}<\Omega_{P_\star}<0$                                                                               & 2     & 2     & 0   & 2     & 2     & Fig.\ref{fig:nD.4} \\ \cline{3-11} 
                               &                                               & $P_\star=\frac{\epsilon_c (1+ \sigma)^2}{4\, \delta}$                                               & $\xi=-1$                  & $\Omega_{P_\star}=\frac{3}{2} \Omega_\epsilon \frac{(1+\sigma)^2}{\Omega_{\partial P}-\sigma}$                                                                                   & 1     & 1     & 0   & 1     & 1     & Fig.\ref{fig:nD.5} \\ \cline{3-11} 
\multicolumn{1}{|c|}{-1}       &                                               & $P_\star<\frac{\epsilon_c (1+ \sigma)^2}{4\, \delta}$                                               & $\xi<-1$                  & $\Omega_{P_\star}<\frac{3}{2} \Omega_\epsilon \frac{(1+\sigma)^2}{\Omega_{\partial P}-\sigma}$                                                                                   & 0     & 0     & 0   & 0     & 0     & Fig.\ref{fig:nD.6} \\ \cline{2-11} 
                               &                                               & $0<P_\star$                                                                                         & $\sigma<\xi$              & $0<\Omega_{P_\star}$                                                                                                                                                             & 1     & 1     & 1   & 1     & 1     & Fig.\ref{fig:nD.1} \\ \cline{3-11} 
\multicolumn{1}{|c|}{}         &                                               & $ \frac{\epsilon_c (1+3 \sigma)^2}{36\, \delta}<P_\star<0$                                          & $-\frac{1}{3}<\xi<\sigma$ & $\frac{1}{6} \Omega_\epsilon \frac{(1+3\,\sigma)^2}{\Omega_{\partial P}-\sigma}<\Omega_{P_\star}<0$                                                                              & 2     & 2     & 2   & 2     & 2     & Fig.\ref{fig:nD.2} \\ \cline{3-11} 
\multicolumn{1}{|c|}{}         & \multicolumn{1}{c|}{$-\frac{1}{3}<\sigma$}    & $ P_\star=\frac{\epsilon_c (1+3 \sigma)^2}{36\, \delta}$                                            & $\xi=-\frac{1}{3}$        & $\Omega_{P_\star}=\frac{1}{6} \Omega_\epsilon \frac{(1+3\,\sigma)^2}{\Omega_{\partial P}-\sigma}$                                                                                & 2     & 2     & 1   & 2     & 2     & Fig.\ref{fig:nD.3} \\ \cline{3-11} 
                               &                                               & $\frac{\epsilon_c (1+\sigma)^2}{4 \delta}< P_\star<\frac{\epsilon_c (1+3 \sigma)^2}{36\, \delta}$   & $-1<\xi<-\frac{1}{3}$     & $\frac{3}{2} \Omega_\epsilon \frac{(1+\sigma)^2}{\Omega_{\partial P}-\sigma}<\Omega_{P_\star}<\frac{1}{6} \Omega_\epsilon \frac{(1+3\,\sigma)^2}{\Omega_{\partial P}-\sigma}$    & 2     & 2     & 0   & 2     & 2     & Fig.\ref{fig:nD.4} \\ \cline{3-11} 
                               &                                               & $ P_\star=\frac{\epsilon_c (1+\sigma)^2}{4 \delta}$                                                 & $\xi=-1$                  & $\Omega_{P_\star}=\frac{3}{2} \Omega_\epsilon \frac{(1+\sigma)^2}{\Omega_{\partial P}-\sigma}$                                                                                   & 1     & 1     & 0   & 1     & 1     & Fig.\ref{fig:nD.5} \\ \cline{3-11} 
                               &                                               & $P_\star<\frac{\epsilon_c (1+\sigma)^2}{4 \delta}$                                                  & $\xi<-1$                  & $\Omega_{P_\star}<\frac{3}{2} \Omega_\epsilon \frac{(1+\sigma)^2}{\Omega_{\partial P}-\sigma}$                                                                                   & 0     & 0     & 0   & 0     & 0     & Fig.\ref{fig:nD.6} \\ \hline
\end{tabular}
\end{table*}
\section{Conclusions}\label{sec:concl}

This work introduces a framework to analyze dynamically systems of barotropic fluids with non-negative energy density in spatially curved FRW spacetimes in absence of the cosmological constant. First we have introduced the new variables and the new evolution parametrization of this framework along with the function $\Gamma$, which includes all the information about the EoS. In this general setup we have identified three critical lines:
\begin{itemize}
    \item two de Sitter for spatially flat FRW,
    \item one static universe for non-negative curvatures,
\end{itemize}
that are independent of the EoS. The stability of these lines depends on the value of the variable $\Omega_{\partial P}$ along the lines themselves. Then we have discussed general features of the function $\Gamma$:
\begin{itemize}
    \item we have shown that $\Gamma=\Gamma\left(\Omega_{\partial P},\frac{\Omega_P}{\Omega_\epsilon}\right)$;
    \item we have shown that the roots of $\Gamma$ are stationary points in time and in the case  $\Gamma=\Gamma(\Omega_{\partial P})$ they define invariant subsets;
    \item we have studied these invariant subsets in the case that there is a single root $\Tilde{\Omega}_{\partial P}$ and have found that:
    \begin{itemize}
        \item there is a pair of new critical points corresponding to one-fluid flat universe, whose stability depends on $\Tilde{\Omega}_{\partial P}$,
        \item there is a pair of Milne critical points for non-positive curvature, whose stability depends on $\Tilde{\Omega}_{\partial P}$. 
    \end{itemize}
\end{itemize}

In the second part of the work we have provided two examples of how the framework we have introduced can be used.
\begin{itemize}
    \item In the first example we have taken a function of $\Gamma$ linear in $\Omega_{\partial P}$ with two free parameters and through physically motivated arguments, like causality, we have trimmed the $\Gamma$ model to a one parameter model with specific value interval. The resulting EoS represents a linear superposition of an exotic fluid with stiff matter. The stiff matter part of EoS dominates for large energy densities, while for low energy densities the exotic fluid part takes over. In this example apart from the dynamical elements identified in the general setup, a new invariant subset and a new pair of critical points exist. The new pair corresponds to Milne-like models. Regarding the invariant subsets, the one coming from the general analysis confines our model to obey causality, while the new one does not allow the EoS to cross the $P=-\epsilon$ line.  
    \item In the second example we have applied our framework to the quadratic EoS studied in \cite{quadratic} and made the comparison with that study. We have identified all the critical points found in \cite{quadratic}, except from a critical point describing the  Minkowski spactime, and additionally we have found
    \begin{itemize}
        \item a pair of fluid-dominated models for the flat case,
        \item two pairs of Milne-like solutions for negative curvature,
        \item two pairs of critical points with Milne-like behavior at $\Omega_{\partial P}\rightarrow \pm \infty$.
    \end{itemize}
\end{itemize}

\begin{acknowledgments}
The authors would like to thank Sante Carloni for his comments. 
\end{acknowledgments}      

\appendix

\section{Function $\Gamma$ linear in $\frac{\Omega_P}{\Omega_\epsilon}$.}\label{sec:gammalinratio}

The linear $\Gamma$ function Eq.~\eqref{eq:linearG} can be written also in terms of the dimensionless combination $\frac{\Omega_P}{\Omega_{\epsilon}}$. By solving  Eq.~\eqref{eq:genomegp} in term of $\Omega_{\partial P}$
\begin{equation}
    \Omega_{\partial P}= \frac{1+\alpha}{3}\frac{\Omega_P-\Omega_{P_\star}}{\Omega_\epsilon}+\beta,
\end{equation}
and then substituting it into the Eq.~\eqref{eq:linearG} we get
\begin{equation}\label{eq:linGratio}
    \Gamma= \hat{\alpha}\,\frac{\hat{\Omega}_{P}}{\Omega_\epsilon}+\hat{\beta},
\end{equation}
where $\displaystyle \hat{\alpha}=\frac{\alpha\,(\alpha+1)}{3}$, $\displaystyle \hat{\beta}=\beta\,(\alpha+1)$ and $$\displaystyle  \hat{\Omega}_{P}=\Omega_P-\Omega_{P_\star}=\frac{P-P_\star}{D^2}.$$ In the case that $P_\star=0$, $\Gamma$ is just a linear function of $\frac{\Omega_P}{\Omega_\epsilon}$.

\section{Invariant subsets for quadratic EoS}

\begin{figure*}[ht]
    \centering
\includegraphics[width=0.43\textwidth]{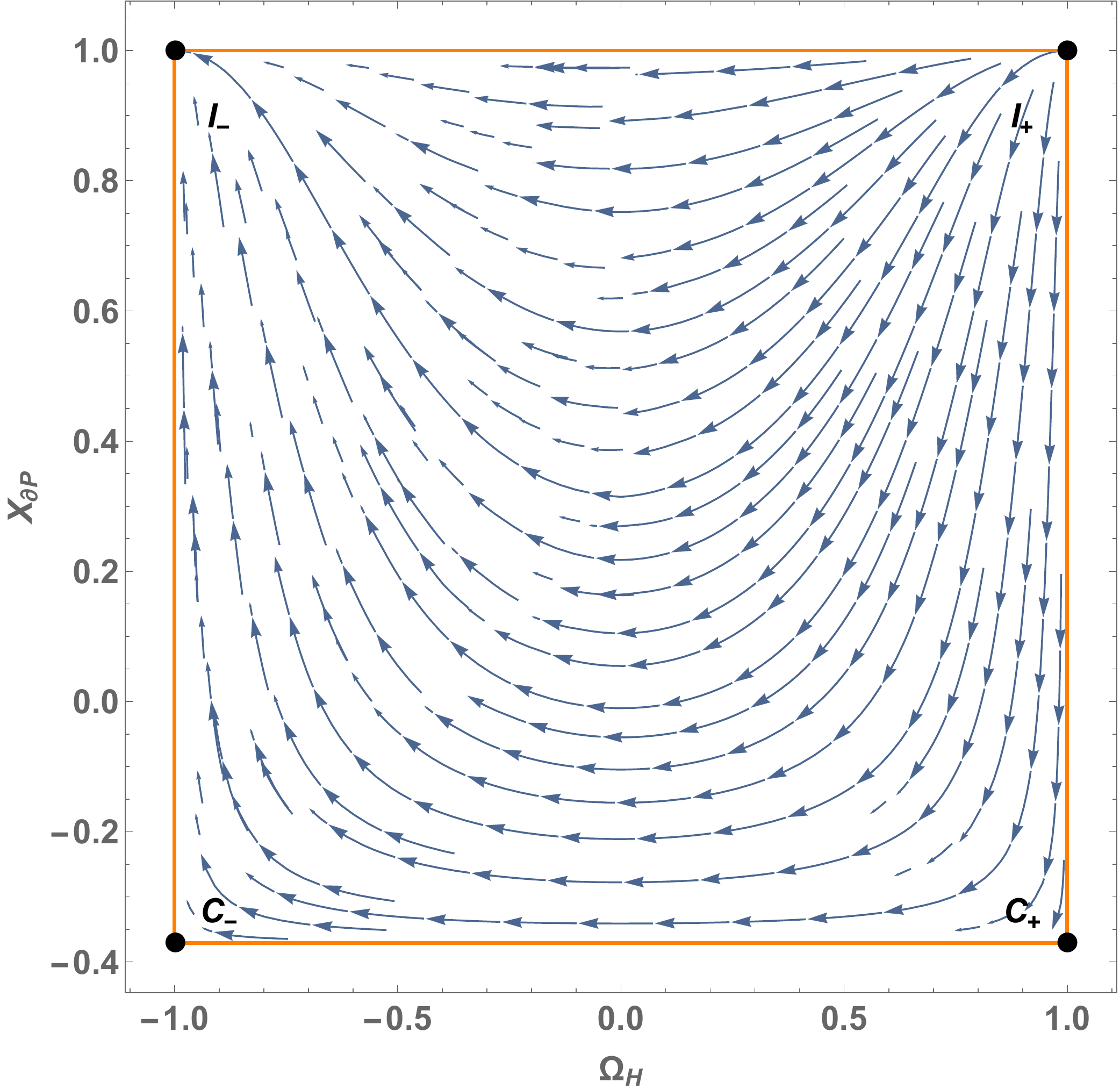}\quad
\includegraphics[width=0.43\textwidth]{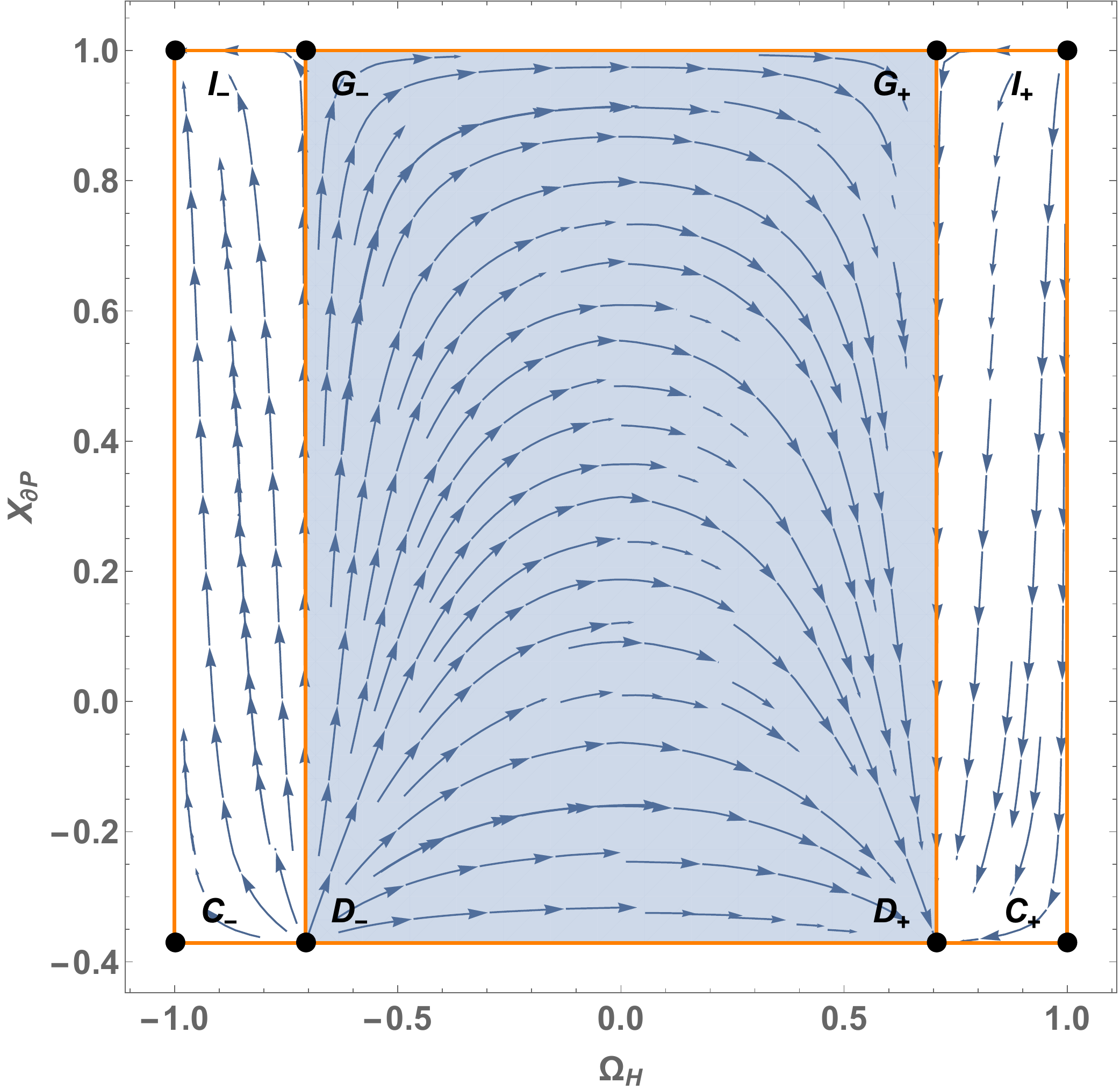}
    \caption{Invariant subsets $\lbrace \Omega_H,X_{\partial P}\rbrace$ for the quadratic EoS with $P_\star\neq 0$. The right panel corresponds to the positive spatial curvature and the left panel corresponds to the negative spatial curvature case. Invariant subsets are plotted for the parameters $\delta=1$, $\sigma=-4$, $\xi=1$ and $\zeta=0.1$ (these figures are topologically similar to the cases with parameters $\delta=1$, $-1<\sigma<-\frac{1}{3}$, $-\frac{1}{3}<\xi$  and also $\delta=1$, $-\frac{1}{3}<\sigma$, $\sigma<\xi$). The orange thick lines are the separatrices of the system, the blue region corresponds to $\Omega_\epsilon<0$. This figure corresponds to Fig.~10 in \cite{quadratic}. }\label{fig:pD.1}
\end{figure*}

\begin{figure*}[ht]
    \centering
\includegraphics[width=0.43\textwidth]{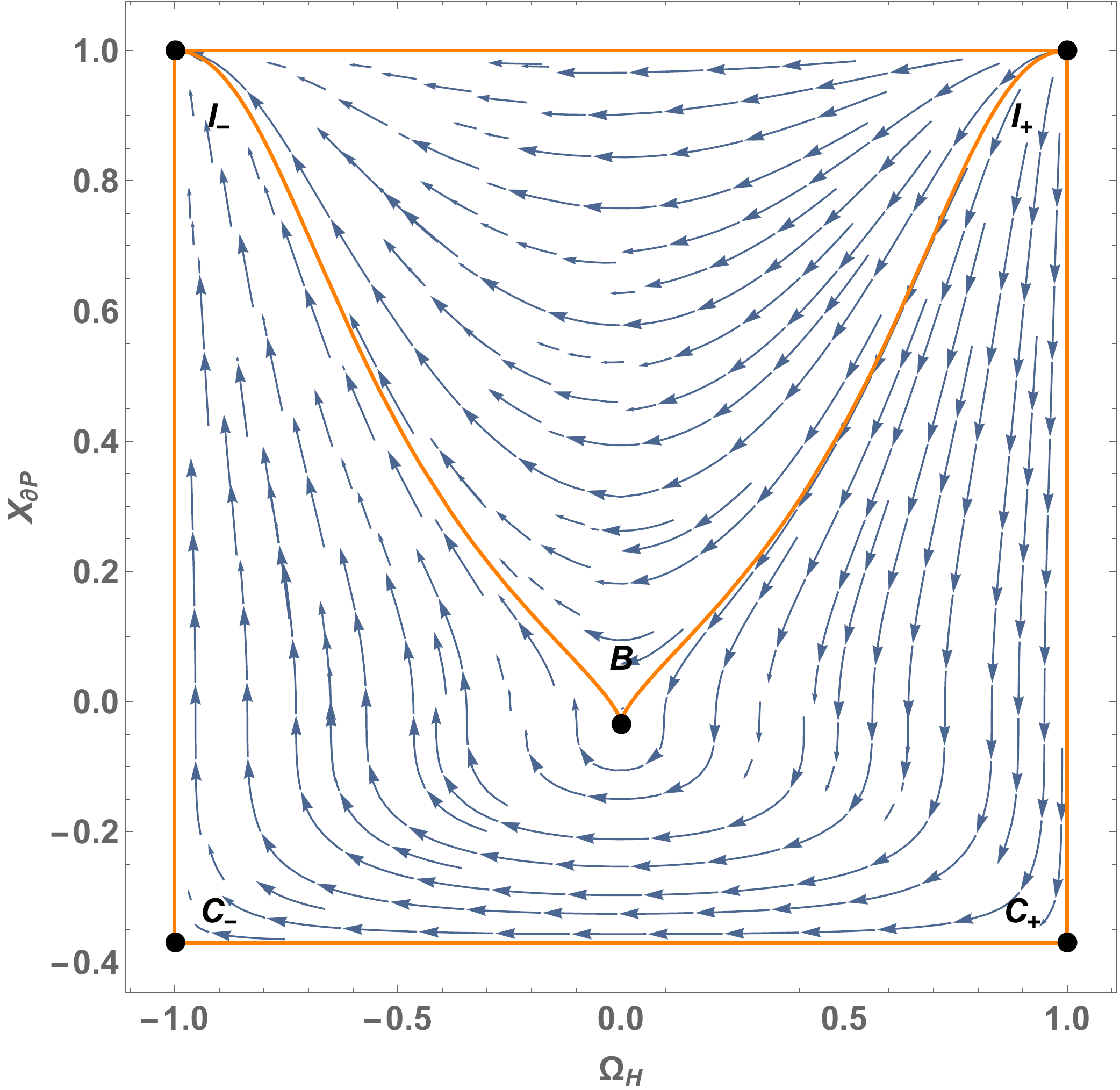}\quad
\includegraphics[width=0.43\textwidth]{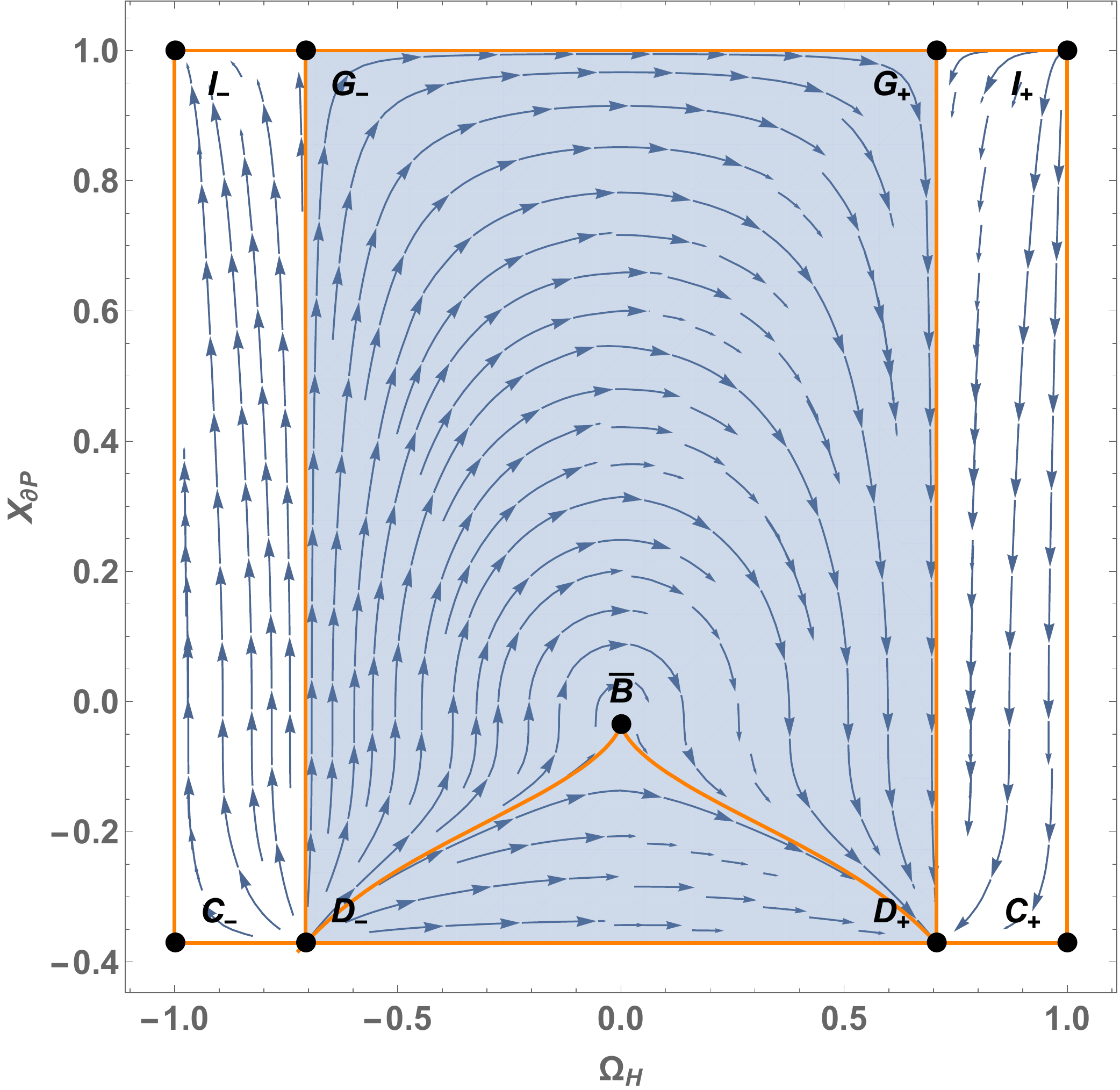}
    \caption{Invariant subsets $\lbrace \Omega_H,X_{\partial P}\rbrace$ for the quadratic EoS with $P_\star\neq 0$. The right panel corresponds to the positive spatial curvature and the left panel corresponds to the negative spatial curvature case. The invariant subsets are plotted for the parameters $\delta=1$, $\sigma=-4$, $\xi=-\frac{1}{3}$ and $\zeta=0.1$ (these figures are topologically similar to the case with parameters $\delta=1$, $-1<\sigma<-\frac{1}{3}$ and $\xi=-\frac{1}{3}$). The orange thick lines are the separatrices of the system, the blue region corresponds to $\Omega_\epsilon<0$.  This figure corresponds to Fig.~14 in \cite{quadratic}.}\label{fig:pD.2}
\end{figure*}

\begin{figure*}[ht]
    \centering
\includegraphics[width=0.43\textwidth]{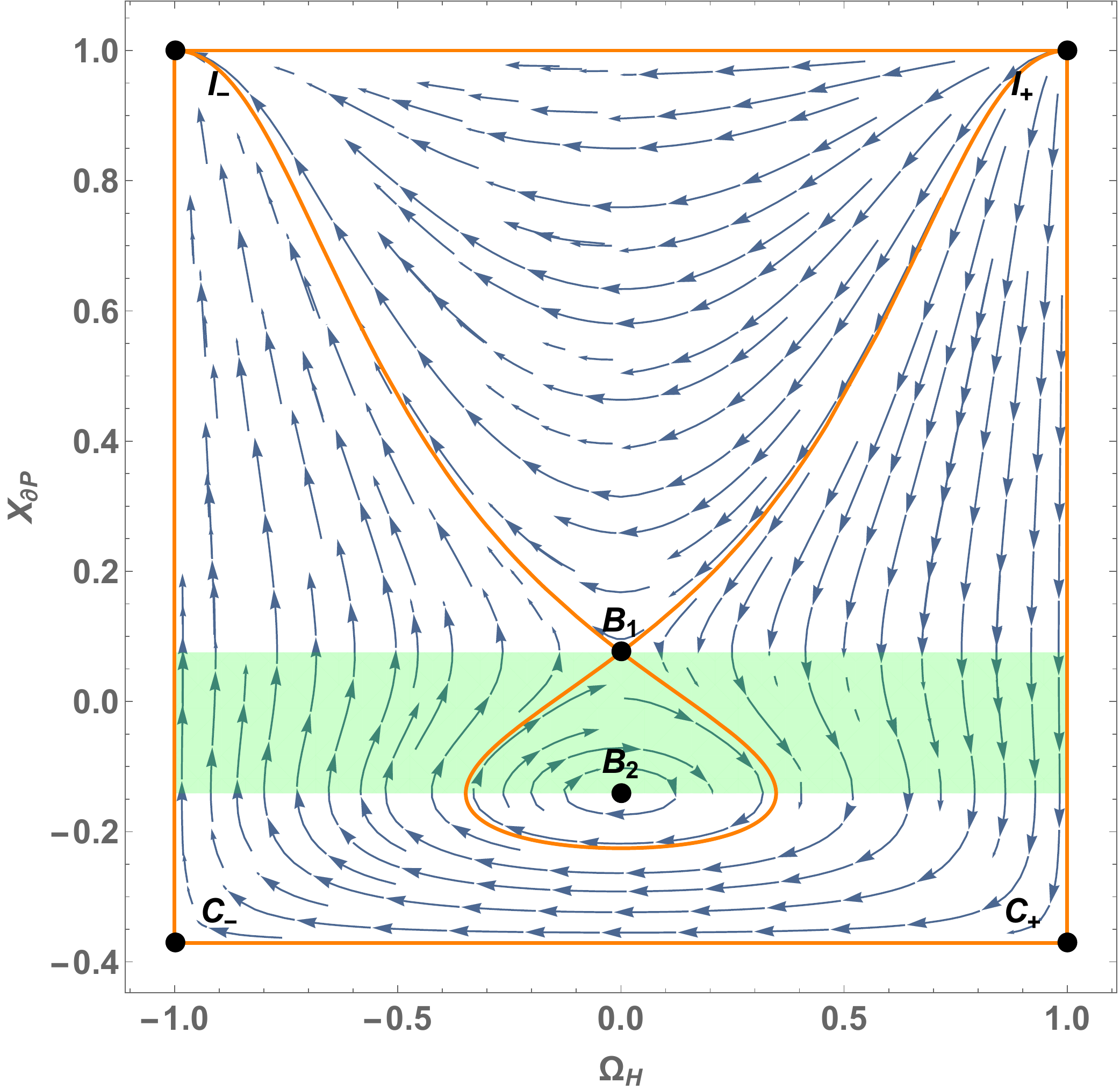}\quad
\includegraphics[width=0.43\textwidth]{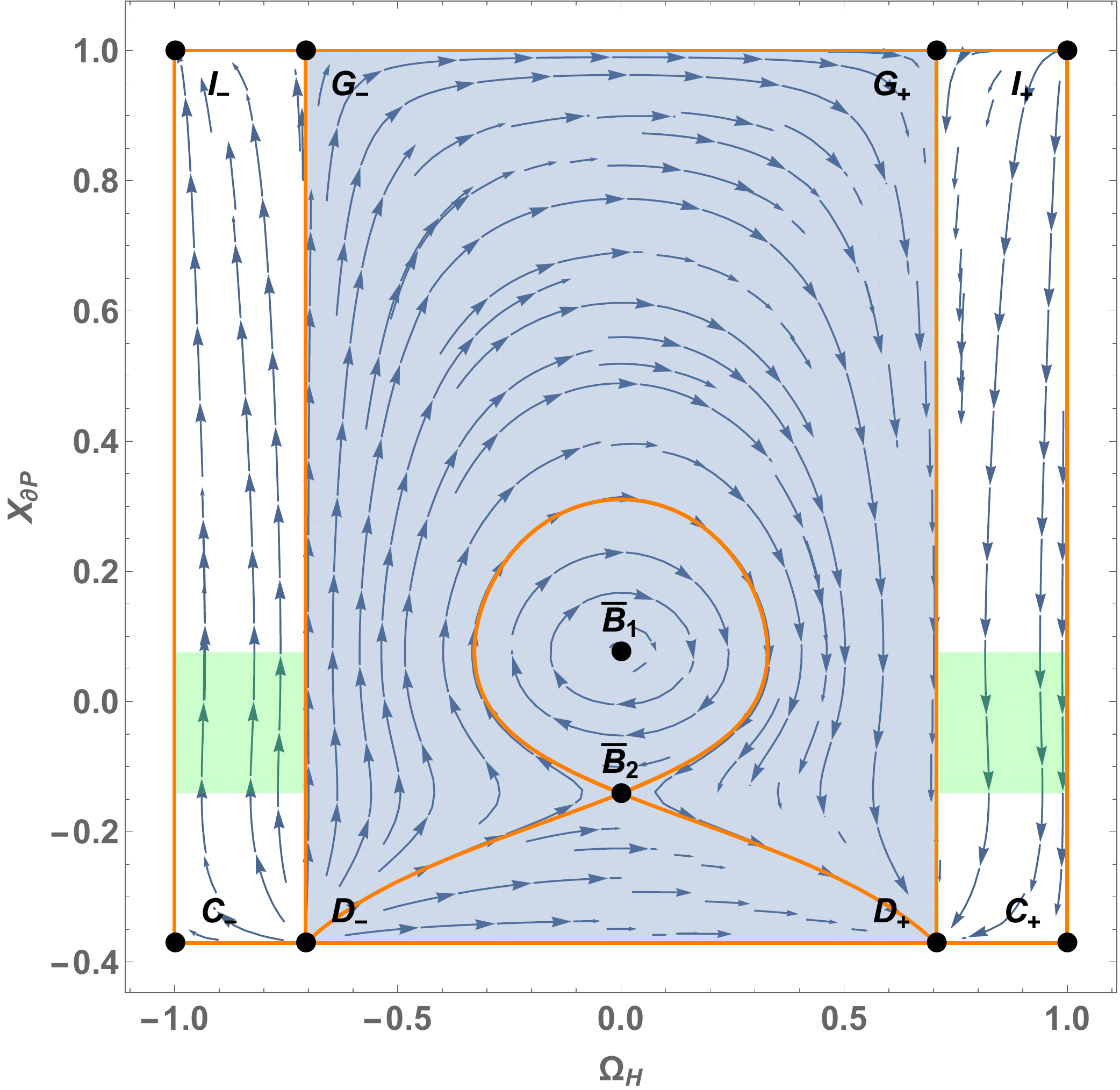}
    \caption{Invariant subsets $\lbrace \Omega_H,X_{\partial P}\rbrace$ for the quadratic EoS with $P_\star\neq 0$. The right panel corresponds to the positive spatial curvature and the left panel corresponds to the negative spatial curvature case. The invariant subsets are plotted for the parameters $\delta=1$, $\sigma=-4$, $\xi=-0.5$ and $\zeta=0.1$  (these figures are topologically similar to the case with parameters $\delta=1$, $-1<\sigma<-\frac{1}{3}$, $\sigma<\xi<-\frac{1}{3}$). The orange thick lines are the separatrices of the system, the blue region corresponds to $\Omega_\epsilon<0$ and the green shaded region corresponds to accelerated dynamics. This figure corresponds to Fig.~15 in \cite{quadratic}. }\label{fig:pD.3}
\end{figure*}

\begin{figure*}[ht]
    \centering
\includegraphics[width=0.43\textwidth]{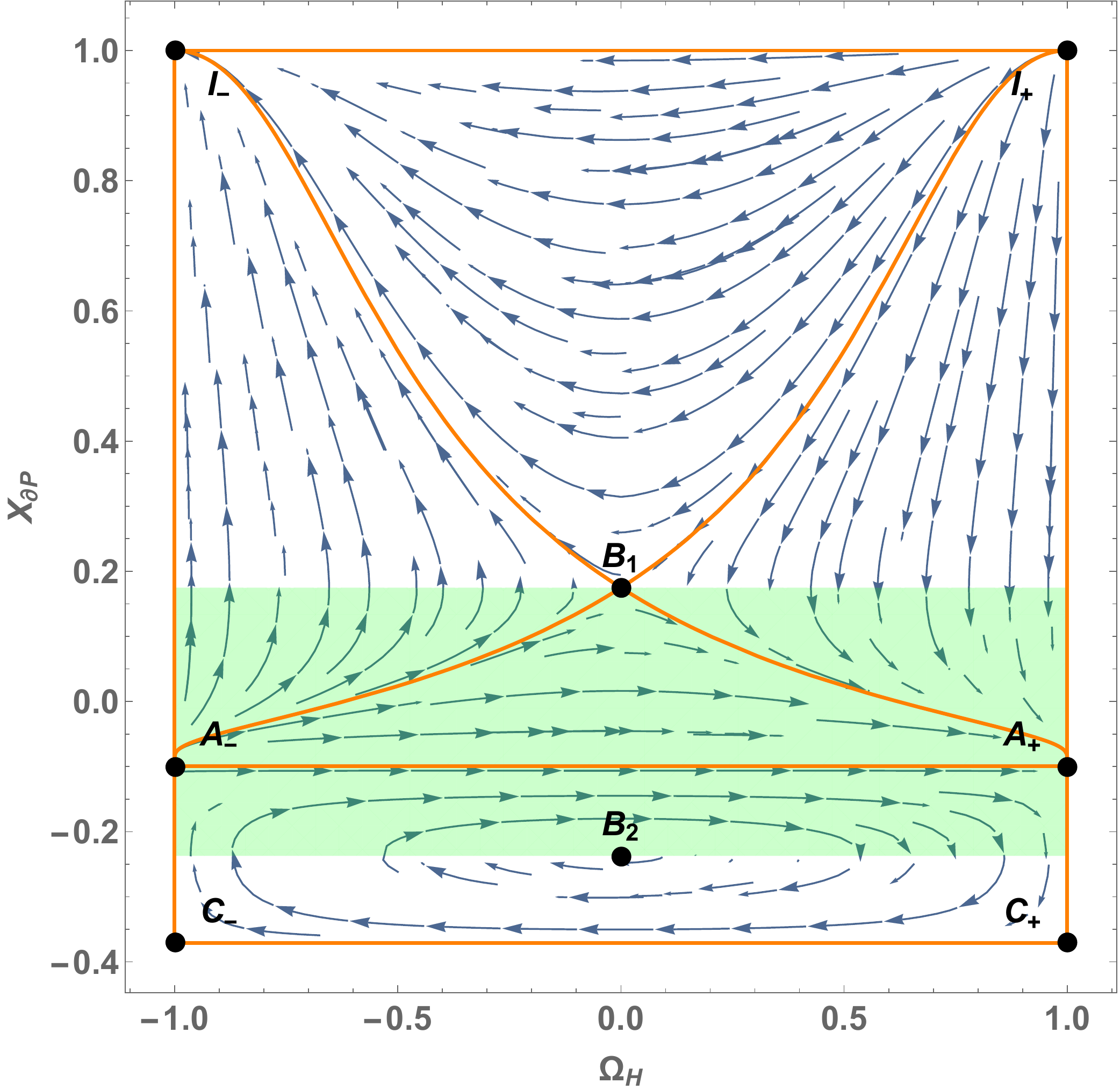}\quad
\includegraphics[width=0.43\textwidth]{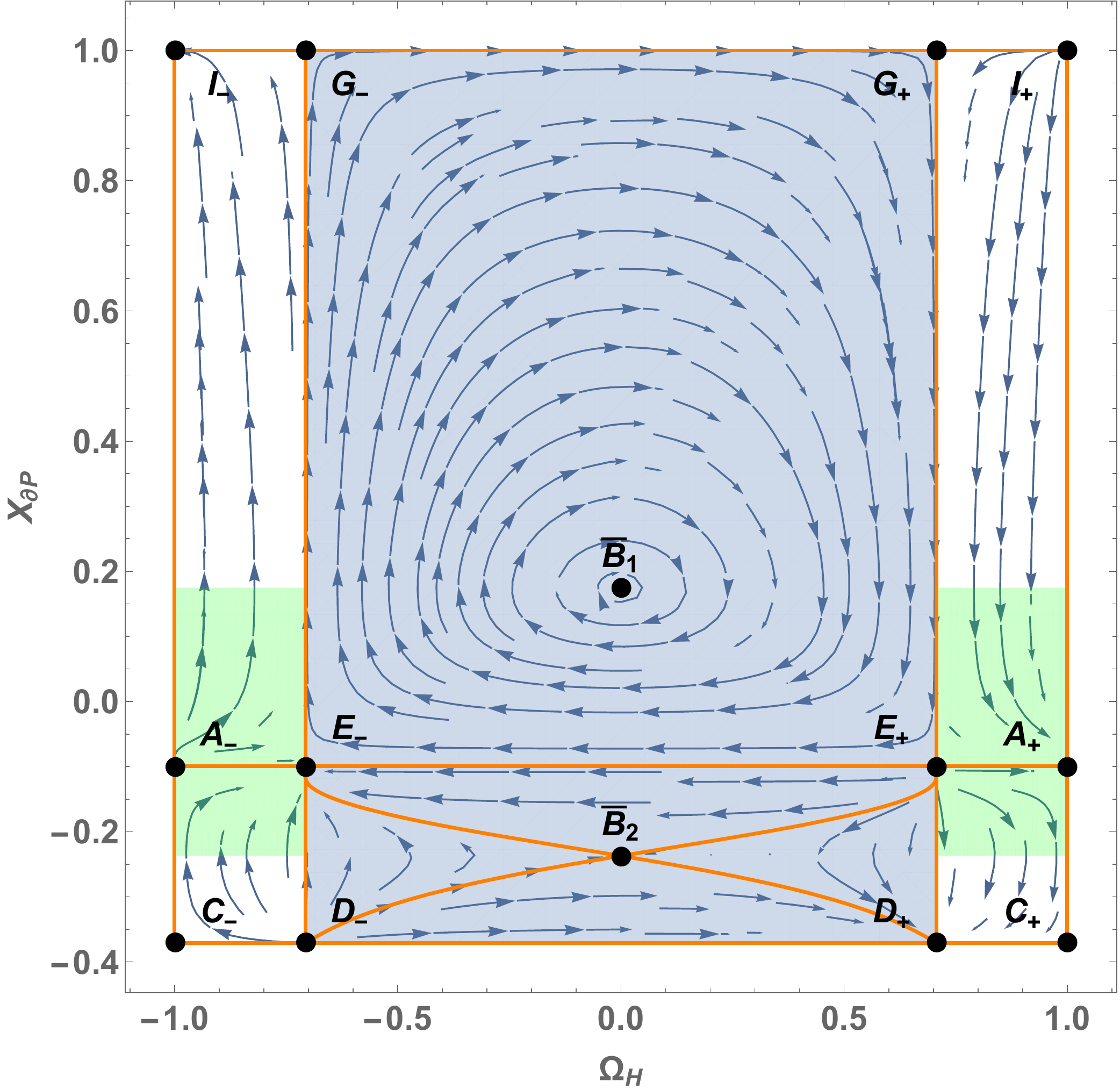}
    \caption{Invariant subsets $\lbrace \Omega_H,X_{\partial P}\rbrace$ for the quadratic EoS with $P_\star\neq 0$. The right panel corresponds to the positive spatial curvature and the left panel corresponds to the negative spatial curvature case. The invariant subsets are plotted for the parameters $\delta=1$, $\sigma=-4$, $\xi=-1$ and $\zeta=0.1$. The orange thick lines are the separatrices of the system, the blue region corresponds to $\Omega_\epsilon<0$ and the green shaded region corresponds to accelerated dynamics. This figure corresponds to Fig.~16 in \cite{quadratic}. }\label{fig:pD.4}
\end{figure*}

\begin{figure*}[ht]
    \centering
\includegraphics[width=0.43\textwidth]{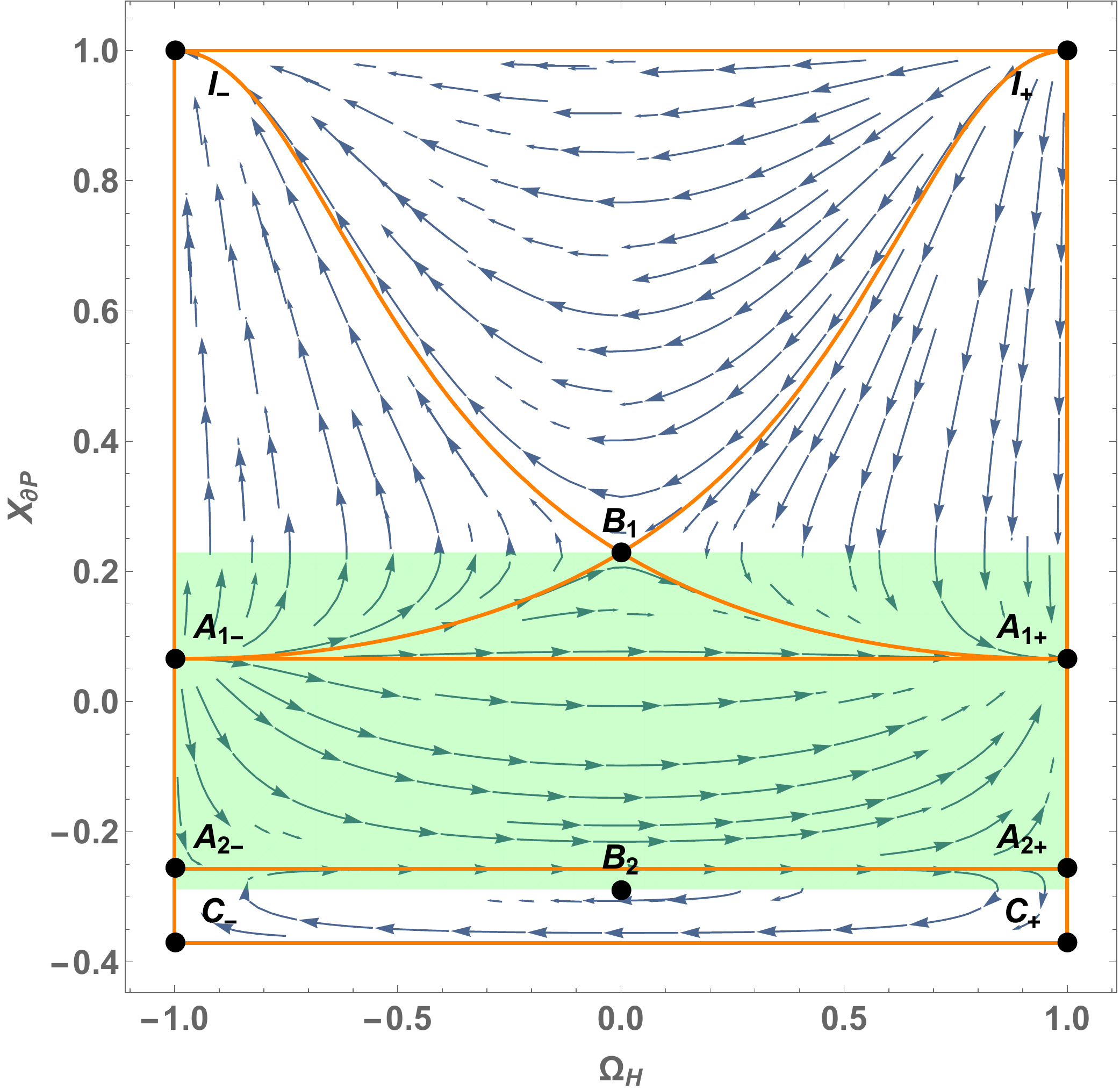}\quad
\includegraphics[width=0.43\textwidth]{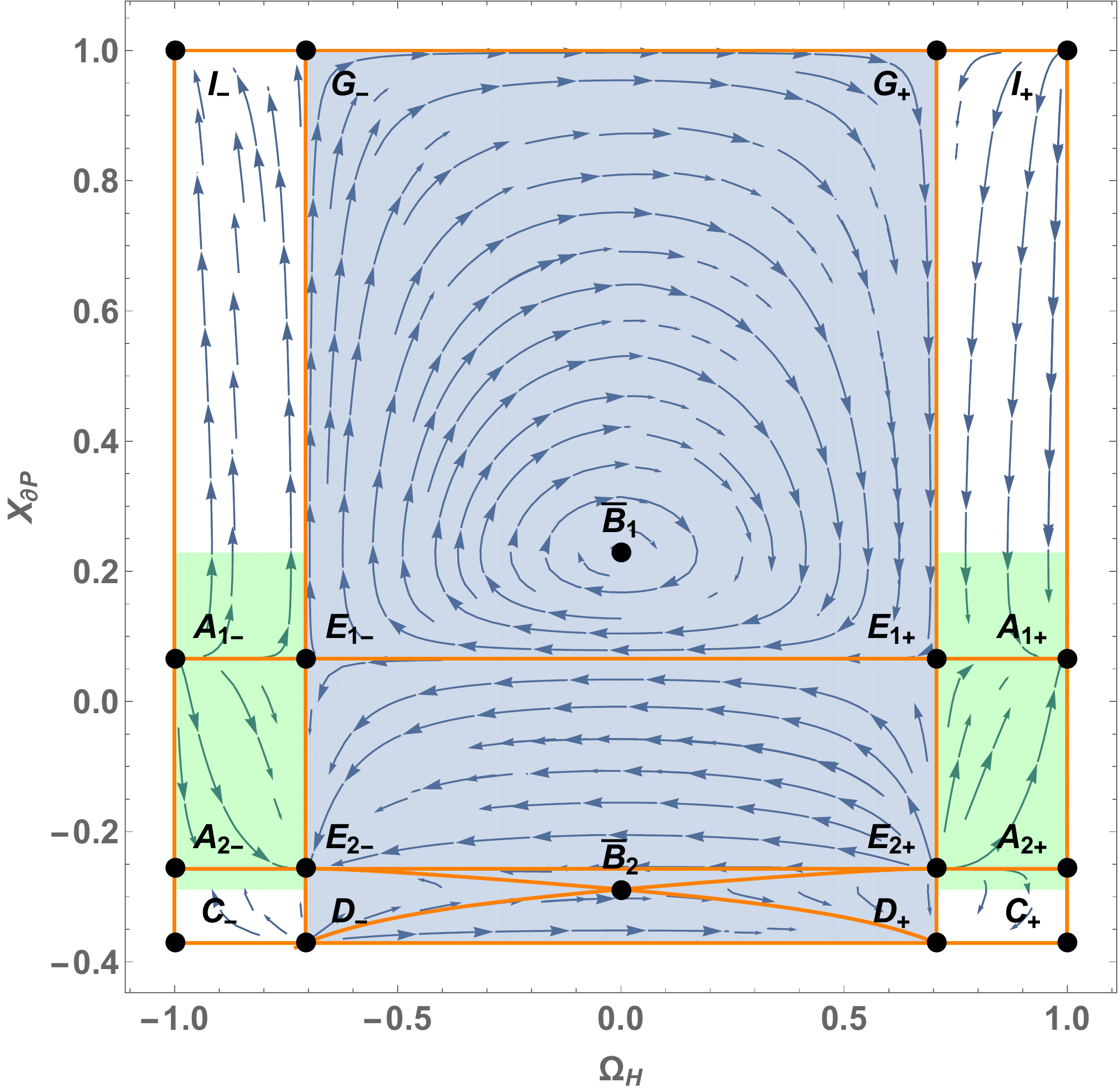}
    \caption{Invariant subsets $\lbrace \Omega_H,X_{\partial P}\rbrace$ for the quadratic EoS with $P_\star\neq 0$. The right panel corresponds to the positive spatial curvature and the left panel corresponds to the negative spatial curvature case. The invariant subsets are plotted for the parameters $\delta=1$, $\sigma=-4$, $\xi=-1.5$ and $\zeta=0.1$. The orange thick lines are the separatrices of the system, the blue region corresponds to $\Omega_\epsilon<0$ and the green shaded region corresponds to accelerated dynamics. This figure corresponds to Fig.~17 in \cite{quadratic}. }\label{fig:pD.5}
\end{figure*}

\begin{figure*}[ht]
    \centering
\includegraphics[width=0.43\textwidth]{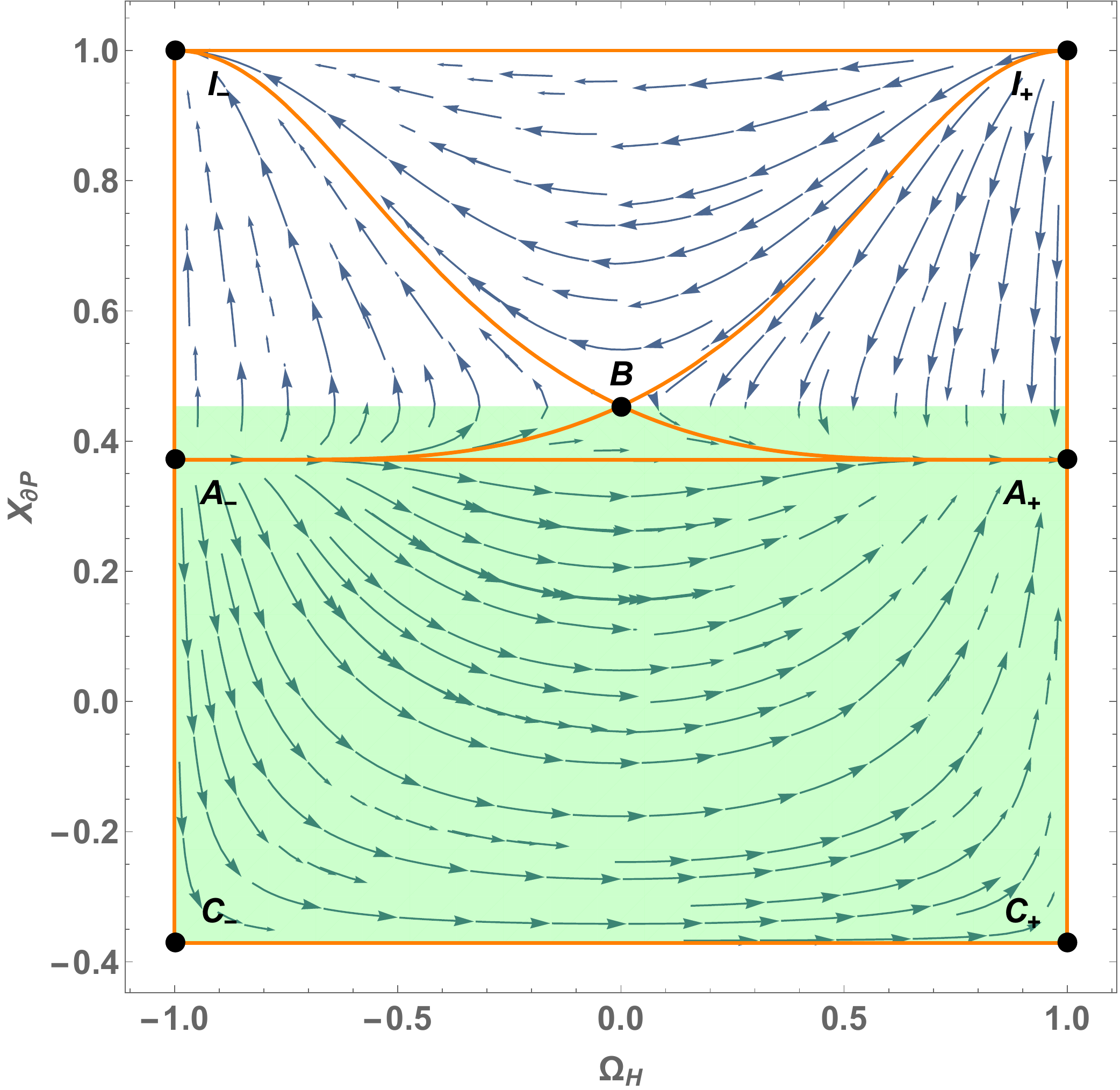}\quad
\includegraphics[width=0.43\textwidth]{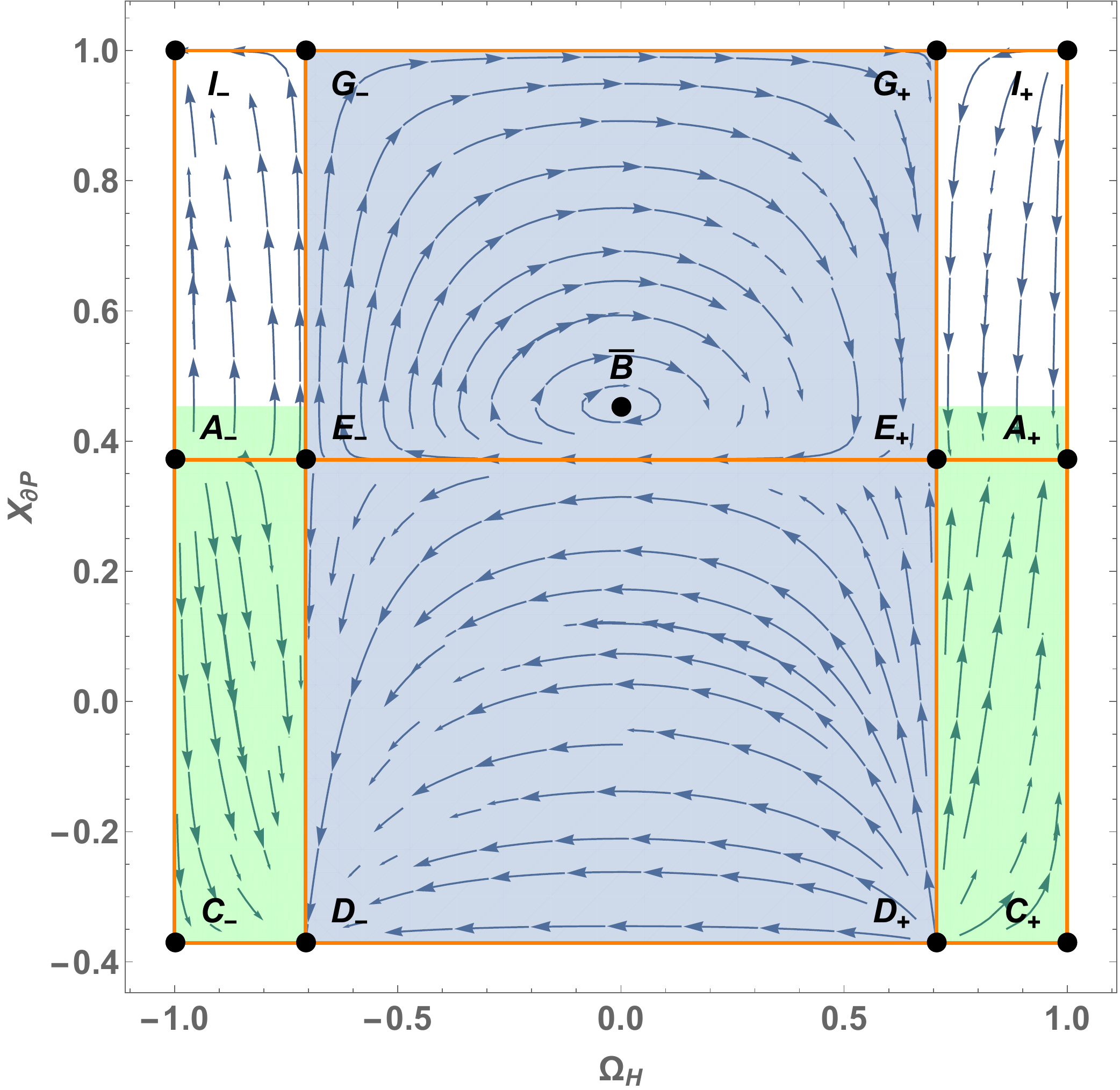}
    \caption{Invariant subsets $\lbrace \Omega_H,X_{\partial P}\rbrace$ for the quadratic EoS with $P_\star\neq 0$. The right panel corresponds to the positive spatial curvature and the left panel corresponds to the negative spatial curvature case. The invariant subsets are plotted for the parameters $\delta=1$, $\sigma=-4$, $\xi=-8$ and $\zeta=0.1$ (these figures are topologically similar to the cases with parameters $\delta=1$, $-1<\sigma<-\frac{1}{3}$, $\xi<\sigma$  and also $\delta=1$, $-\frac{1}{3}<\sigma$, $\xi<\sigma$). The orange thick lines are the separatrices of the system, the blue region corresponds to $\Omega_\epsilon<0$ and the green shaded region corresponds to accelerated dynamics. This figure corresponds to Fig.~13 in \cite{quadratic}.}\label{fig:pD.6}
\end{figure*}

\begin{figure*}[ht]
    \centering
\includegraphics[width=0.43\textwidth]{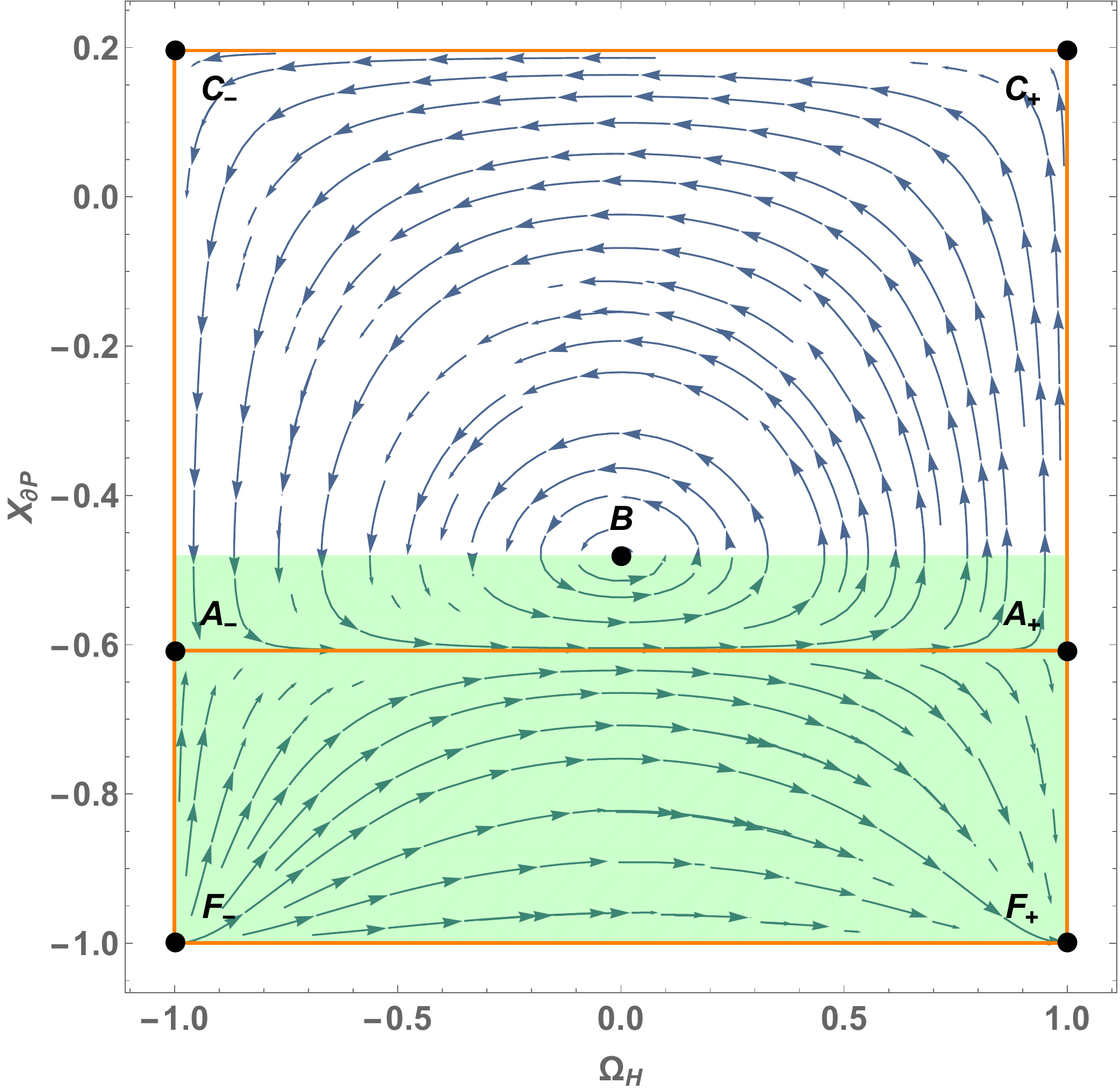}\quad
\includegraphics[width=0.43\textwidth]{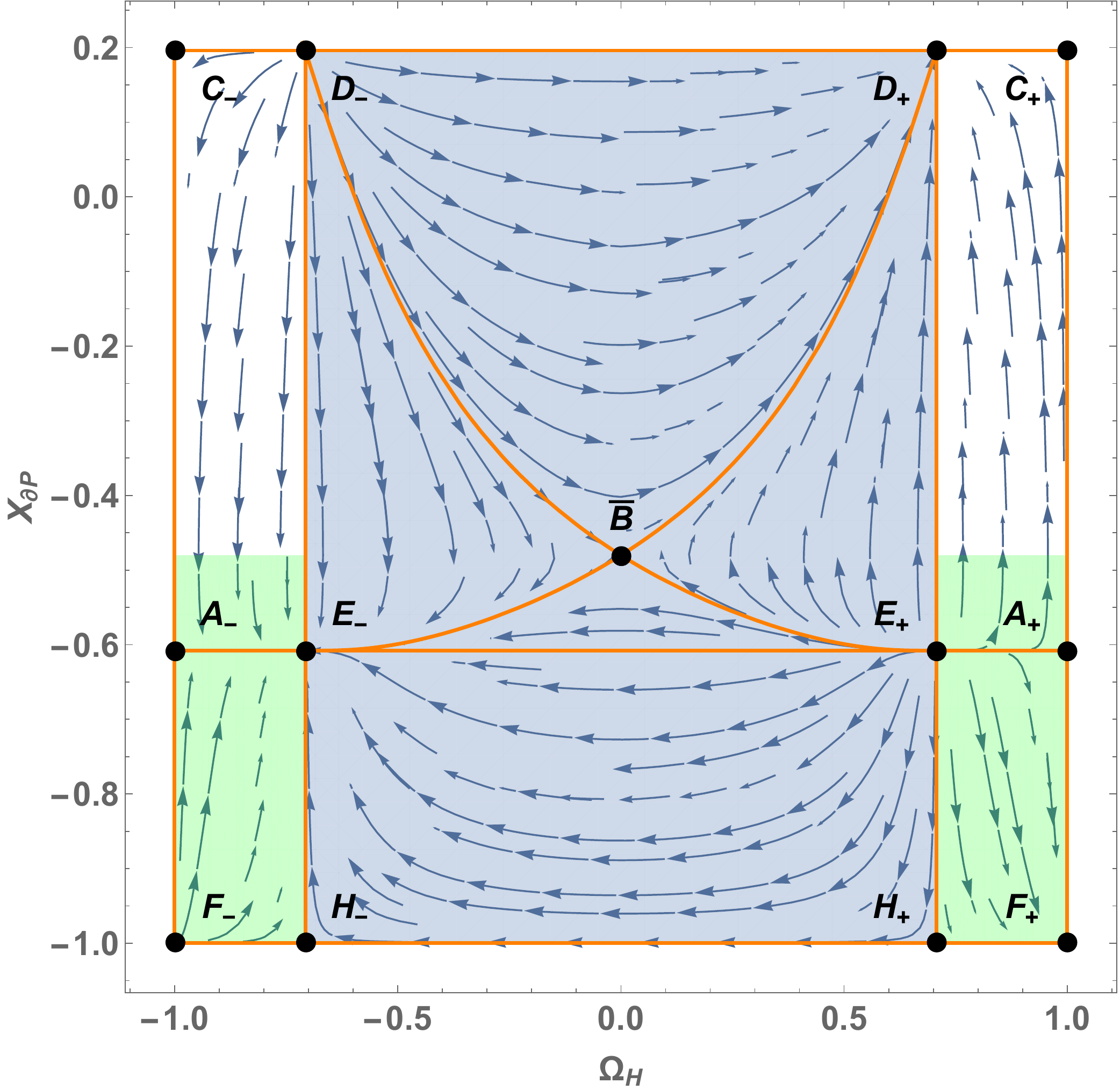}
    \caption{Invariant subsets $\lbrace \Omega_H,X_{\partial P}\rbrace$ for the quadratic EoS with $P_\star\neq 0$. The right panel corresponds to the positive spatial curvature and the left panel corresponds to the negative spatial curvature case. The invariant subsets are plotted for the parameters $\delta=-1$, $\sigma=1$, $\xi=3$ and $\zeta=0.2$ (these figures are topologically similar to the cases with parameters $\delta=-1$, $\sigma<-1$, $\sigma<\xi$  and also $\delta=-1$, $-1<\sigma<-\frac{1}{3}$, $\sigma<\xi$). The orange thick lines are the separatrices of the system, the blue region corresponds to $\Omega_\epsilon<0$ and the green shaded region corresponds to accelerated dynamics. This figure corresponds to Fig.~8 in \cite{quadratic}.}\label{fig:nD.1}
\end{figure*}

\begin{figure*}[ht]
    \centering
\includegraphics[width=0.43\textwidth]{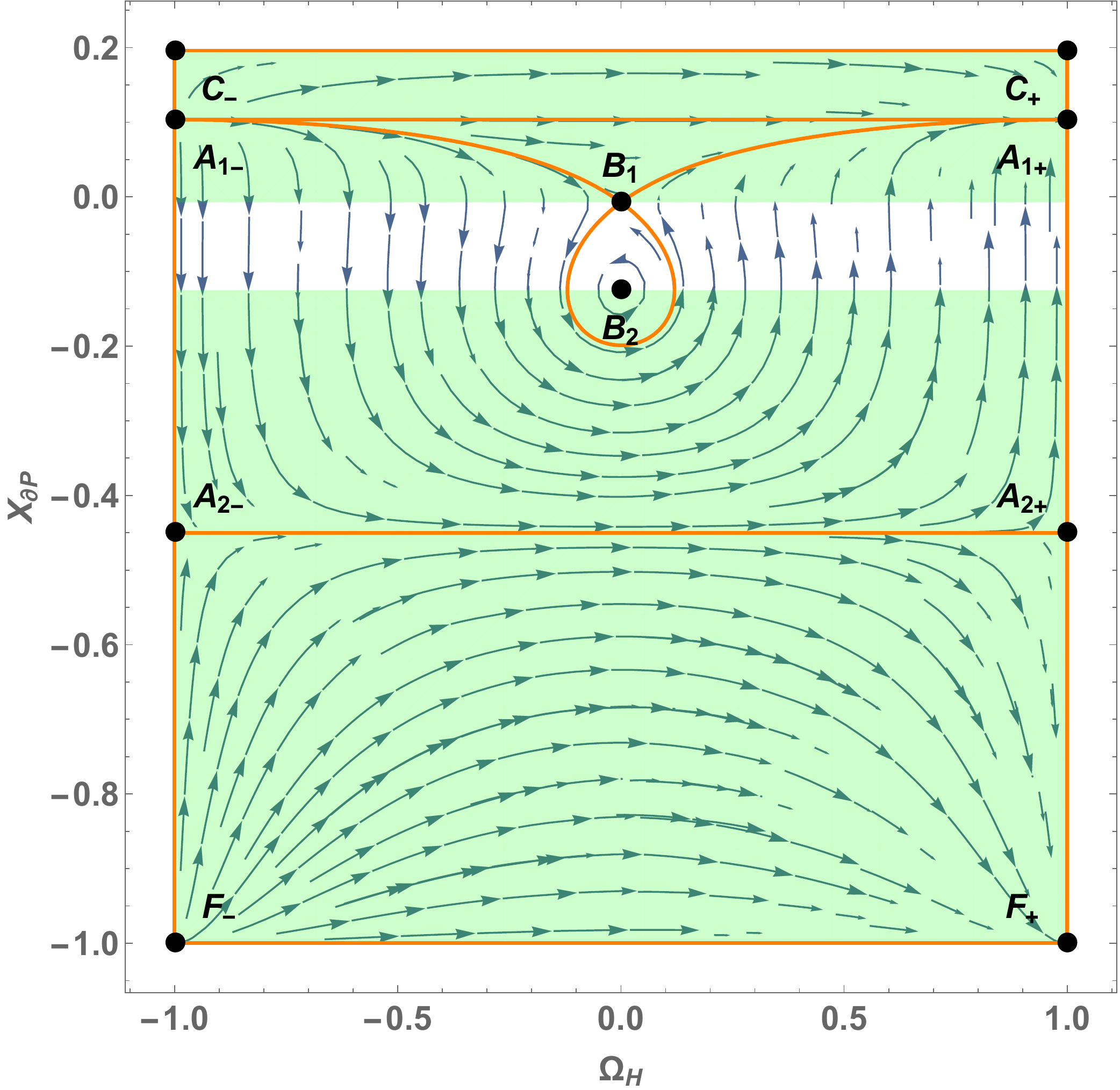}\quad
\includegraphics[width=0.43\textwidth]{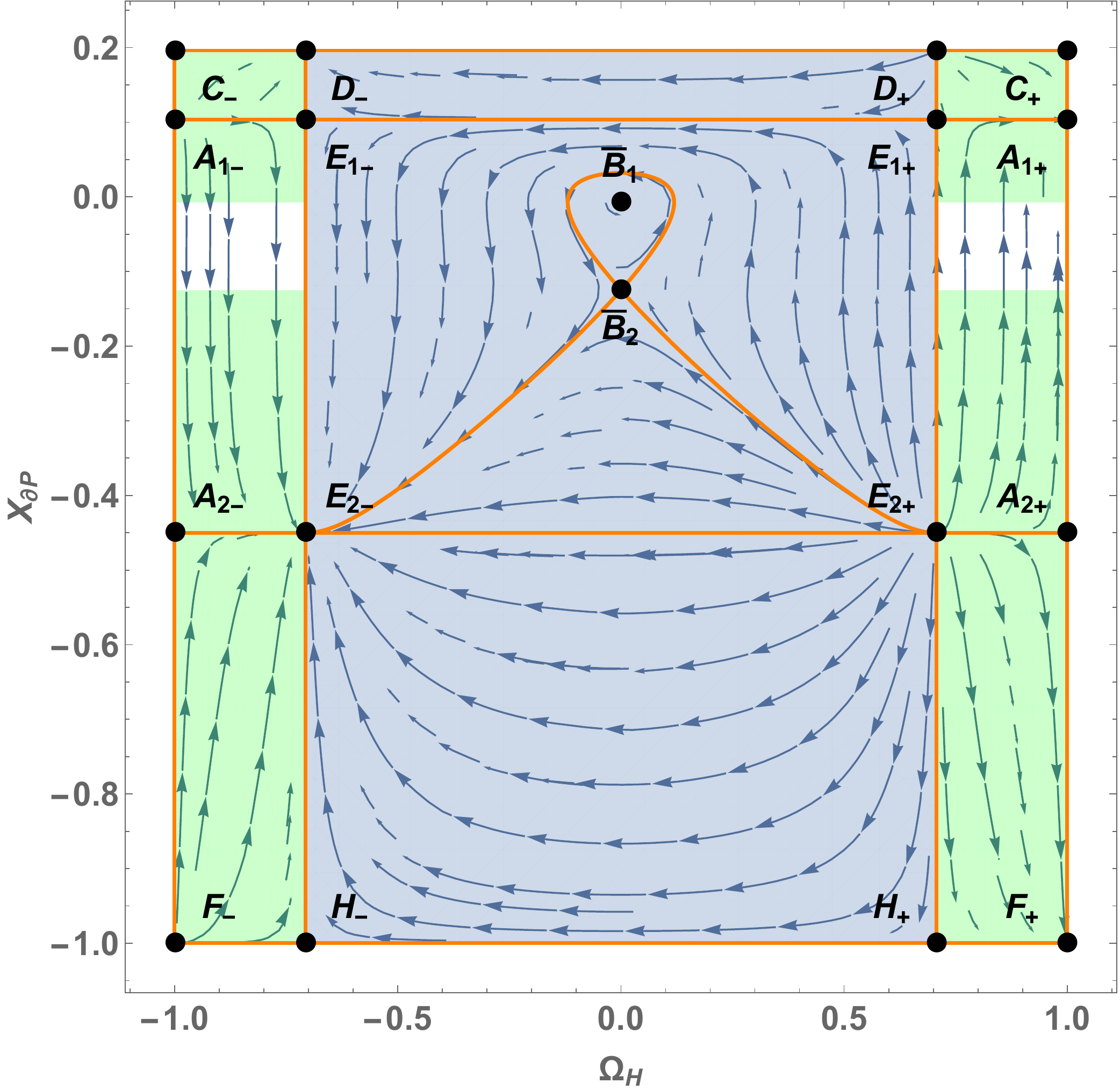}
    \caption{Invariant subsets $\lbrace \Omega_H,X_{\partial P}\rbrace$ for quadratic EoS with $P_\star\neq 0$. The right panel corresponds to the positive spatial curvature and the left panel corresponds to the negative spatial curvature case. The invariant subsets are plotted for the parameters $\delta=-1$, $\sigma=1$, $\xi=-0.3$ and $\zeta=0.2$. The orange thick lines are the separatrices of the system, the blue region corresponds to $\Omega_\epsilon<0$ and the green shaded region corresponds to accelerated dynamics the blue regions are not covered by the analysis. This figure corresponds to Fig.~18 in \cite{quadratic}. }\label{fig:nD.2}
\end{figure*}

\begin{figure*}[ht]
    \centering
\includegraphics[width=0.43\textwidth]{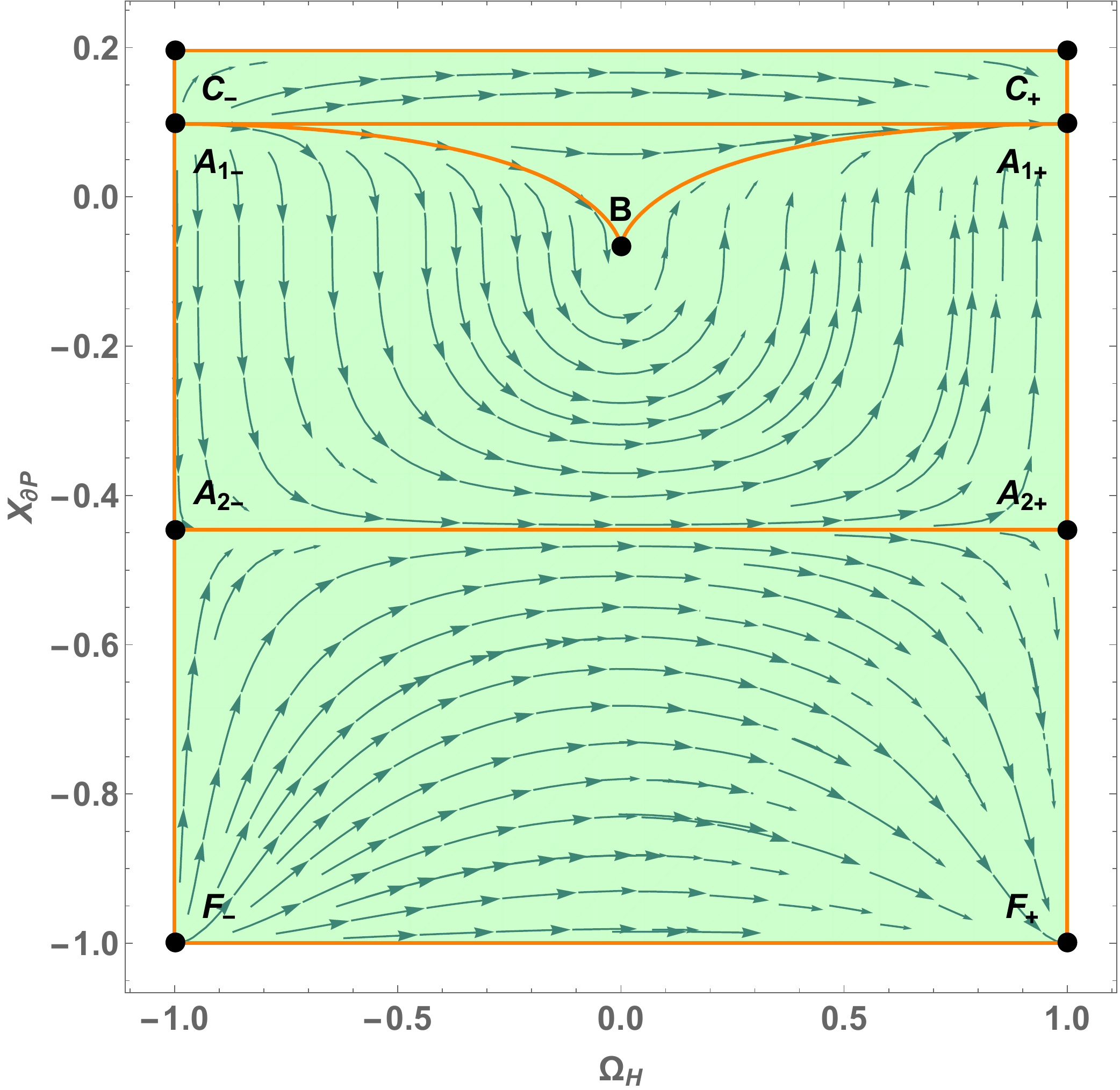}\quad
\includegraphics[width=0.43\textwidth]{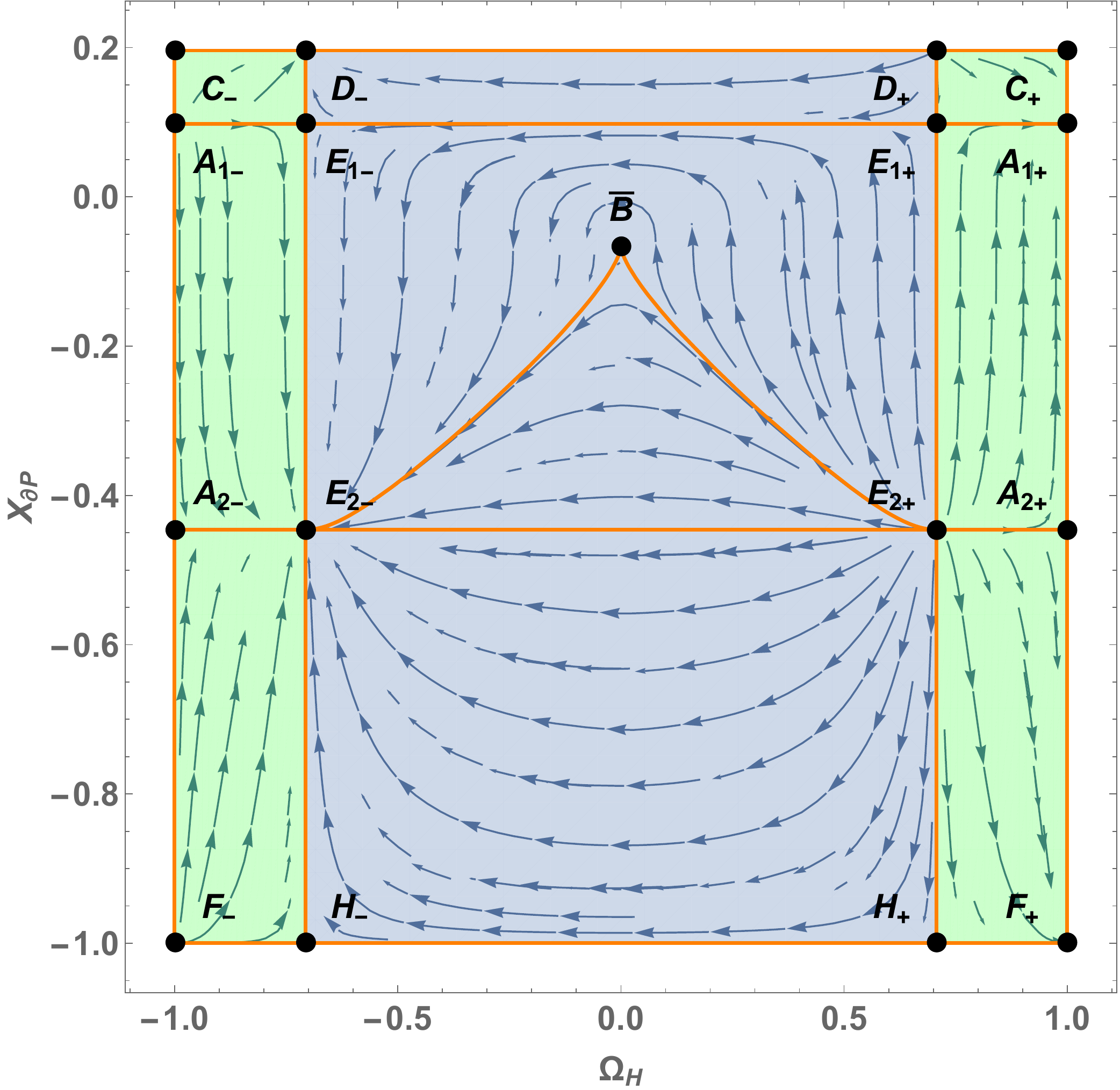}
    \caption{Invariant subsets $\lbrace \Omega_H,X_{\partial P}\rbrace$ for quadratic EoS with $P_\star\neq 0$. The right panel corresponds to the positive spatial curvature and the left panel corresponds to the negative spatial curvature case. Invariant subsets are plotted for the parameters $\delta=-1$, $\sigma=1$, $\xi=-\frac{1}{3}$ and $\zeta=0.2$. The orange thick lines are the separatrices of the system, the blue region corresponds to $\Omega_\epsilon<0$ and the green shaded region corresponds to accelerated dynamics. This figure corresponds to Fig.~19 in \cite{quadratic}.}\label{fig:nD.3}
\end{figure*}

\begin{figure*}[ht]
    \centering
\includegraphics[width=0.43\textwidth]{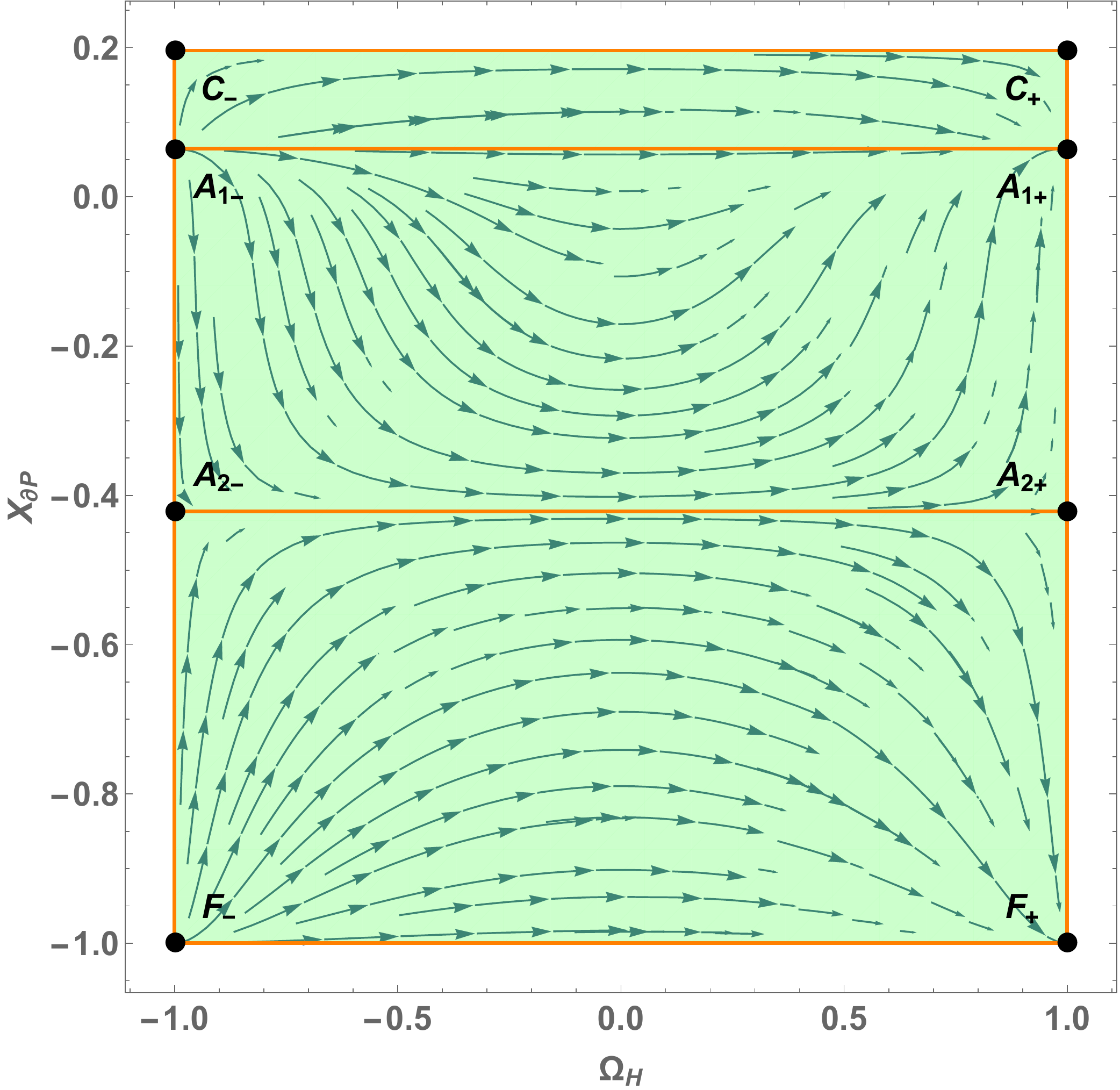}\quad
\includegraphics[width=0.43\textwidth]{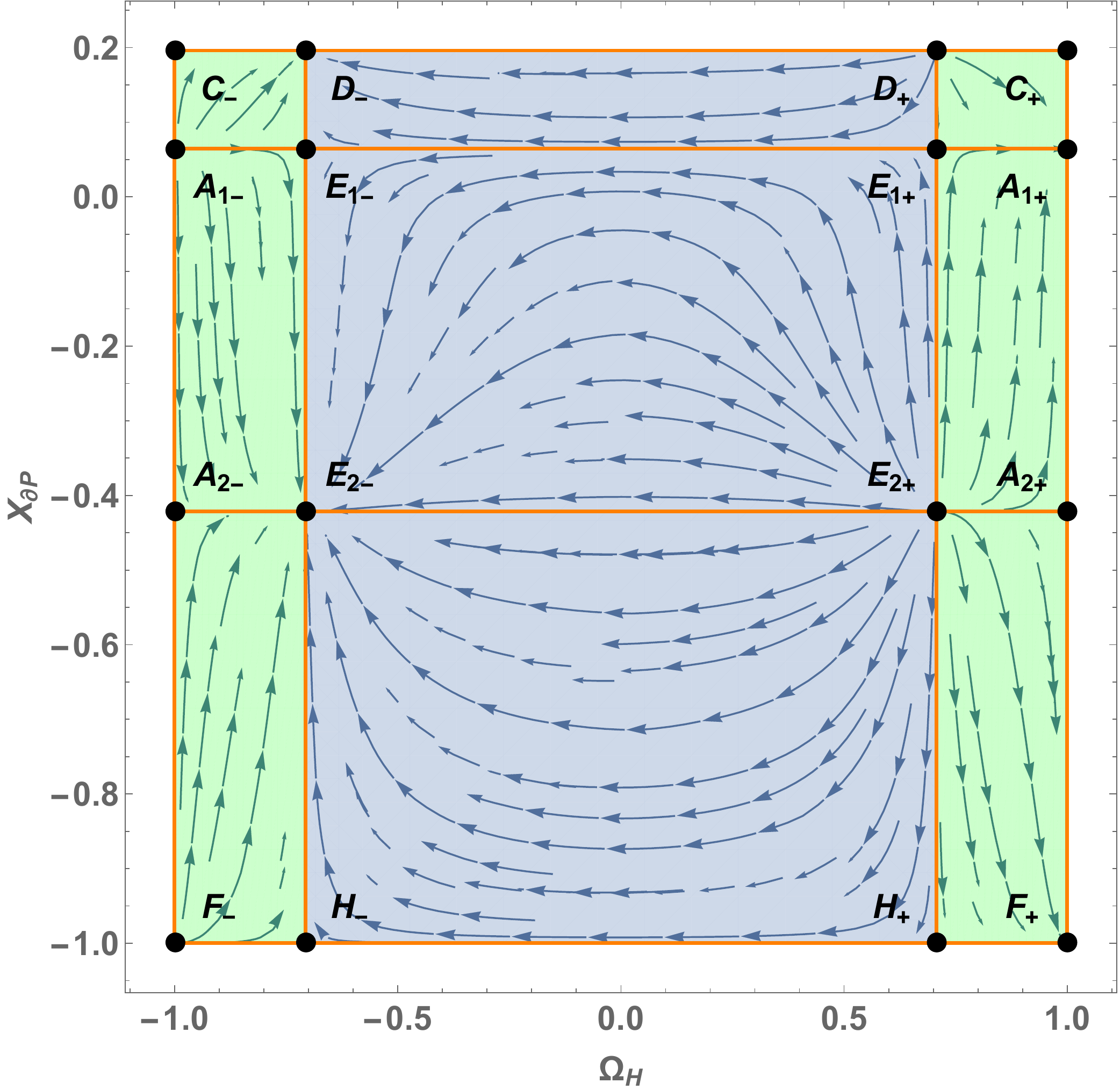}
    \caption{Invariant subsets $\lbrace \Omega_H,X_{\partial P}\rbrace$ for the quadratic EoS with $P_\star\neq 0$. The right panel corresponds to the positive spatial curvature and the left panel corresponds to the negative spatial curvature case. Invariant subsets are plotted for the parameters $\delta=-1$, $\sigma=1$, $\xi=-0.5$ and $\zeta=0.2$ (these figures are topologically similar to the case with parameters $\delta=-1$, $-1<\sigma<-\frac{1}{3}$, $-1<\xi<\sigma$). The orange thick lines are the separatrices of the system, the blue region corresponds to $\Omega_\epsilon<0$ and the green shaded region corresponds to accelerated dynamics. This figure corresponds to Fig.~20 in \cite{quadratic}. }\label{fig:nD.4}
\end{figure*}

\begin{figure*}[ht]
    \centering
\includegraphics[width=0.43\textwidth]{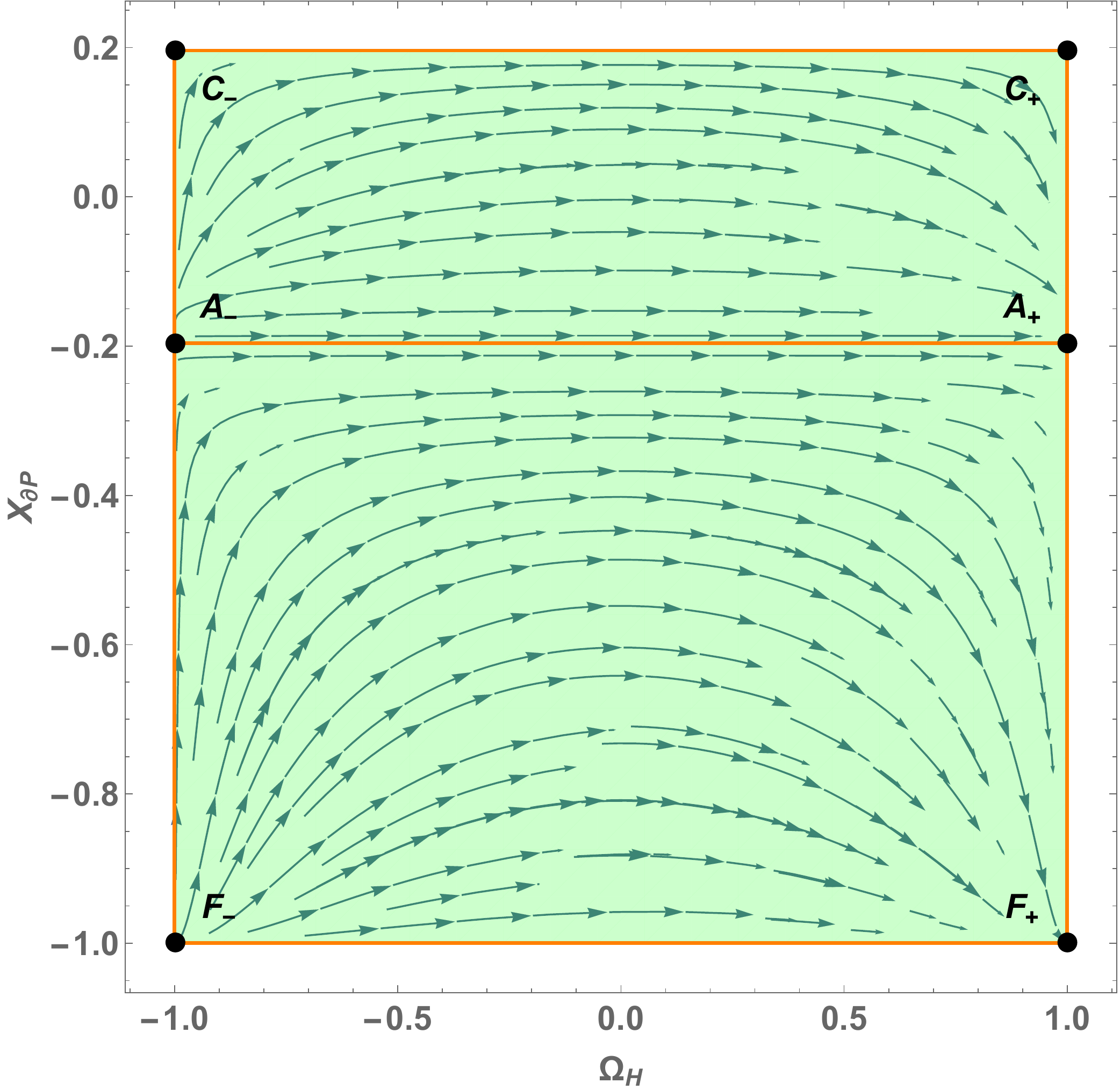}\quad
\includegraphics[width=0.43\textwidth]{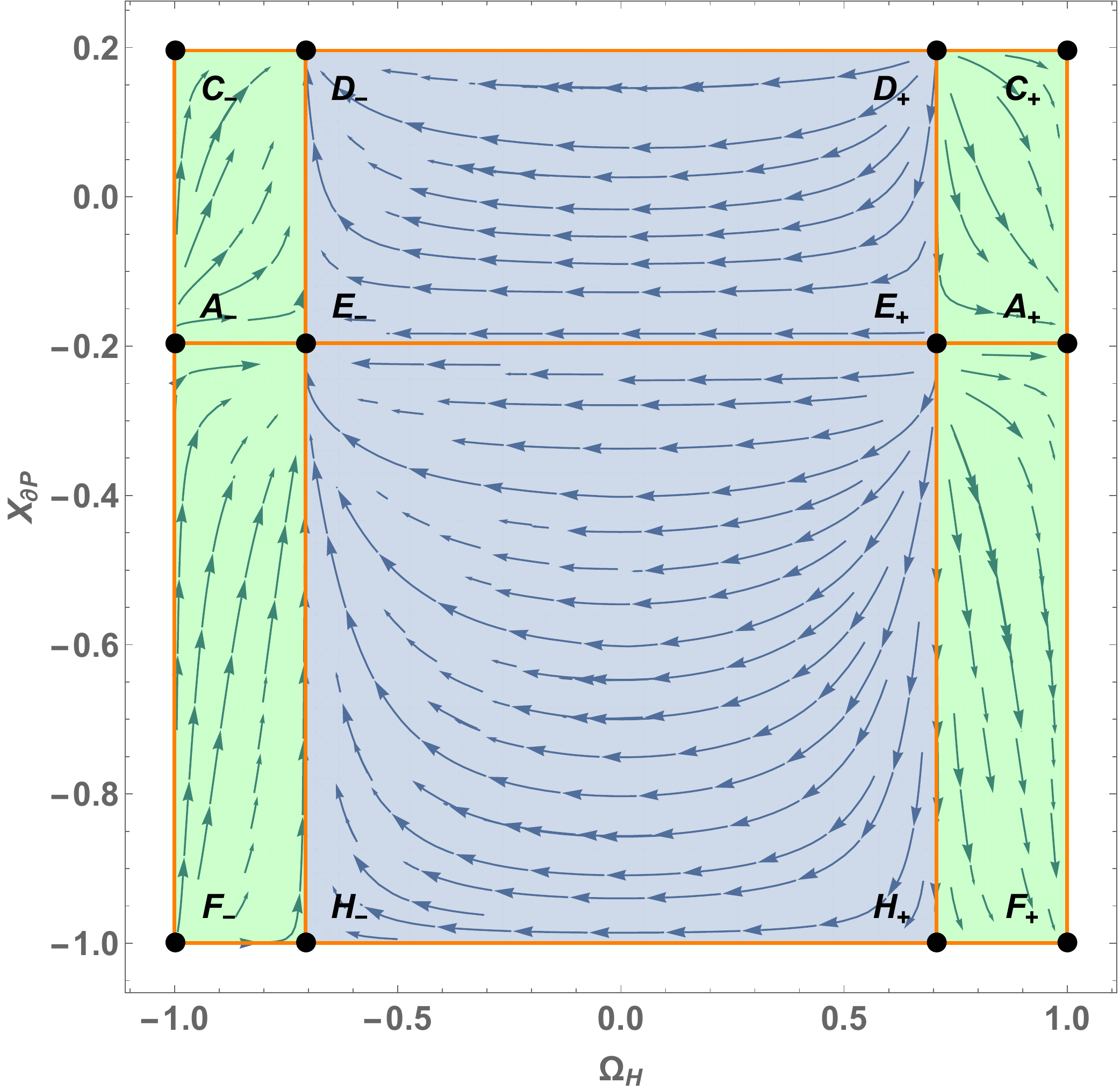}
    \caption{Invariant subsets $\lbrace \Omega_H,X_{\partial P}\rbrace$ for the quadratic EoS with $P_\star\neq 0$. The right panel corresponds to the positive spatial curvature and the left panel corresponds to the negative spatial curvature case. The invariant subsets are plotted for the parameters $\delta=-1$, $\sigma=1$, $\xi=-1$ and $\zeta=0.2$ (these figures are topologically similar to the case with parameters $\delta=-1$, $-1<\sigma<-\frac{1}{3}$, $\xi=-1$). The orange thick lines are the separatrices of the system, the blue region corresponds to $\Omega_\epsilon<0$ and the green shaded region corresponds to accelerated dynamics. This figure corresponds to Fig.~21 in \cite{quadratic}.}\label{fig:nD.5}
\end{figure*}

\begin{figure*}[ht]
    \centering
\includegraphics[width=0.43\textwidth]{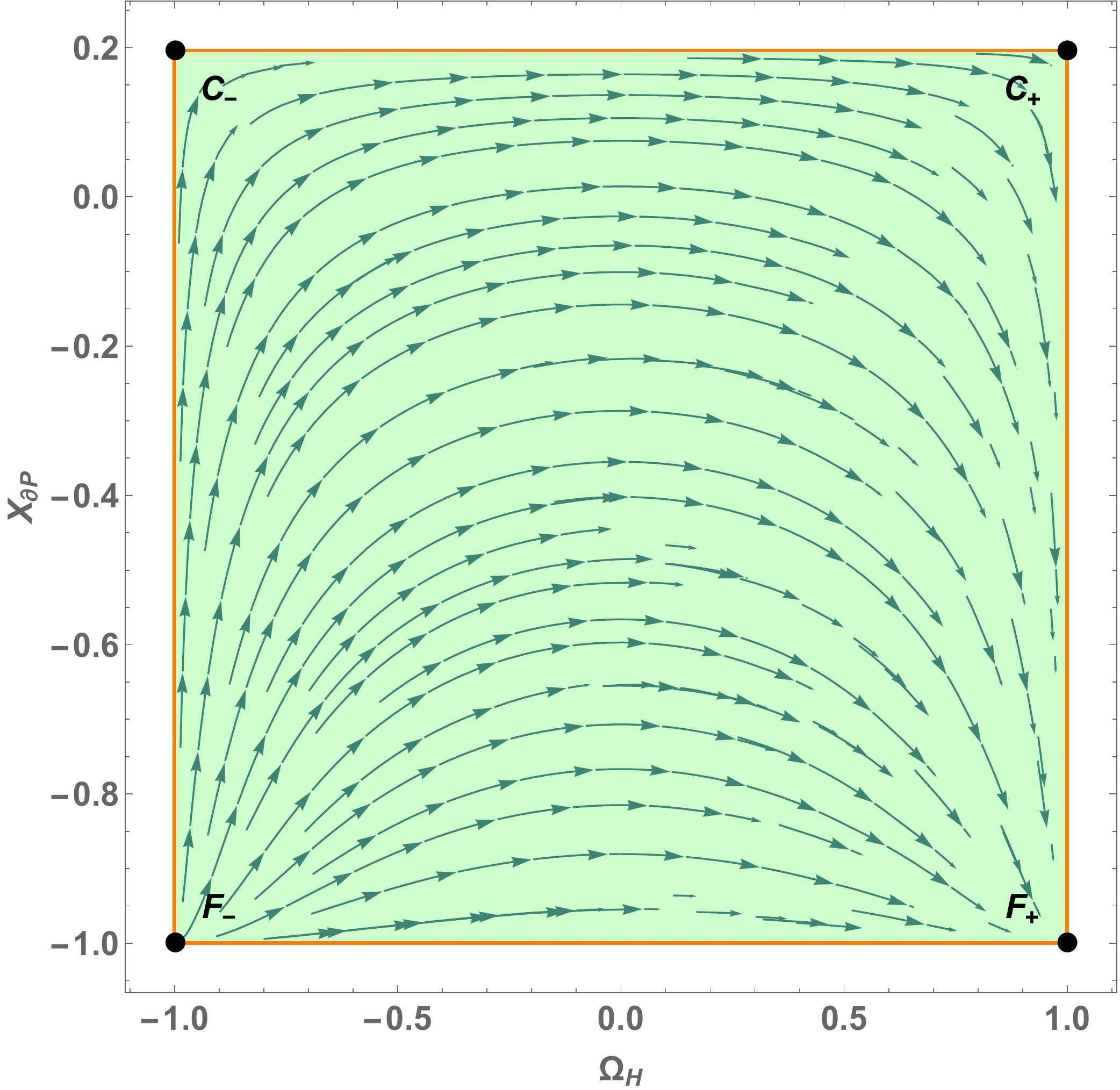}\quad
\includegraphics[width=0.43\textwidth]{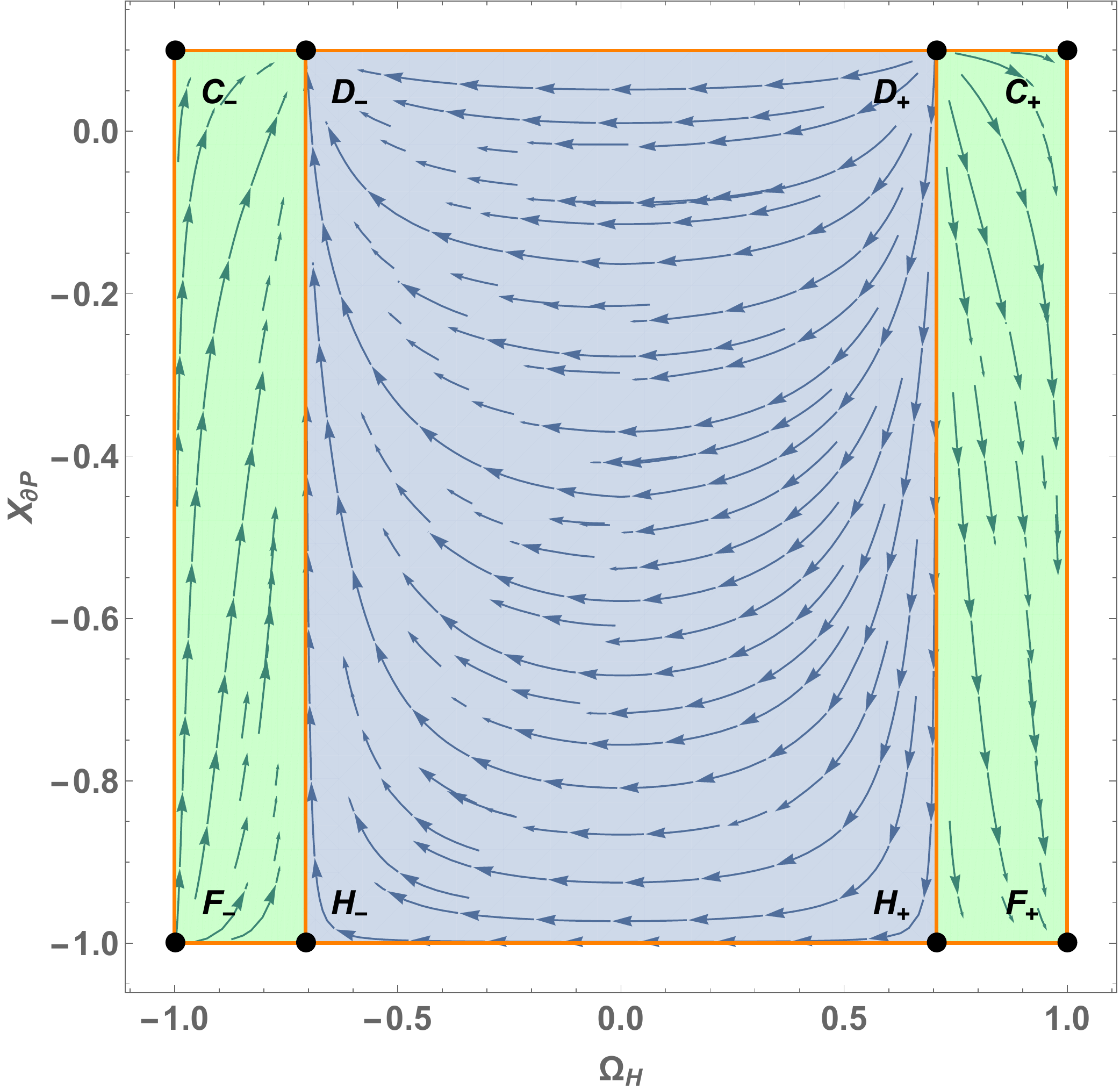}
    \caption{Invariant subsets $\lbrace \Omega_H,X_{\partial P}\rbrace$ for the quadratic EoS with $P_\star\neq 0$. The right panel corresponds to the positive spatial curvature and the left panel corresponds to the negative spatial curvature case. The invariant subsets are plotted for the parameters $\delta=-1$, $\sigma=1$, $\xi=-2$ and $\zeta=0.2$ (these figures are topologically similar to the cases with the parameters $\delta=-1$, $\sigma<-1$, $\xi<\sigma$  and also $\delta=-1$, $-1<\sigma<-\frac{1}{3}$, $\xi<-1$). The orange thick lines are the separatrices of the system, the blue region corresponds to $\Omega_\epsilon<0$ and the green shaded region corresponds to accelerated dynamics. This figure corresponds to Fig.~11 in \cite{quadratic}.}\label{fig:nD.6}
\end{figure*}

In Figs.~\ref{fig:pD.1}-\ref{fig:nD.6} we compare the results of our analysis in different parameter cases with respect to the ones obtained in \cite{quadratic}. Parameter cases that are topologically analogous are represented by a single figure for each case -- see Table~\ref{tab:gentab} and the captions of the respective figures for details.

\bibliographystyle{unsrt}
\bibliography{refs}
\end{document}